\documentclass[journal, 10pt]{IEEEtran}
\IEEEoverridecommandlockouts
\usepackage{booktabs} % For formal tables
\usepackage{url} % Referencing URLs in document
\usepackage{graphicx} % Add graphics to the document
\usepackage{amsmath,amssymb,amsfonts} % Insert equations
\usepackage{xcolor} % Import varied colors into the document
\usepackage[font=small,labelfont=bf]{caption} % Option to customize captions for figures, and floats
\usepackage{textcomp} % Used for baht, bullet (lists), copyright, etc.

\usepackage{dblfloatfix}
\usepackage{cite}
\usepackage{soul} % Used for spacing out, underlining, striking out, highlighting, etc/
\usepackage{enumitem} % Complete control of three list environments: enumerate, itemize, and description
\usepackage{lipsum} % Generates garbage text for checking
\usepackage[hang,flushmargin]{footmisc}
\usepackage{etoolbox}
\usepackage{float}
\usepackage{multirow}
\usepackage{hhline}
\usepackage{multicol}
\newlist{inlinelist}{enumerate*}{1}
\setlist[inlinelist]{label=(\roman*)}

\newcommand{\changed}[1]{\textcolor{black}{#1}}

\usepackage{algorithmicx}
\usepackage{algorithm}
\usepackage{algpseudocode}
\usepackage{wrapfig}

\algnewcommand\algorithmicinput{\textbf{Input:}}
\algnewcommand\Input{\item[\algorithmicinput]}

\algnewcommand\algorithmicconst{\textbf{Constraints:}}
\algnewcommand\Const{\item[\algorithmicconst]}

\algnewcommand\algorithmicoutput{\textbf{Output:}}
\algnewcommand\Output{\item[\algorithmicoutput]}

\algnewcommand{\algorithmicgoto}{\textbf{go to}}%
\algnewcommand{\Goto}[1]{\algorithmicgoto~\ref{#1}}%

\algrenewcommand\algorithmicindent{0.5em}

\renewcommand{\thefigure}{\arabic{figure}}

\usepackage{tikz}
\newcommand*\circled[1]{\tikz[baseline=(char.base)]{
		\node[shape=circle,draw,inner sep=1pt] (char) {#1};}}

\usepackage{array}
\newcolumntype{L}[1]{>{\raggedright\let\newline\\\arraybackslash\hspace{0pt}}m{#1}}
\newcolumntype{C}[1]{>{\centering\let\newline\\\arraybackslash\hspace{0pt}}m{#1}}
\newcolumntype{R}[1]{>{\raggedleft\let\newline\\\arraybackslash\hspace{0pt}}m{#1}}
\newcolumntype{M}[1]{>{\centering\arraybackslash}m{#1}}
\newcolumntype{O}[1]{>{\raggedleft\arraybackslash}m{#1}}
\usepackage{makecell}
\usepackage{diagbox}

\newcommand{\midtilde}{\raisebox{-0.27\baselineskip}{\textasciitilde}}

\usepackage[hidelinks, bookmarks=false]{hyperref}

\def\BibTeX{{\rm B\kern-.05em{\sc i\kern-.025em b}\kern-.08em
    T\kern-.1667em\lower.7ex\hbox{E}\kern-.125emX}}
\begin{document}

\title{\huge BioNetExplorer: Architecture-Space Exploration of Bio-Signal Processing Deep Neural Networks for Wearables}

\author{Bharath~Srinivas~Prabakaran\IEEEauthorrefmark{1},~\IEEEmembership{Student~Member,~IEEE,}
        Asima~Akhtar\IEEEauthorrefmark{1}, 
        Semeen~Rehman,
        Osman~Hasan,~\IEEEmembership{Senior~Member,~IEEE},
        and~Muhammad~Shafique,~\IEEEmembership{Senior~Member,~IEEE}% <-this % stops a space
\thanks{\IEEEauthorrefmark{1}~These two authors have contributed to this work equally.}
\thanks{B. S. Prabakaran and S. Rehman are with Technische Universit{\"a}t Wien\\ (TU Wien), Vienna, Austria.
\\E-mail: \{\href{mailto:bharath.prabakaran@tuwien.ac.at}{bharath.prabakaran}, \href{mailto:semeen.rehman@tuwien.ac.at}{semeen.rehman}\}@tuwien.ac.at.}% <-this % stops a space
\thanks{A. Akhtar and O. Hasan are with National University of Sciences and Technology (NUST), Islamabad, Pakistan.
\\E-mail: \{\href{mailto:aakhtar.msee17seecs@seecs.edu.pk}{aakhtar.msee17seecs}, \href{mailto:osman.hasan@seecs.edu.pk}{osman.hasan}\}@seecs.edu.pk.}% <-this % stops a space
\thanks{M. Shafique is with the Division of Engineering, New York University Abu Dhabi (NYU-AD), Abu Dhabi, United Arab Emirates.
\\E-mail: \href{mailto:muhammad.shafique@nyu.edu}{muhammad.shafique@nyu.edu}}% <-this % stops a space
\thanks{Manuscript received September 07, 2020; revised January 24, 2021.}% <-this % stops a space
\thanks{Copyright \copyright~2021 IEEE. Personal use of this material is permitted. However, permission to use this material for any other purposes must be obtained from the IEEE by sending a request to \href{mailto:pubs-permissions@ieee.org}{pubs-permissions@ieee.org}.}\vspace{-1cm}}

% \author{
%     \IEEEauthorblockN{
%     Bharath Srinivas Prabakaran\IEEEauthorrefmark{1}\textsuperscript{,}\IEEEauthorrefmark{3}\thanks{\IEEEauthorrefmark{3}~These two authors have contributed to this work equally.}, Asima Akhtar\IEEEauthorrefmark{2}\textsuperscript{,}\IEEEauthorrefmark{3}, Semeen Rehman\IEEEauthorrefmark{1}, Osman Hasan\IEEEauthorrefmark{2}, Muhammad Shafique\IEEEauthorrefmark{1}}
    
%     \IEEEauthorblockA{\IEEEauthorrefmark{1}Technische Universit{\"a}t Wien (TU Wien), Vienna, Austria
%     \\\{bharath.prabakaran, semeen.rehman, muhammad.shafique\}@tuwien.ac.at}
    
%     \IEEEauthorblockA{\IEEEauthorrefmark{2}National University of Sciences and Technology (NUST), Islamabad, Pakistan
%     \\\{aakhtar.msee17seecs, osman.hasan\}@seecs.edu.pk}
% \vspace{-1cm}}

\maketitle

\begin{abstract}
Deep Learning (DL) has been shown to be highly effective in solving various problems across numerous applications and domains, such as autonomous driving and image recognition. 
Due to the advent of DL, plenty of research works have explored the applicability of DL, more specifically Deep Neural Networks (DNNs), \changed{to solve pattern recognition and computer vision challenges.}
More recently, researchers have focused on the topic of automated generation and exploration of DNN architectures, which tend to mostly focus on image recognition or visual datasets, primarily, due to the computer vision-related DL advancements.

In this work, we propose the \textit{BioNetExplorer} framework to systematically generate and explore multiple DNN architectures for bio-signal processing in wearable devices.
Our framework \changed{varies} key neural architecture parameters to search for an embedded DNN architecture with a low hardware overhead, which can be deployed in wearable edge devices to analyze the bio-signal data and to extract the relevant information, such as arrhythmia and seizure. 
Furthermore, BioNetExplorer reduces the exploration time by deploying genetic algorithms, such as NSGA-II, SPEA-2, etc. 
Our framework also enables hardware-aware DNN architecture search by imposing user requirements and hardware constraints (storage, FLOPs, etc.) during the exploration stage, thereby limiting the number of networks explored. 
Moreover, \textit{BioNetExplorer} can also be used to search for DNNs based on the user-required output classes; for instance, a user might require a specific output class, attributed towards ventricular fibrillation, due to genetic predisposition or a pre-existing heart condition. 

The use of genetic algorithms reduces the exploration time, on average, by $9\times$, compared to exhaustive exploration.
We are successful in identifying Pareto-optimal designs, which can reduce the storage overhead of the DNN by \midtilde$30$MB for a quality loss of less than $0.5\%$. 
To enable low-cost embedded DNNs, \textit{BioNetExplorer} also employs different model compression techniques to further reduce the storage overhead of the network by up to $53\times$ for a quality loss of $<0.2\%$.

\end{abstract}


\bstctlcite{IEEEexample:BSTcontrol}

\section{Introduction}
\label{sec:Intro}

According to recent estimates from the International Data Corporation (IDC), by the year $2025$, $41.6$ billion IoT (Internet-of-Things) devices are expected to generate nearly $79.4$ Zettabytes ($\times10^{12}$ Gigabytes) of data annually~\cite{IDCGrowth}.
This includes the data collected by autonomous cars, home-automation environments, smart wearables, etc.
Recent estimates suggest that the number of smart wearables and fitness trackers in the market will rise to \midtilde$356$ million by the end of $2020$, with each device generating roughly $560$ Megabytes of data every month (averaging \midtilde$185$ Petabytes, monthly)~\cite{Cisco}.
These trends are expected to rapidly escalate due to the current Covid-19 outbreak.
In such situations, low-cost wearables can be widely used for personalized monitoring and prediction to contain any current/future pandemics~\cite{world2020coronavirus}~\cite{reutersGermany}.
The data collected from these smart edge-devices is typically transmitted to the cloud for filtering, processing, and
mining to extract valuable information, which can benefit the end-user or the system developer to provide health and lifestyle recommendations to improve the users' quality-of-life.
However, transmitting such a large amount of data to a fog device incurs significant time, power, and energy overheads on the wearable device~\cite{nia2015energy}. 
Besides the overheads associated with these systems, transmitting data to the fog or cloud may create security and privacy concerns for the users who might not wish to transfer their physiological information over untrusted networks or store their data on third-party cloud platforms and data-servers~\cite{stergiou2018secure}.

\begin{figure}[b]
    \centering
    \captionsetup{singlelinecheck=false}
    \includegraphics[width = \linewidth]{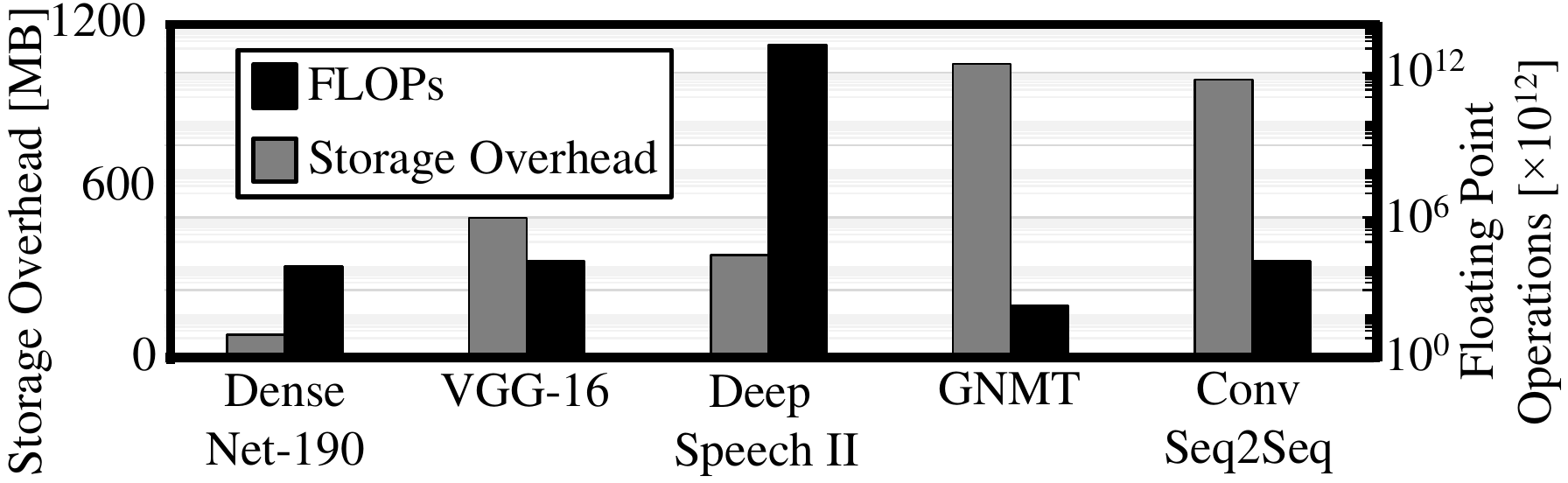}
    \caption{\textbf{Storage Overhead and Number of Floating Point Operations of State-of-the-Art Deep Neural Networks~(based on data from~\cite{amodei2016deep,wu2016google,li2018harmonious,huang2018condensenet,gehring2017convolutional}).} These networks have a very high hardware overhead and perform \midtilde$10^{12}$ floating point operations.}
    \label{fig:InitAnalysis}
\end{figure}

To address these issues, the \textit{Edge Computing} paradigm has emerged~\cite{mao2017survey}, which brings the computing layer closer to the sensor-nodes, \textit{i.e.}, the edge devices where the data is collected, to improve response times and to reduce energy consumption.
For example, such a system model is also adopted in the so-called Near-Sensor Computing paradigm~\cite{zhou2020near}.
However, these edge devices are typically not capable of executing compute- and memory-intensive data processing algorithms \changed{like} complex Deep Neural Networks (DNNs), which are currently used for various purposes such as classification, long-/short-term predictions, and anomaly detection in bio-signal data.
Fig.~\ref{fig:InitAnalysis} presents an analysis of the storage overhead and number of floating point operations required for a \textit{single} inference of different state-of-the-art DNNs~\cite{amodei2016deep,wu2016google,li2018harmonious,huang2018condensenet,gehring2017convolutional}.

\noindent \textit{From these results, one can make the following \textbf{observations}}:
\begin{itemize}[leftmargin=*]
    \item DNNs are application dependent, \textit{i.e.}, each application, based on its requirement, needs a specific trained DNN architecture that could be very different from the architecture used by another application. For example, DenseNet-$190$ is used for image classification, whereas Deep Speech-II is considered to be state-of-the-art for speech recognition;
    \item DNNs incur a large on-chip memory overhead for storing the network's parameters, which are required for executing the DNN;
    \item DNNs require a large number of floating-point operations to be executed in a second, to enable the system to work and process data in real-time;
    % \item the edge devices or wearables that are required to process these DNNs are resource-constrained and cannot meet the storage or computational requirements of these applications.
\end{itemize}

Based on these observations, the following \textbf{\textit{research challenges}} need to be addressed to enable embedded DNNs for resource-constrained wearables:
\begin{itemize}[leftmargin=*]
    \item Based on the number of DNN parameters and their possible values, the design space of DNNs can explode, leading to a very large number of networks in the architecture-space that need to be exhaustively trained and evaluated.
    This is a complex problem to solve; requiring a huge amount of resources for network search and training. These kinds of resources may only be affordable by large-scale enterprises. 
    The key scientific questions, that we target in this paper are:
    \begin{itemize}[leftmargin=2.5em]
        \item[\textbf{[Q1]}] \textit{How do we traverse the architecture-space of DNNs to effectively reduce the exploration time, while identifying DNNs that trade-off between output quality and hardware overhead?}
        \item[\textbf{[Q2]}] \textit{Can we effectively and simultaneously optimize both the output quality and the hardware overhead, instead of just one of these factors, while traversing the DNN architecture-space?}
    \end{itemize}
    \item Besides exploring the conventional DNN architecture-space, model compression techniques, such as Pruning and Quantization, can also be used to reduce the hardware overhead of the network. 
    In this regard, the question is:
    \begin{itemize}[leftmargin=2.5em]
        \item[\textbf{[Q3]}] \textit{How would different pruning and quantization techniques impact the quality of the bio-signal processing application and minimize the hardware overhead of the DNN?}
    \end{itemize}
\end{itemize}

To address these research challenges, we propose the following \textbf{\textit{novel contributions}}:
\begin{itemize}[leftmargin=*]
    \item We propose a novel \textit{BioNetExplorer} framework, which systematically varies the \changed{key} neural architecture parameters of a DNN to construct an architecture-space that can be subsequently explored and evaluated to identify networks that satisfy the user requirements and hardware constraints of the target platform.
    \item We investigate four well-established genetic algorithms to identify near-optimal points/designs without exhaustively exploring the complete search-space and study their outcome when applied \changed{to} our architecture-space.
    \item We propose a weighted cost function that can be used to simultaneously optimize output quality and hardware overhead. The weights allocated to these two factors will determine the primary optimization goal of the genetic algorithm-based search and its priority.
    \item Our \textit{BioNetExplorer} framework effectively constructs a new architecture-space of the near-optimal DNN by varying the key pruning and quantization parameters. It then explores this architecture-space to identify networks that minimize the hardware overhead with minimal or no quality loss.
\end{itemize}

\textbf{Open-source Contributions:} The complete framework along with the Pareto-optimal networks for all the explored DNN architecture-spaces will be made open-source and available online at \textcolor{blue}{\url{https://bionetexplorer.sourceforge.io}}.

\begin{figure}[h]
    \centering
    \captionsetup{singlelinecheck=false}
    \includegraphics[width = \linewidth]{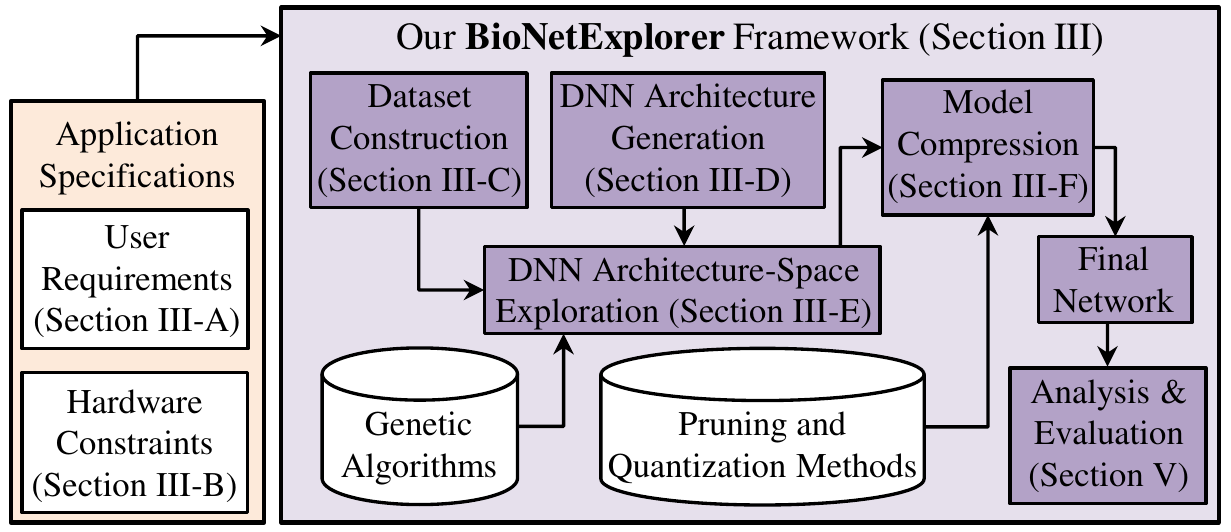}
    \caption{\textbf{An Overview of the Proposed Contributions (Dark Highlighted Boxes) in Our \textit{BioNetExplorer} Framework.}}
    \label{fig:UCDiag}
\end{figure}

Fig.~\ref{fig:UCDiag} illustrates an overview of the proposed contributions integrated in our novel \textit{BioNetExplorer} framework.

\section{Related Work \& Preliminaries}
\label{sec:RW}

Wearables are highly pervasive smart electronic devices that are used to improve the \changed{users' quality-of-life}.
This includes a \changed{wide range} of devices like smart-watches, such as the ones manufactured by Apple~\cite{AppleWatch} and Samsung~\cite{SamsungWatch}, fitness trackers, like Fitbit~\cite{Fitbit}, \changed{sports} watches, smart clothing, bio-implants, etc.
One of the most important functions offered by these devices is physiological signal monitoring, which can be used to track the health of its users~\cite{seneviratne2017survey}. 
For example, the Apple Watch and Fitbit can be used to monitor the electrical signals \changed{associated with a heartbeat} and record them in the form of an electrocardiogram (ECG).
The data gathered from these devices are transmitted to the cloud for further processing and detecting anomalies such as \textit{Atrial Fibrillation}~\cite{turakhia2019rationale}.
There has been a wide range of research works that have focused on the adoption of wearables for monitoring the user's health~\cite{fu2018blood,sopic2018real,amirshahi2019ecg}.
For instance, Sun~\textit{et~al.}~\cite{sun2020lightweight} have proposed an optimized vector transformation approach that reduces the overhead of key generation, thereby enabling lightweight encryption and decryption in smart-health IoT infrastructure.
Similarly, to enable real-time feedback to the user, a coding-free control system for an Industrial IoT (IIoT) infrastructure was presented in~\cite{liu2019wireless}.

There has been a wide range of works on Neural Architecture Search (NAS), from the use of basic algorithms to Reinforcement learning for the sole purpose of designing DNNs. 
A comprehensive survey of the neural architecture search approaches based on the search space, \changed{the} strategy of search, and performance of the approach is presented in~\cite{elsken2018neural}.
One of the most popular techniques for hyper-parameter optimization is Bayesian Optimization (BO)~\cite{snoek2012practical}.
However, it has not yet been extensively explored for NAS.
This is mainly because the BO toolboxes are based on Gaussian processes and tend to focus on low-dimensional continuous optimization problems, and is, thus, not suitable for architecture-space exploration~\cite{elsken2018neural}.
A recurrent neural network (RNN), which can be used to generate DNN architectures for classifying objects in the CIFAR-10 image dataset, is presented in~\cite{zoph2016neural}. The RNN is trained using a reinforcement learning approach to maximize the accuracy of the networks generated.
Liu \textit{et al.}~\cite{liu2018progressive} presented a sequential model-based optimization strategy, which can be used to generate convolutional neural networks $5\times$
as efficiently as the approach presented in~\cite{zoph2016neural}.
In $2019$, Liu \textit{et al.}~\cite{liu2019towards} proposed a \changed{network-level} structure search combined with the traditional cell-level structure search to compose a hierarchical architecture space of state-of-the-art DNNs.
Typically, the state-of-the-art approaches for NAS focus on image classification and image segmentation~\cite{deng2009imagenet} without much focus on other application domains such as bio-signal processing. 
Recently, Tan \textit{et al.}~\cite{tan2019mnasnet} proposed a platform-aware neural architecture search approach, which focuses solely on image classification and object recognition.
Similarly, several other recent works presented platform-aware DNN architecture-space exploration approaches that are targeted for image classification~\cite{wu2019fbnet,dai2019chamnet,howard2019searching}.
These approaches do not offer the same accuracy-hardware trade-offs for other application domains, such as bio-signal processing, as illustrated in Section~\ref{subsec:SoA}.

Typically, \textit{Model Compression} techniques such as \textit{Pruning} and \textit{Quantization} have been shown to be very effective in reducing the hardware overheads and computational requirements of DNNs.
Han \textit{et al.}~\cite{han2015deep} proposed a layer-wise approach to prune and quantize DNNs that are used for image classification.
Anwar \textit{et al.}~\cite{anwar2017structured} presented a technique that can be used to introduce structured sparsity in the network to reduce the representation effort and maximize parallel computation.
As an alternative approach to structured pruning, Luo \textit{et al.}~\cite{luo2017thinet} presented a method that eliminates complete filters based on data from subsequent layers to reduce the number of computations required for inference and accelerate DNN execution.
Marchisio \textit{et al.}~\cite{marchisio2018prunet} improved upon the work presented in~\cite{han2015deep} by introducing an iterative class-blind approach for pruning weights in DNNs to further reduce the hardware overhead of the network.
Lin \textit{et al.}~\cite{lin2017runtime} proposed a framework that can be used to prune a DNN dynamically at runtime, based on the input image and feature maps,  to preserve the full ability of the original DNN.
Similarly, there is a wide range of techniques focused on quantizing DNNs to reduce their hardware overhead~\cite{wang2019haq,jacob2018quantization,zhu2016trained,zhang2018lq,jain2018compensated}.
Besides NAS and model compression, Kumar \textit{et al.}~\cite{kumar2017resource} and Gural \textit{et al.}~\cite{gural2019memory} presented effective ML-based approaches that can be deployed on resource-constrained IoT devices to perform search predictions and image classification.

In this work, we primarily focus on the automatic generation and exploration of bio-signal processing DNNs that can be deployed on resource-constrained wearable devices.
Plenty of research works have used \textit{manually-designed} DNNs (\textit{i.e.}, not obtained through automatic neural architecture search) for classification of ECG signal components and/or arrhythmia~\cite{xia2018automatic,yildirim2018arrhythmia,hannun2019cardiologist}.
The DNN architecture proposed by Hannun \textit{et al.}~\cite{hannun2019cardiologist} can classify $12$ rhythm classes on their custom ECG dataset and is considered to be the state-of-the-art for arrhythmia classification.
Our \textit{BioNetExplorer} framework adopts the basic building block used in~\cite{hannun2019cardiologist} to generate and explore DNN architectures for specific arrhythmia categories, as per user requirement, which can be deployed on wearables like smart-watches or fitness trackers.
Note, we have to modify the input and output layers of this DNN to ensure that the network can process data from the available training dataset and classify them into the required output classes based on the target application. 
We have also included a \textit{Long Short-Term Memory} (LSTM)~\cite{hochreiter1997long} layer at the end of the DNN to improve accuracy in worst-case scenarios, \textit{i.e.}, lower number of feature extraction layers.
Our ECG processing DNN architecture, modified from the DNN presented in~\cite{hannun2019cardiologist}, is illustrated in Fig.~\ref{fig:ECGDNNSOA}.
Given such a DNN architecture, the challenge is to explore the following architectural parameters:
\begin{inlinelist}
    \item \textit{Number of ResNet Blocks};
    \item \textit{Number of Filters}; and
    \item \textit{Number of LSTM Cells}.
\end{inlinelist}
This requires us to generate and explore different DNN variants considering these network parameters to identify a set of Pareto-optimal configurations for the DNN architecture.
This provides a low-complexity solution instead of employing a full-scale NAS from scratch.

\setcounter{figure}{2}
\begin{figure}[h]
    \centering
    \captionsetup{singlelinecheck=false}
    \includegraphics[width = \linewidth]{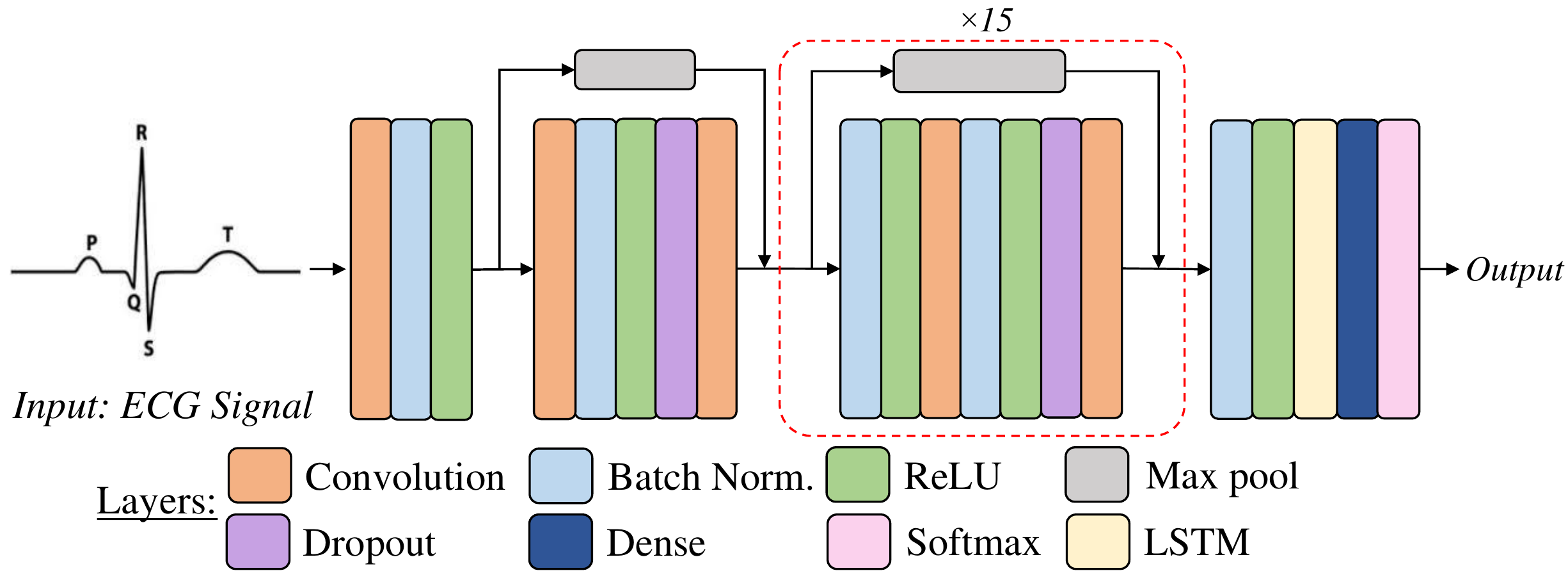}
    \caption{\textbf{Our ECG Processing DNN Architecture to Detect Anomalies (modified from the DNN presented in~\cite{hannun2019cardiologist}).}}
    \label{fig:ECGDNNSOA}
\end{figure}

\setcounter{figure}{3}
\begin{figure*}[!t]
    \centering
    \captionsetup{singlelinecheck=false}
    \includegraphics[width = \linewidth]{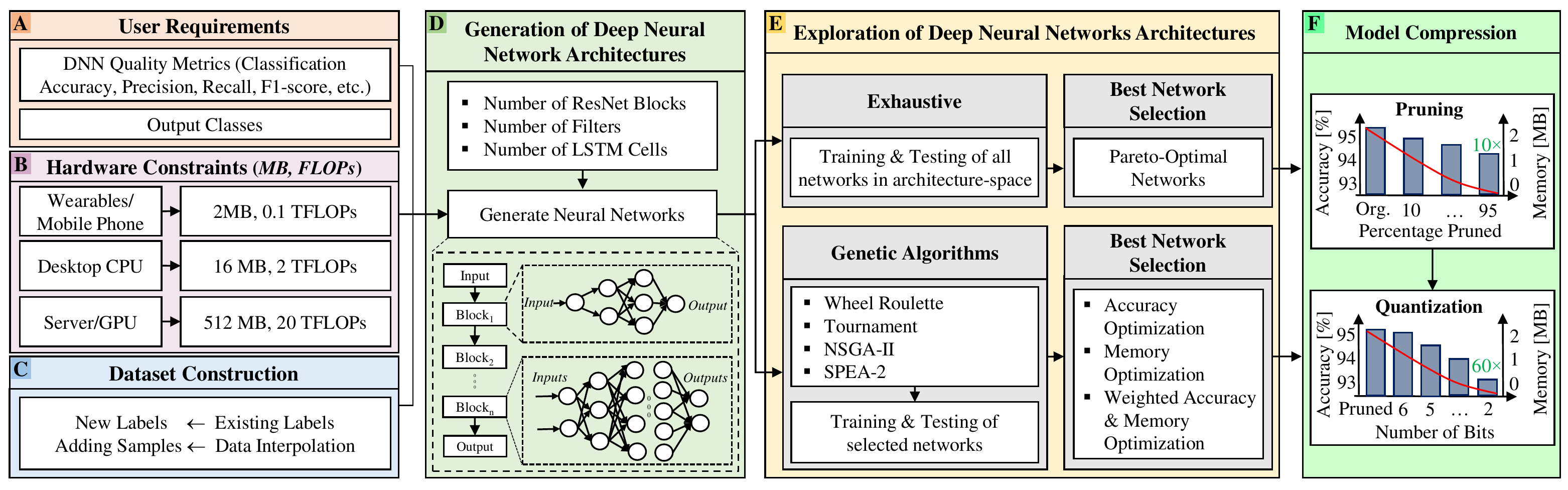}
    \caption{\textbf{An Overview of Different Key Steps in Our \textit{BioNetExplorer} Framework.} We consider the applications' quality requirements and hardware constraints of the system as inputs to our framework. Based on these requirements, we construct and explore the architecture-space of DNNs. These DNNs are trained using a custom dataset, which we construct for the target application from existing open-source datasets. Afterwards, the DNNs are further compressed to obtain a DNN with the minimum-possible hardware overhead and maximum-possible output quality, under the imposed requirements and constraints.}
    \label{fig:Meth}
\end{figure*}

% \begin{itemize}
%     \item \todored{TODO}
%     \item NAS
%     \item Healthcare-DNNs
%     \item Hardware associated Problems oriented Research
% \end{itemize}
\section{The BioNetExplorer Framework}
\label{sec:Meth}

An overview of our \textit{BioNetExplorer} framework is presented in Fig.~\ref{fig:Meth}.
It considers the user requirements (\textit{e.g.}, output classes and quality) to construct and label the comprehensive bio-signal data-set, and hardware constraints (\textit{e.g.}, \changed{Floating-point} Operations (FLOPs) and memory storage overhead (MB)) to generate and explore the architecture-space of bio-signal processing DNNs for wearables.

% \subsection{User Requirements, Hardware Constraints, \& Dataset Construction}

\subsection{User Requirements}
\label{subsec:User}
The \textit{BioNetExplorer} framework restricts the output classes of the DNN being explored.
For instance, in case the user requires another output class besides the two base cases of normal and anomalous, due to pre-existing conditions, another output class and the associated label can be included by our framework to search for an efficient DNN architecture in the new architecture-space.
Note, the labeled data associated with the new output class has to be present in the dataset.
We consider four metrics for evaluating the quality ($Q_{DNN}$) of DNNs, namely, \textit{Accuracy}~($A_{DNN}$), \textit{Recall} ($R_{DNN}$), \textit{Precision} ($P_{DNN}$), and \textit{F1-score} ($F1_{DNN}$).
These metrics are defined as:
\begin{equation}
\label{eq1}
    A_{DNN} = \frac{\text{True Positives + True Negatives}}{\text{Total Number of Classifications}}
\end{equation}
\begin{equation}
\label{eq2}
    R_{DNN} = \frac{\text{True Positives}}{\text{True Positives + False Negatives}}
\end{equation}
\begin{equation}
\label{eq3}
    P_{DNN} = \frac{\text{True Positives}}{\text{True Positives + False Positives}}
\end{equation}
\begin{equation}
\label{eq4}
    F1_{DNN} = 2*\frac{\text{Precision * Recall}}{\text{Precision + Recall}}
\end{equation}
\textit{Recall} is the ability of the DNN under consideration to find all relevant cases of a specific class within the dataset, while \textit{Precision} defines the ability of the DNN to identify relevant points in the dataset. \textit{F1-score} is defined as the harmonic mean of \textit{Precision} and \textit{Recall}.
\textit{The user can define a minimum quality constraint using any of the aforementioned metrics to ensure that the final network achieved by the exploration satisfies this constraint}.

\subsection{Hardware Constraints}
\label{subsec:HW}
Besides the quality constraint, a hardware-level constraint on the networks being explored ensures that these DNNs do not require compute or memory resources beyond the configuration of the target wearable platform.
We can estimate the hardware requirements of a DNN using metrics such as \textit{Storage Overhead} ($S_{DNN}$) based on the number of parameters (weights and biases) of the network, and \textit{\changed{Floating-point} Operations} ($F_{DNN}$) that determines the execution time of the DNN on the target platform.
Due to the hardware constraints of the wearable device, an efficient DNN that can be executed on such devices needs to offer the \changed{best possible} quality with minimum-possible $S_{DNN}$ and $F_{DNN}$.
This typically creates a trade-off, wherein the quality of the DNN typically decreases when we reduce the number of parameters to curtail the memory footprint/FLOPs, and vice-versa.

\subsection{Dataset Construction}
\label{subsec:Data}
Our framework constructs a dataset for each specific DNN exploration by combining labels and creating the desired output classes from the primary bio-signal dataset.
Each of the labels/annotations in the primary dataset should correspond to one of the output classes of the explored DNN in the final constructed dataset.
For example, in our case study, we create a new dataset by grouping together a $256$ bio-signal sample window and assigning a specific label to this window based on the original labels of the signals present.
Each of the $41$ \textit{beat} and \textit{non-beat} annotations in the dataset are eventually classified into one of the output labels in each DNN exploration. 
We use $70\%$ of the constructed datasets for training the DNN architecture, $10\%$ for validating the network, and the remaining $20\%$ for testing and estimating the output quality of the DNN.
We have explained this scenario with help of a few different examples in Section~\ref{sec:ECG}.

\subsection{Generation of DNN Architectures}
\label{subsec:GenDNN}
After collecting information regarding all the relevant constraints and requirements that can be used to restrict the architectural-space, and the dataset required for the target application, we generate the set of possible DNN architectures ($\psi$) by varying three key \changed{neural architecture parameters}, namely,
\begin{inlinelist}[label=(\arabic*)]
    \item \#ResNet Blocks, 
    \item \#Filters, and 
    \item \#LSTM Cells.
\end{inlinelist}
These three parameters can potentially take up any value in the domain of ${\rm I\!R^{+}}$, thereby, creating an unbounded design space that is impossible to explore.
Therefore, in this work, we consider the parameters of the current state-of-the-art DNN architectures as the upper-bounds to our architecture-space exploration.
% Similarly, to reduce the number of architectures explored, we explore a subset of the values that can be taken up by these parameters.
Precisely, in our case-study, the ranges of these parameters are as follows:
\begin{enumerate}[leftmargin=*]
    \item The \#ResNet Blocks~\cite{he2016identity} can vary from $0$ to $15$, with each block being composed of two 1-D convolution layers, followed by batch normalization~\cite{ioffe2015batch}, Rectified Linear Unit (ReLU) activation, and dropout~\cite{srivastava2014dropout};
    \item The \#Filters (size 16) in each convolution layer is defined as $32\times2^y$ where the value of $y$, starting from $0$, is incremented by $1$ after every $x$ ResNet Blocks where $x \in [1,4]$; and
    \item The \#LSTM cells can vary as a function of $2^z$ where $z \in [4,8]$.
\end{enumerate}
Due to the upper-bound on the number of parameters and storage size imposed by the state-of-the-art networks, we can reduce the number of networks to be explored to realistic limits.
For example, in our case study (see Section~\ref{sec:ECG}), the number of explored networks is reduced from $320$ to $135$.
It is important to note that, since the exploration of the architecture-space is limited by the current generation of state-of-the-art architectures, any major upgrades in the architectures of the state-of-the-art will, in turn, affect the architecture-space of the explored DNNs.

\subsection{Exploration of DNN Architectures}
\label{subsec:ExploreDNN}
Each of the DNNs obtained from varying the architectural parameters \changed{needs} to be trained using the constructed bio-signal data-set to detect anomalies and/or specific conditions.
Since training is a compute-intensive task, it is important to 
\begin{inlinelist}
    \item generate fewer DNN models (as discussed above), and 
    \item explore the search-space fast.
\end{inlinelist}
At this stage, there are two ways to explore the architecture-space:

(1) \textbf{Exhaustive Search}, which involves \textit{training all the DNNs in the architecture-space} to identify the set of Pareto-optimal DNN architectures that trade-off QoS for $S_{DNN}$/$F_{DNN}$ and vice-versa. 
Training all the networks in the architecture-space requires hundreds of GPU-hours depending upon the size of the networks being explored.
Due to the practical constraints imposed by hardware resources present on wearables and the upper bound imposed by state-of-the-art DNNs, we have a curtailed architecture-space, for which we can perform an exhaustive search.
Therefore, for benchmarking the efficiency of the genetic algorithms and random search, we additionally implemented the exhaustive search.
However, when the network complexity, number of parameters, parameter instances, and the number of hyper-parameters increase, thereby leading to exponentially large architecture-spaces, we usually rely on genetic-algorithm based architecture-search methods.

% due to practical constraints on XYZ and XYZ, we already have a curtailed design space, for which we can do exhaustive search in TODO hours. Therefore, for benchmarking the efficiency of genetic algorithms and quality of optimal solutions, we additionally implemented the exhaustive search. However, when the network complexity and the number of hyper-parameters grow, we have to rely on the genetic-algorithm based DSE only."
(2) \textbf{Genetic Algorithm-Based Search}, which involves effectively \textit{training a small subset of the networks in the architecture-space} to reduce the training time to only tens of GPU-hours.
Genetic algorithms have proved to be quite effective in utilizing the given cost function to obtain near-optimal solutions while reducing the time required for exploration~\cite{sastry2005genetic}.
Towards this, we leverage the Genetic Algorithms that rely on the biological principles of reproduction and evolution to create a new generation of networks that can potentially perform better than the previous generation to optimize the cost function.
The scope of our work is limited to the investigation of genetic algorithms, a small subset of meta-heuristics that encompasses other techniques like Tabu Search~\cite{glover1998tabu}, Simulated Annealing~\cite{van1987simulated}, Ant Colony Optimizations~\cite{dorigo2006ant}, etc., for the exploration of our DNN architecture-spaces.

\setcounter{figure}{4}
\begin{figure}[t]
    \centering
    \captionsetup{singlelinecheck=false}
    \includegraphics[width = 0.85\linewidth]{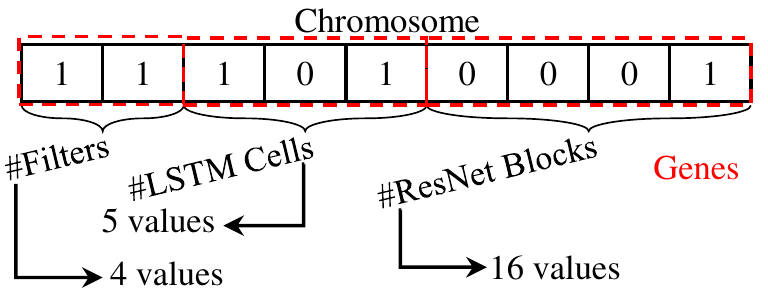}
    \caption{\textbf{Overview of our Constructed Chromosome.} The three network parameters are encoded as genes, which are combined together to form our chromosome. The number of bits for each gene is selected based on the possible number of values for each parameter. This leads to $2^9$ possible DNN architectures.}
    \label{fig:GeneLen}
\end{figure}

\setcounter{figure}{5}
\begin{figure}[b]
    \centering
    \captionsetup{singlelinecheck=false}
    \includegraphics[width = \linewidth]{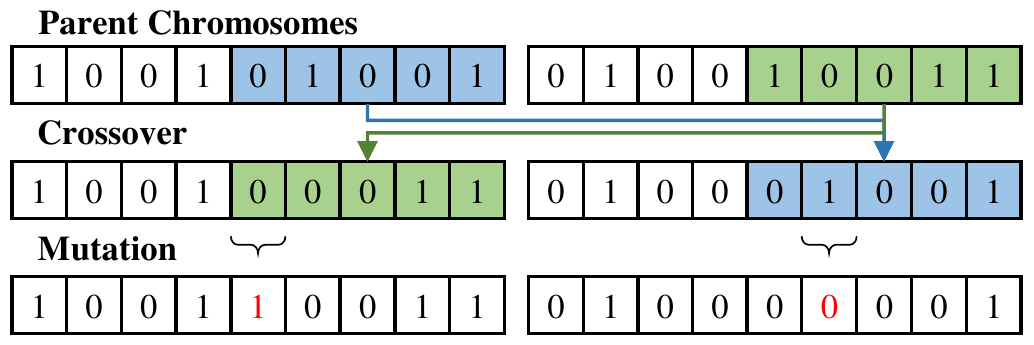}
    \caption{\textbf{Example of Crossover and Mutation in our Chromosome.}}
    \label{fig:Cross}
\end{figure}

Initially, we start with a set of random individuals/DNNs in the architecture-space, which we refer to as the \textit{Initial Population}. 
Based on our experiments and recommendation from previous works~\cite{reeves2002genetic}, we have identified that a population size of $30$ offers the best results in terms of designs obtained for our architecture-space $\psi$.
Each of the three key network parameters (\#Filters, \#ResNet Blocks, \#LSTM cells) that can be varied in the DNN architecture-space is encoded as a \textit{Gene}. These genes are joined together in the form of a string to generate a \textit{Chromosome}, which can be decoded to construct the DNN architecture.
Based on the number of possible values for each gene, our chromosome is a binary string of length $9$, as illustrated in Fig.~\ref{fig:GeneLen}.
The next step is to evaluate the viability of an individual (\textit{i.e.}, a DNN architecture) based on its ability to compete with the other individuals. 
This is known as the \textit{Fitness Value}, which is subsequently evaluated for each individual in the population and depends on the cost function $\phi$.
The fitness value is computed either as the cost function in scenarios where the decoded individual (\texttt{ARCH}) corresponds to a DNN in the existing architecture-space $\psi$ or is taken to be $0$ otherwise and discarded from the search.
For example, even though the \#LSTM cells is encoded using $3$-bits, the gene only has $5$ possible values. This leads to $3 \times 2^5$ potential DNN architectures that are not present in the architecture-space $\psi$, \changed{which} can be immediately discarded.
Next, through a process called \textit{Selection}, we pick two individuals, based on their fitness values, and have them pass their genes to the next generation.
At this point, two important factors come into picture, namely \textit{Crossover} and \textit{Mutation}.
For each pair of parents selected for mating, an ordered crossover occurs with a probability of $0.4$, and a crossover point in the parents' chromosomes are selected at random.
The offspring are created by exchanging the genes of parents among themselves from the start of the chromosome until the crossover point. 
The new offspring generated by this crossover mechanism are added to the population of the next generation.
Furthermore, the offspring's genes can undergo a shuffle-index mutation with a probability of $0.11$, which would flip the bit, thereby introducing ``diversity'' in the population and prevent premature convergence.
An example illustrating the mechanism of crossover and mutation is depicted in Fig.~\ref{fig:Cross}.
We determine a population of size $30$, based on the individuals' fitness values, and re-iterate to produce $5$ generations of offspring to explore and identify the individuals that offer the best fitness value.
Fig.~\ref{fig:GA} presents a flow chart depicting the stages for our genetic algorithm-based DNN architecture-space exploration.
% Based on recommendations from previous works~\cite{reeves2002genetic} and initial experiments, we have observed that the probability values used in this framework offer the best exploration results.
% Similarly, even though the \#ResNet blocks is encoded using $4$-bits, it does not utilize all $4$-bits for all configurations of \#Filters, as discussed in the earlier section.
% In such scenarios, when the encoded the DNN architecture, does not correspond to an \texttt{ARCH} in $\psi$, the fitness value of the \texttt{ARCH} is determined to be $0$ and discarded from the search.

% The Genetic Algorithm-based approach identifies near-optimal DNNs that offer similar QoS to the Pareto-optimal DNNs with an insignificant increase in $S$/$F$ and vice-versa.s

\setcounter{figure}{6}
\begin{figure}[t]
    \centering
    \captionsetup{singlelinecheck=false}
    \includegraphics[width = 0.95\linewidth]{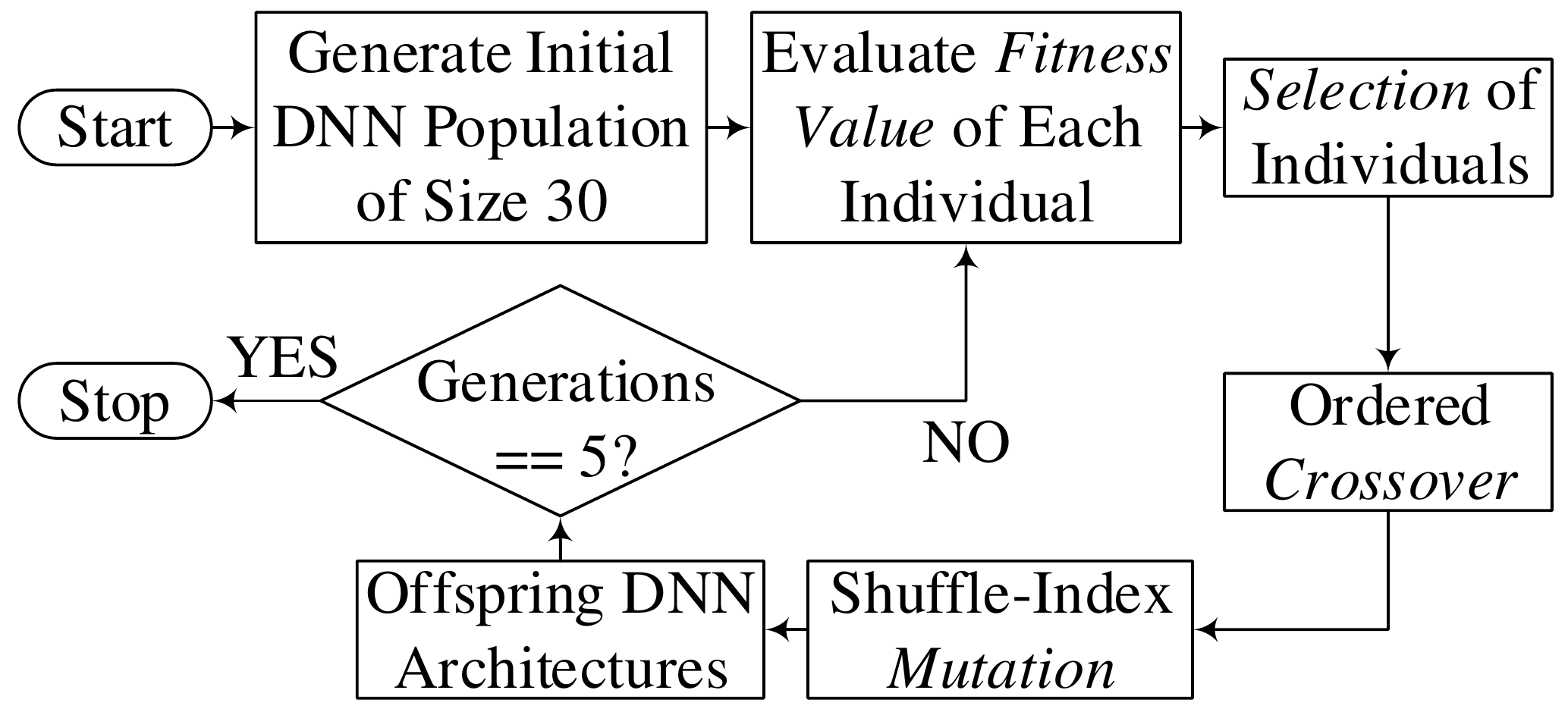}
    \caption{\textbf{Flowchart Depicting the Stages for our Genetic Algorithm-Based DNN Architecture-Space Exploration.} \textit{BioNetExplorer} evaluates the \textit{fitness value} of each individual by training them to determine the cost function $\phi$, if they are a part of the architecture-space $\psi$, and presetting the others to zero.}
    \label{fig:GA}
\end{figure}

Without loss of generality, in this work, we consider 4 well-known genetic algorithms for exploring our architectural-space, namely, Roulette Wheel~\cite{zames1981genetic}, Tournament Selection~\cite{miller1995genetic}, NSGA-II~\cite{deb2002fast}, and SPEA-2~\cite{zitzler2001spea}.
The time complexity of our \textit{BioNetExplorer} framework depends on the size of the architecture-space ($N$) and the computational complexity of the genetic algorithms used for exploring the architecture-space.
Roulette Wheel has a time complexity of $O(N*\log N)$ when using binary search in the selection process. Similarly, Tournament Selection, NSGA-II, and SPEA-2 have a time complexity of $O(N)$, $O(N^2)$, and $O(N^2*\log N)$, respectively.
Since each of these algorithms select and train different DNN architectures as part of their exploration, we will also compare the efficiency of these algorithms, as illustrated in Section~\ref{sec:ECG}.
These algorithms require us to propose a cost function ($\phi$), which the genetic algorithm can optimize (minimize/maximize) to obtain a set of near-optimal DNN architectures ($\omega$) for the given exploration.
To ensure that the obtained design offers the best quality ($Q_{DNN}$) with minimum storage overhead ($S_{DNN}$) or $F_{DNN}$, we propose the following weighted cost functions:
\begin{equation}
\label{eq5}
    \phi = \alpha*Q + \beta*\left[1-\frac{S_{DNN}}{S_{MAX}}\right]
\end{equation}
where $\alpha \in [0,1]$ and $\beta \in [0,1]$ denote the weights for quality $Q_{DNN}$ and storage overhead $S_{DNN}$, respectively, and $S_{MAX}$ denotes the maximum storage overhead of the state-of-the-art DNN architecture.
Without loss of generality, in this work, we model the hardware constraint as the storage overhead of the DNN. Similarly, $F$ and $Energy$ of DNN can also be modeled and constrained.
The system designer can specify the importance of the two factors by varying the weight parameters ($\alpha,\beta$) and if both are equally important, then we set both values to be equal.
The outcome of this weighted optimization technique is illustrated in Section~\ref{sec:ECG}.
The Weighted DNN Architecture Search is depicted in Algorithm~\ref{Algo1}, where the \textit{ExplorationAlgorithm} (line~\ref{Mark}) function call employs the search algorithm (exhaustive or one of the four aforementioned genetic algorithms) under the quality ($Q_{Const}$) and hardware constraints ($S_{Const}$).
An overview of the symbols and denotations used in our work is provided in Table~\ref{tab:Symbols}.

\begin{algorithm}[t]
    \normalsize
	\caption{Weighted DNN Architecture Search}
	\label{Algo1}
	\begin{algorithmic}[1]
	\Input $\psi,~\alpha,~\beta,~S_{MAX}$
	\Const $Q_{Const},~S_{Const}$
	\Output $\omega$
	\State $S_{DNN} = [];$
	\For{\texttt{ARCH} \textbf{in} $\psi$}
	\If{$StorageOverhead(\texttt{ARCH})\leq S_{Const}$}
	\State $S_{DNN}.append(StorageOverhead(\texttt{ARCH}));$
	\Else
	\State $\psi.remove(\texttt{ARCH});$
	\EndIf
	\EndFor
	\State $\phi = \alpha*Q_{DNN} + \beta*\left[1-\frac{S_{DNN}}{S_{MAX}}\right]$
	\State $\omega=ExplorationAlgorithm(\phi,\psi);$\label{Mark}
	\For{\texttt{DNN} \textbf{in} $\omega$}
	\If{$Q_{DNN} < Q_{Const}$}
	\State $\omega.remove(\texttt{DNN});$
	\EndIf
	\EndFor
	\end{algorithmic}
\end{algorithm}

\begin{table}[h]
    \centering
    \caption{\protect\centering\textbf{Symbols and Denotations}}
    \begin{tabular}{|C{1.5cm}|L{4.6cm}|}
        \cline{1-2}
        \textbf{Symbol} & \multicolumn{1}{c|}{\textbf{Denotation}} \\ \hline
        $Q_{DNN}$ & Quality of the DNN \\ \hline
        $Q_{Const}$ & Quality Constraint of the Application \\ \hline
        $A_{DNN}$ & Accuracy of the DNN (Eq.~\ref{eq1}) \\ \hline
        $R_{DNN}$ & Recall of the DNN (Eq.~\ref{eq2}) \\ \hline
        $P_{DNN}$ & Precision of the DNN (Eq.~\ref{eq3}) \\ \hline
        $F1_{DNN}$ & F1-score of the DNN (Eq.~\ref{eq4}) \\ \hline
        $S_{DNN}$ & Storage Overhead of the DNN \\ \hline
        $S_{MAX}$ & Maximum Storage Overhead of State-of-the-Art DNN Architecture \\ \hline
        $S_{Const}$ & Storage Overhead Constraint of the Target Platform \\ \hline
        $F_{DNN}$ & Number of \changed{Floating-point} Operations Reqd. for the DNN \\ \hline
        $\psi$ & Set of possible DNN Architectures (Architecture-space) \\ \hline
        \texttt{ARCH} & DNN in $\psi$ \\ \hline
        $N$ & Size of the Architecture-space $\psi$ \\ \hline
        $\phi$ & Cost Function for the Genetic Algorithms (Eq.~\ref{eq5})\\ \hline
        $\omega$ & Set of Near-optimal DNN Architectures (Output of Genetic Algorithm) \\ \hline
        $\alpha$ & Weight for Output Quality $Q_{DNN}$ in the cost function $\phi$ \\ \hline
        $\beta$ & Weight for Hardware Overhead $S_{DNN}$ (or $F_{DNN}$) in the cost function $\phi$ \\ \hline
        \end{tabular}
    \label{tab:Symbols}
\end{table}

\subsection{Model Compression}
Besides the use of architecture-space exploration to identify networks with similar output quality and low hardware overhead, we include the ability to further compress the network size using model compression techniques, like Pruning and Quantization, in our framework, and study their impact on the quality of bio-signal processing applications.

\textbf{Pruning:} This technique involves eliminating redundant and less important weights/kernels, or at times layers in the network, to further reduce the DNN's hardware overhead, thereby increasing their deployability in wearables.
Furthermore, eliminating weights and network connections results in reducing the number of \changed{floating-point} operations ($F$), which also decreases the execution time and energy required for inference. 
We retrain the network with the remaining weights to achieve an output quality similar to that of the original network obtained from the architecture-space exploration.
We integrate different pruning techniques from~\cite{anwar2017structured,luo2017thinet,han2015deep,marchisio2018prunet,lin2017runtime} in our framework, from which the system designer can select the appropriate technique based on DNN design and application requirements.
For instance, the techniques presented in~\cite{han2015deep}~and~\cite{marchisio2018prunet} prune the weights based on their magnitude.
The approach presented by Han \textit{et al.}~\cite{han2015deep} eliminates the lowest $x\%$ of the absolute weights in each layer of the DNN and retrains the network to achieve the original network accuracy.
On the other hand, the technique presented in~\cite{marchisio2018prunet} is class-blind, \textit{i.e.}, it eliminates $x\%$ of the lowest absolute weights in the entire DNN, irrespective of which layer the weight corresponds to, and retrains the network to improve its accuracy.
We evaluate the effectiveness of these two pruning techniques in our case-study as well, see Section~\ref{sec:ECG}.
Note, other pruning techniques can also be integrated into our framework as long as they comply \changed{with} the standard input/output interfaces of pruning.

\textbf{Quantization:} The weights and biases in the network are typically stored as $32$-bit \changed{floating-point} numbers in memory, which leads to a large storage overhead, access latency, and energy consumption.
Similarly, the energy required for performing a $32$-bit \changed{floating-point} ADD operation is $9\times$ greater than the energy required for a $32$-bit integer ADD operation~\cite{han2015learning}.
Therefore, by limiting the set of possible values through a process known as \textit{quantization}, we can reduce the hardware overhead of the DNN parameters from $32$-bit \changed{floating-point} values to $16$, $8$, or as illustrated in our work, even lesser number of bits.
The key is to analyze the impact of quantization on the quality of the resulting DNN for our targeted bio-signal processing application.

Based on our experiments and previous work~\cite{han2015deep}, the optimal approach to achieve maximum reduction in hardware overhead of the network involves pruning the network to eliminate redundant or less important weights, followed by DNN parameter quantization.
This involves the formation of $2^q$ (where $q$ is the number of quantized bits) clusters based on the DNN parameters in each layer of the DNN using the k-means clustering algorithm. 
Next, we allocate equally spaced values to each cluster of weights in the layer, starting with the minimum value allocated to all zeros up to the maximum value allocated to all ones.
For the purpose of uniformity, we use a uniform number of bits to quantize all layers (both convolutional and fully-connected) in the network.
Note, different quantization techniques can be integrated \changed{into} our \textit{BioNetExplorer} framework if they comply \changed{with} the input/output interfaces of standard quantization techniques.

% \begin{itemize}
%     \item Overview of the \textit{W-NAS} methodology
%     \item Multi-stage approach
%     \item User Requirements: Output classes, accuracy, etc.
%     \item Hardware Constraints: FLOPS, energy, etc.
%     \item Search parameters and constraints
%     \item Exhaustive Search
%     \item Genetic Algorithm Search
%     \item Constrained Search
%     \item proposed weighted optimization technique storage + accuracy
% \end{itemize}
\section{Experimental Setup}
\label{sec:ES}

% \begin{table}[!b]
%     \caption{Global caption}
%     \begin{minipage}{.5\linewidth}
%         \caption{}
%         \centering
%         \begin{tabular}{L{2cm}|L{2cm}}
%             \multicolumn{1}{l|}{\textbf{Parameter}} & \textbf{Values} \\
%             \hline
%             Weights Initialized           & He \textit{et al.}~\cite{he2015delving} \\
%             \multirow{2}{*}{Adam Optimizer~\cite{kingma2014adam}} & $\beta_1=0.9$ \\
%                               & $\beta_2=0.999$\\
%             Dropout                       & $0.2$    \\
%             Batch Size                    & $128$    \\
%             Learning Rate                 & $0.001$ 
%         \end{tabular}
%     \end{minipage}%
%     \begin{minipage}{.5\linewidth}
%         \centering
%         \caption{}
%         \begin{tabular}{L{2cm}|L{2cm}}
%             \multicolumn{1}{l|}{\textbf{Parameter}} & \textbf{Values} \\
%             \hline
%             Chromosome Length            & $9$ \\
%             Population Size        & $30$ \\
%             Generation Size        & $5$ \\
%             % Individual Probability & $0.5$ \\
%             Crossover Probability  & $0.4$ \\
%             Mutation Probability   & $0.11$ 
%         \end{tabular}
%     \end{minipage} 
% \end{table}

\setcounter{figure}{7}
\begin{figure}[t]
    \centering
    \captionsetup{singlelinecheck=false}
    \includegraphics[width = \linewidth]{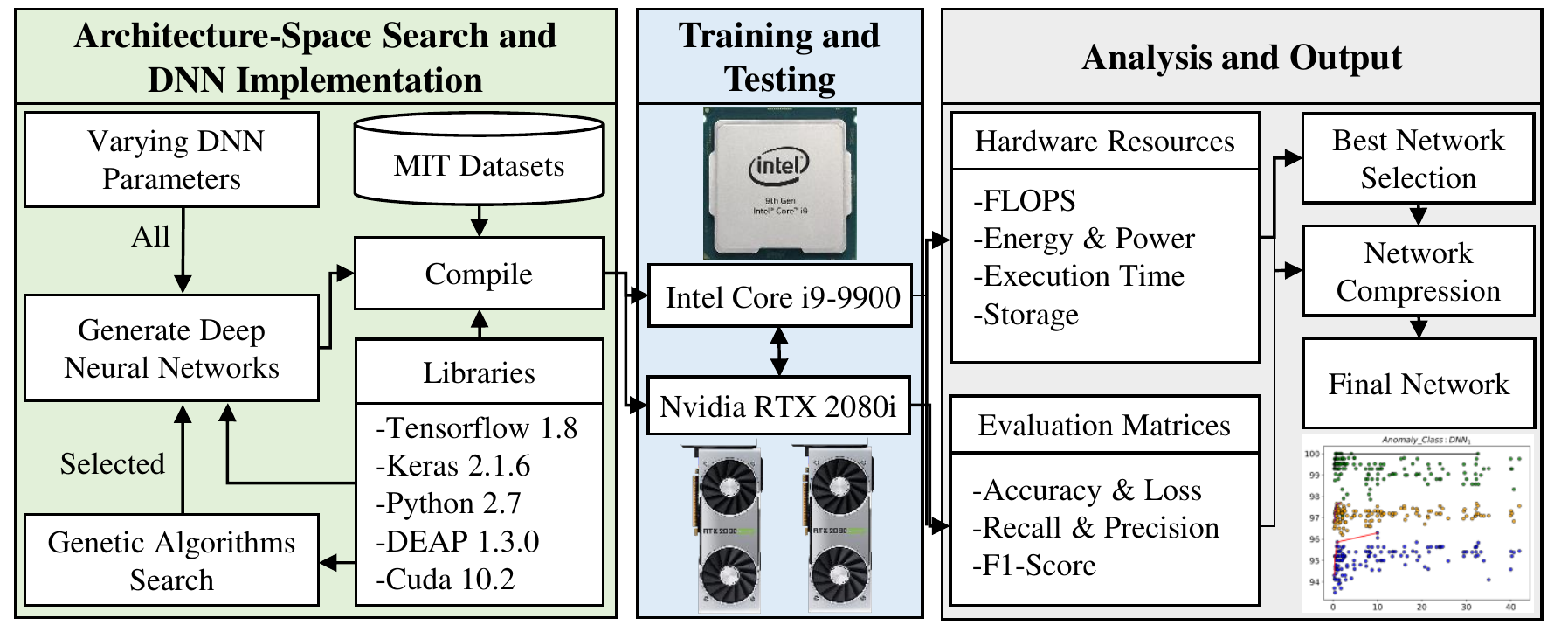}
    \caption{\textbf{Overview of Our Experimental Tool-Flow.}}
    \label{fig:ES}
\end{figure}

\begin{table}[b]
\centering
\caption{\textbf{Hyper-Parameter Values For Training DNNs.}}
\label{tab:HP}
\begin{tabular}{l|l}
\multicolumn{1}{l|}{\textbf{Parameter}} & \textbf{Values} \\
\hline
Weights Initialized           & He \textit{et al.}~\cite{he2015delving} \\
\multirow{2}{*}{Adam Optimizer~\cite{kingma2014adam}} & $\beta_1=0.9$ \\
                              & $\beta_2=0.999$\\
Dropout                       & $0.2$    \\
Batch Size                    & $128$    \\
Learning Rate                 & $0.001$ \vspace{0.1cm}
\end{tabular}
\end{table}

\begin{table}[b]
\centering
\caption{\textbf{Genetic Algorithms' Parameters \& Values.}}
\label{tab:EA}
\begin{tabular}{l|l}
\multicolumn{1}{l|}{\textbf{Parameter}} & \textbf{Values} \\
\hline
Chromosome Length            & $9$ \\
Population Size        & $30$ \\
Generation Size        & $5$ \\
% Individual Probability & $0.5$ \\
Crossover Probability  & $0.4$ \\
Mutation Probability   & $0.11$ 
\end{tabular}
\end{table}

Fig.~\ref{fig:ES} presents an overview of our experimental tool-flow for implementation and validation of the \textit{BioNetExplorer} framework.
The dataset used in this work is openly accessible and is made available online by the Physionet data bank~\cite{goldberger2000physiobank}.
Our framework uses the MIT-BIH ECG data~\cite{moody2001impact} with $41$ beat and non-beat annotations to construct new datasets and labels based on the application's requirement.
Furthermore, to illustrate the benefits of including a unique output class, we include the CU Ventricular Arrhythmia dataset~\cite{nolle1986crei} that contains instances of ECG recordings with labels of \textit{ventricular tachycardia}, \textit{ventricular flutter}, and \textit{ventricular fibrillation}.
We use the Keras environment, which includes the \changed{Tensorflow} machine learning platform, in Python to implement the DNN architectures that we explore in this work.
The hyper-parameters are crucial for evaluating the accuracy of \changed{a} DNN architecture. 
Therefore, we train the DNN architecture over multiple iterations for various values of the hyper-parameters in each iteration and select the hyper-parameter values that offer the best accuracy.
The hyper-parameters that offer the maximum accuracy for these networks are presented in Table~\ref{tab:HP}.
The four genetic algorithms that are used in this work are implemented using the DEAP library~\cite{rainville2012deap}. The parameters used by these algorithms and their values are presented in Table~\ref{tab:EA}.
The exploration algorithm is implemented on top of the network training stage using multiple GPU servers with Core i$9$-$9900$ CPUs integrated with two Nvidia RTX $2080$i GPUs, each.
Furthermore, we enable the early-stopping mechanism during the network training stage to terminate execution when the loss of the DNN does not change between two consecutive epochs.
The trained networks are then evaluated \changed{based on} the number of parameters they require, \changed{which} determines the storage overhead and the number of \changed{floating-point} operations required to perform a single inference, \changed{which, in turn,} determines the execution time of the network on the target platform.

% \begin{itemize}
%     \item Tool-flow
%     \item Deap \cite{rainville2012deap}
%     \item hardware resources
%     \item simulation environment
%     \item Dataset:MIT BIH Arrhythmia dataset \cite{moody2001impact}
%     \& Cu Ventricular Arrhythmia dataset \cite{nolle1986crei}
%     \item Networks
%     \item etc.
% \end{itemize}
\section{Case-Study: Anomaly Detection in ECG Signals}
\label{sec:ECG}

We illustrate the benefits of using the proposed \textit{BioNetExplorer} framework with the help of an ECG signal processing DNN, which is used to classify the following cases:

{\raggedright
\begin{itemize}[leftmargin=*]
    \item \textbf{DNN$_1$:} Normal and Anomaly Classes;
    \item \textbf{DNN$_2$:} Normal, Premature Ventricular Contraction (PVC), and Other Anomalies;
    \item \textbf{DNN$_3$:} Normal, Bundle Branch Block (BB), and\\ Other Anomalies;
    \item \textbf{DNN$_4$:} Normal, Atrial \changed{Anomaly}, Ventricular Anomaly, and Other Anomalies;
    \item \textbf{DNN$_5$:} Normal, Ventricular Fibrillation, and\\ Other Anomalies.%\vspace{-0.4cm}
\end{itemize}
}

\subsection{Effectiveness of State-of-the-Art Neural Architecture Search Approaches in Bio-Signal Processing}
\label{subsec:SoA}

First, we evaluate the effectiveness of two state-of-the-art Neural Architecture Search approaches for our bio-signal processing DNN architecture-space, namely, 
\begin{inlinelist}
    \item MnasNet~\cite{tan2019mnasnet} and 
    \item MobileNetV3~\cite{howard2019searching}.
\end{inlinelist}
The purpose of this experimental study is to corroborate our claims that 
\begin{inlinelist}
    \item traditional NAS approaches designed for image classification applications do not work effectively for bio-signal processing applications and
    \item there is a need and lack of NAS frameworks for bio-signal processing applications, like the \textit{BioNetExplorer} framework proposed in this work.
\end{inlinelist} 
Unlike image classification applications that have \changed{two-dimensional} inputs, bio-signal data is typically present in \changed{a} 1-dimensional format.
Therefore, to accommodate this requirement, we replaced the traditional 2D normal and depth-wise convolution layers with 1D normal and separable convolution, respectively.
We have explored five variants of the MnasNet discussed in~\cite{tan2019mnasnet} and the small and large variants of the MobileNetV3 discussed in~\cite{howard2019searching}, with various filter sizes ([$3$, $5$, $7$, $8$, $16$]) and depth multipliers ([$1$, $0.5$, $0.25$]).

The results of deploying these NAS techniques on the \textbf{DNN$_1$} and \textbf{DNN$_4$} architecture-spaces are presented in Figs.~\ref{fig:DNN1_SoA} and~\ref{fig:DNN4_SoA}, respectively.
The default blocks used in the MnasNet framework are more suited for images and/or 2-dimensional data because of which they do not perform well for bio-signal processing applications, as illustrated in Figs.~\ref{fig:DNN1_SoA}(a)~and~\ref{fig:DNN4_SoA}(a).
The best performing networks obtained using the MnasNet framework (labeled \circled{A} in both figures) consistently perform worse \changed{than} the networks obtained by our \textit{BioNetExplorer} framework, especially in cases where the number of output classifications \changed{is} greater than two (\circled{A} in Fig.~\ref{fig:DNN4_SoA}).
The networks obtained by MobileNetV3-NAS (labeled \circled{B} in both figures) perform much better than MnasNet for both illustrated cases, which is still non-optimal when compared to the networks obtained by our framework.
The basic ResNet block, adopted from~\cite{hannun2019cardiologist} by our \textit{BioNetExplorer} framework, is more suited for bio-signal processing applications as compared to the blocks used in state-of-the-art~\cite{tan2019mnasnet,wu2019fbnet,dai2019chamnet,howard2019searching}. \textit{BioNetExplorer} is successful in obtaining networks with better output quality for the same or reduced hardware overhead and vice-versa when compared to other existing approaches.

\setcounter{figure}{8}
\begin{figure}[t]
    \centering
    \captionsetup{singlelinecheck=false}
    \includegraphics[width = \linewidth]{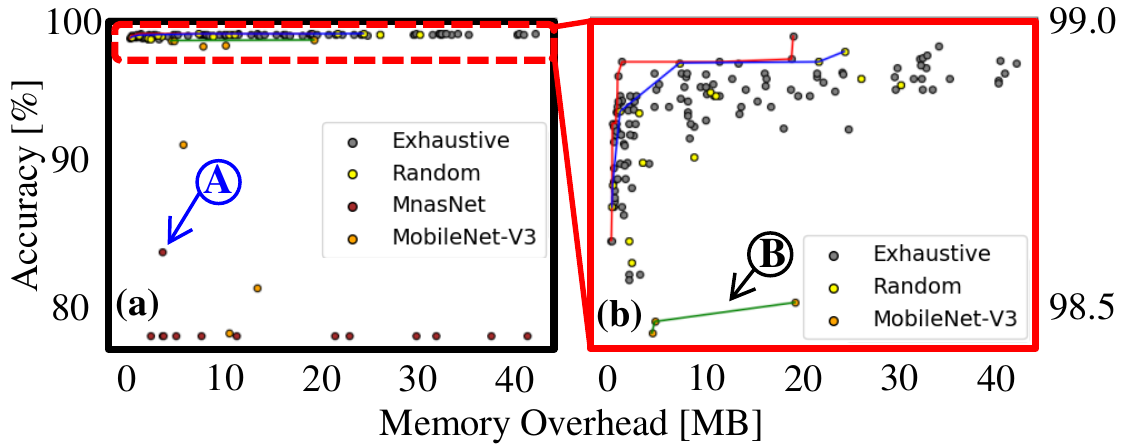}
    \caption{\textbf{Applicability of MnasNet and MobileNetV3 on the \textbf{DNN$_1$} Architecture-Space.}}
    \label{fig:DNN1_SoA}
\end{figure}
\setcounter{figure}{9}
\begin{figure}[t]
    \centering
    \captionsetup{singlelinecheck=false}
    \includegraphics[width = \linewidth]{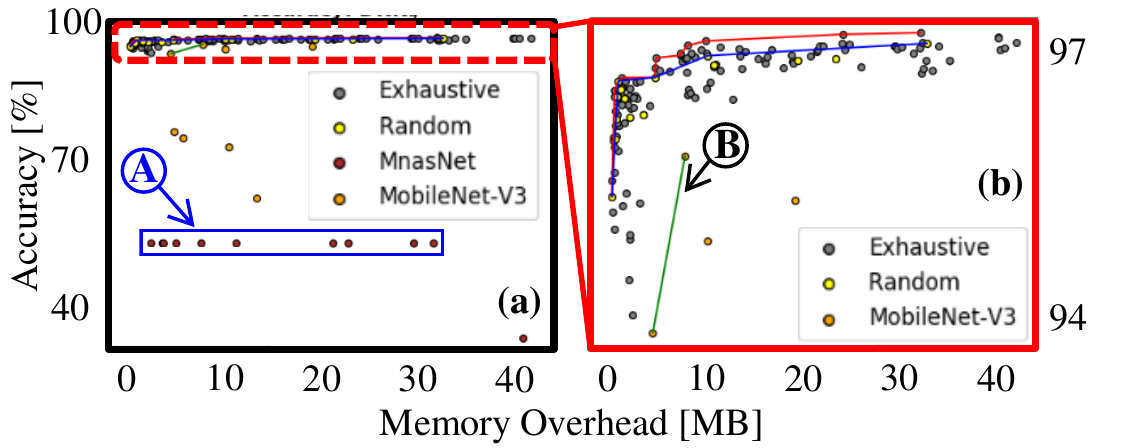}
    \caption{\textbf{Applicability of MnasNet and MobileNetV3 on the \textbf{DNN$_4$} Architecture-Space.}}
    \label{fig:DNN4_SoA}
\end{figure}

The approaches presented in~\cite{wu2019fbnet}~and~\cite{dai2019chamnet} use a basic block structure that is replicated in different combinations, for exploring the DNN architecture-space, which is similar to the block structure and approaches presented in MnasNet~\cite{tan2019mnasnet} and MobileNetV3~\cite{howard2019searching}.
Therefore, these approaches would also behave \changed{similarly} to the state-of-the-art techniques presented in Figs.~\ref{fig:DNN1_SoA}~and~\ref{fig:DNN4_SoA}.

\subsection{Training Time Benefits of Genetic Algorithm-Based DNN Architecture-Space Exploration}

\setcounter{figure}{11}
\begin{figure*}[!b]
\captionsetup{singlelinecheck=false}
\begin{minipage}{\textwidth}
\centering
\includegraphics[scale=0.36]{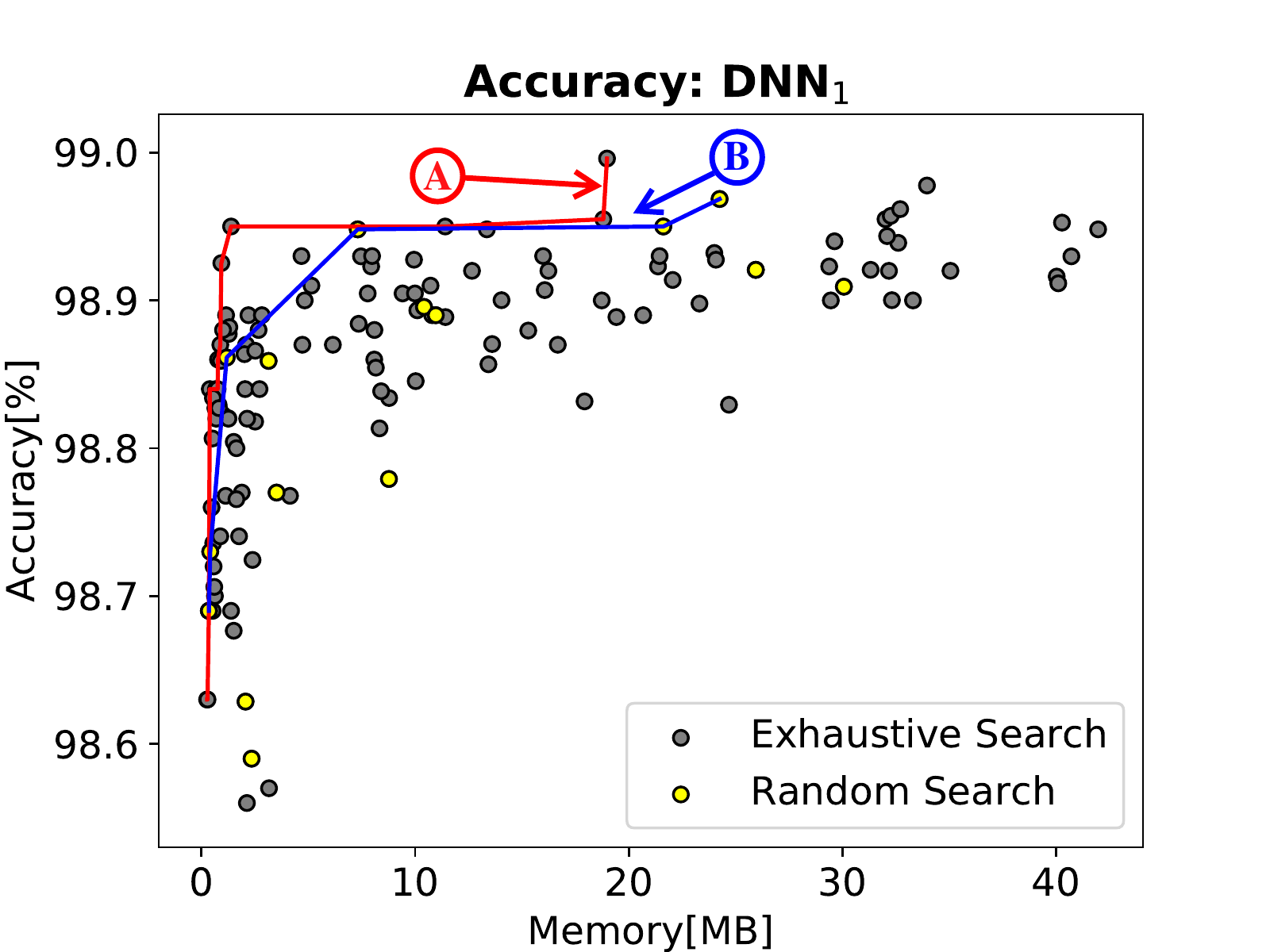}
\includegraphics[scale=0.36]{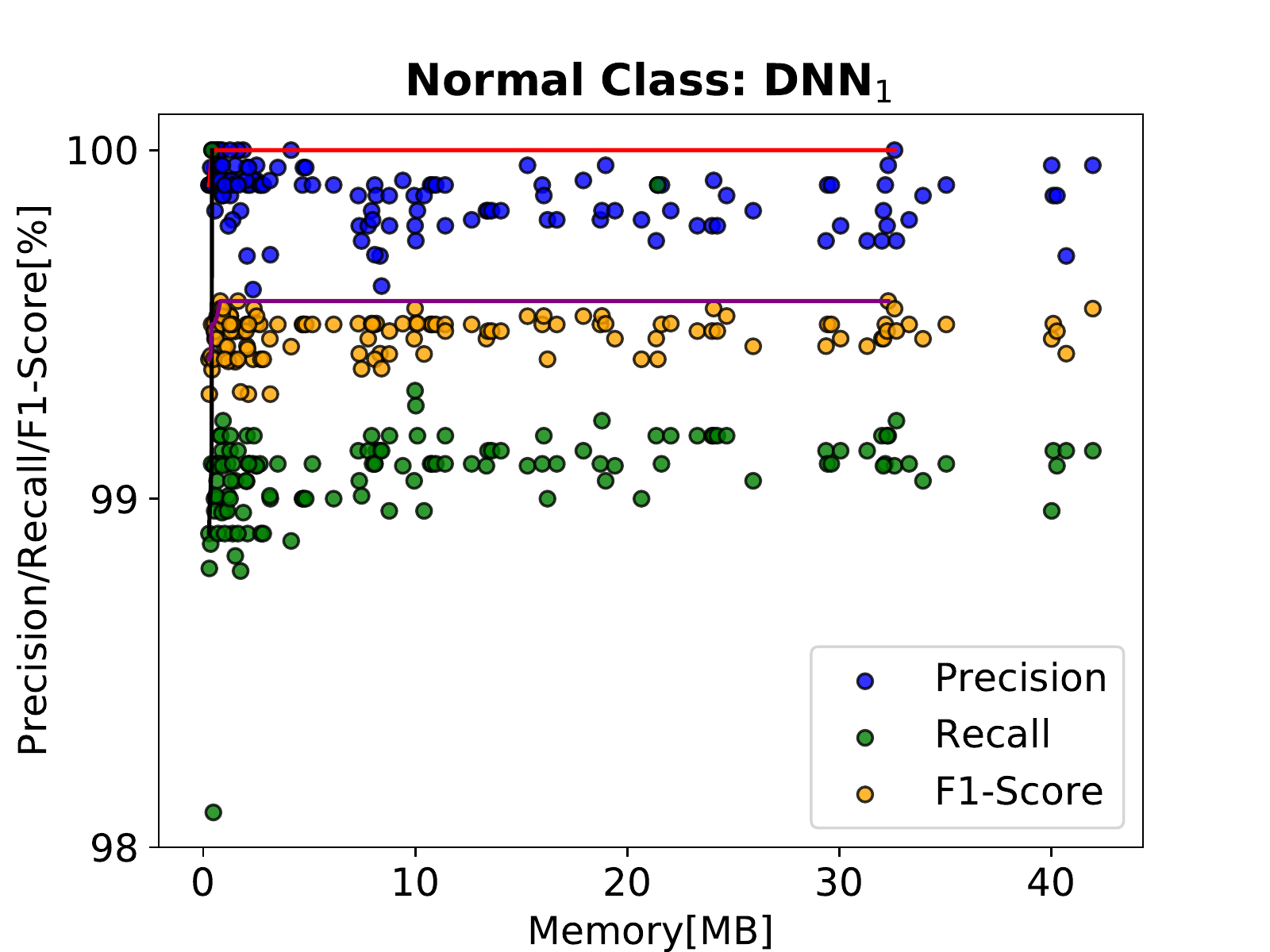}
\includegraphics[scale=0.36]{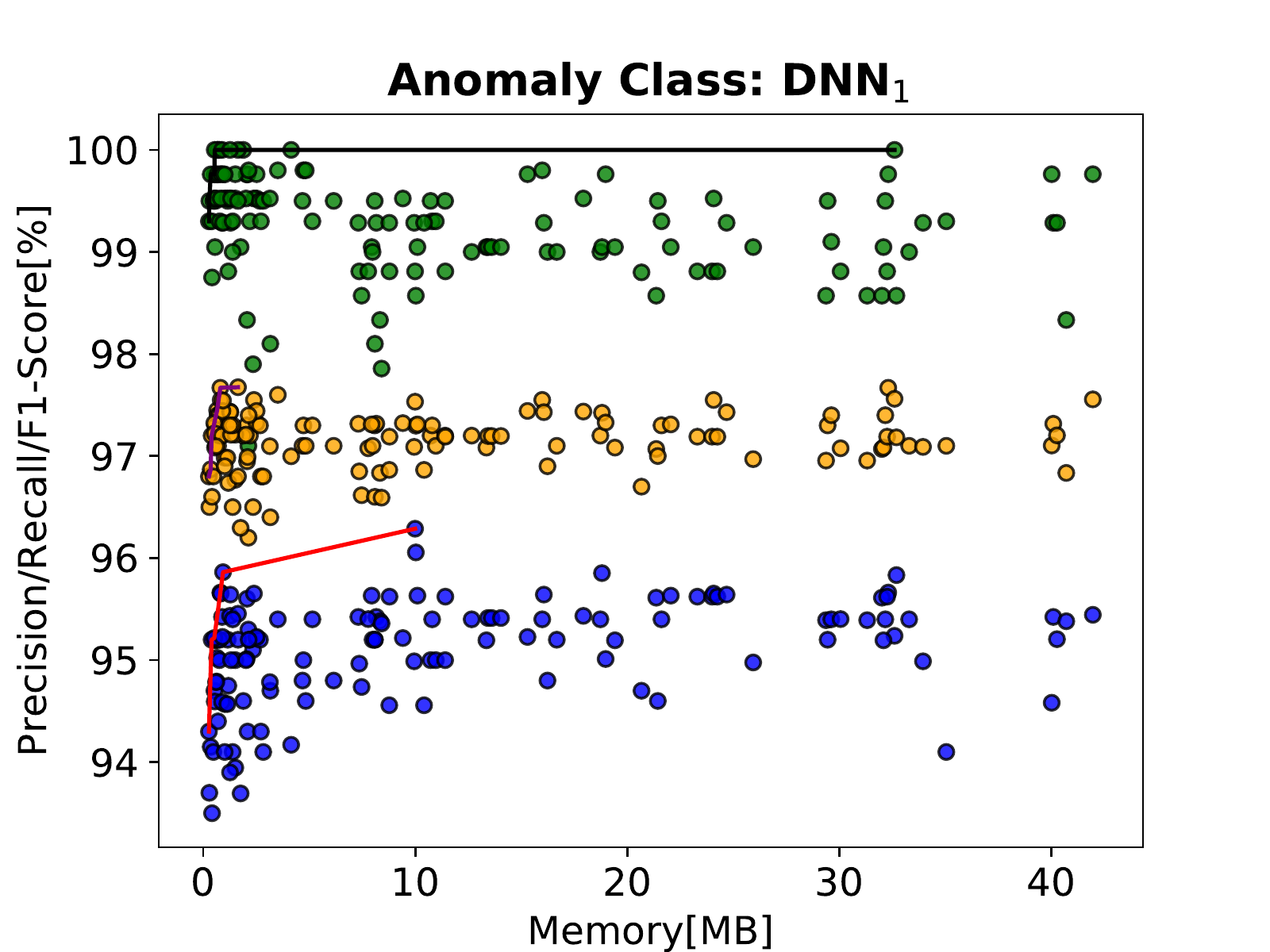}
\end{minipage}
\caption{\textbf{Exhaustive Exploration and Analysis of DNN$_1$ Architectures.}}
\label{fig:DNN1Ex}
\end{figure*}

\setcounter{figure}{12}
\begin{figure*}[!b]
\captionsetup{singlelinecheck=false}
\begin{minipage}{\textwidth}
\centering
\includegraphics[scale=0.36]{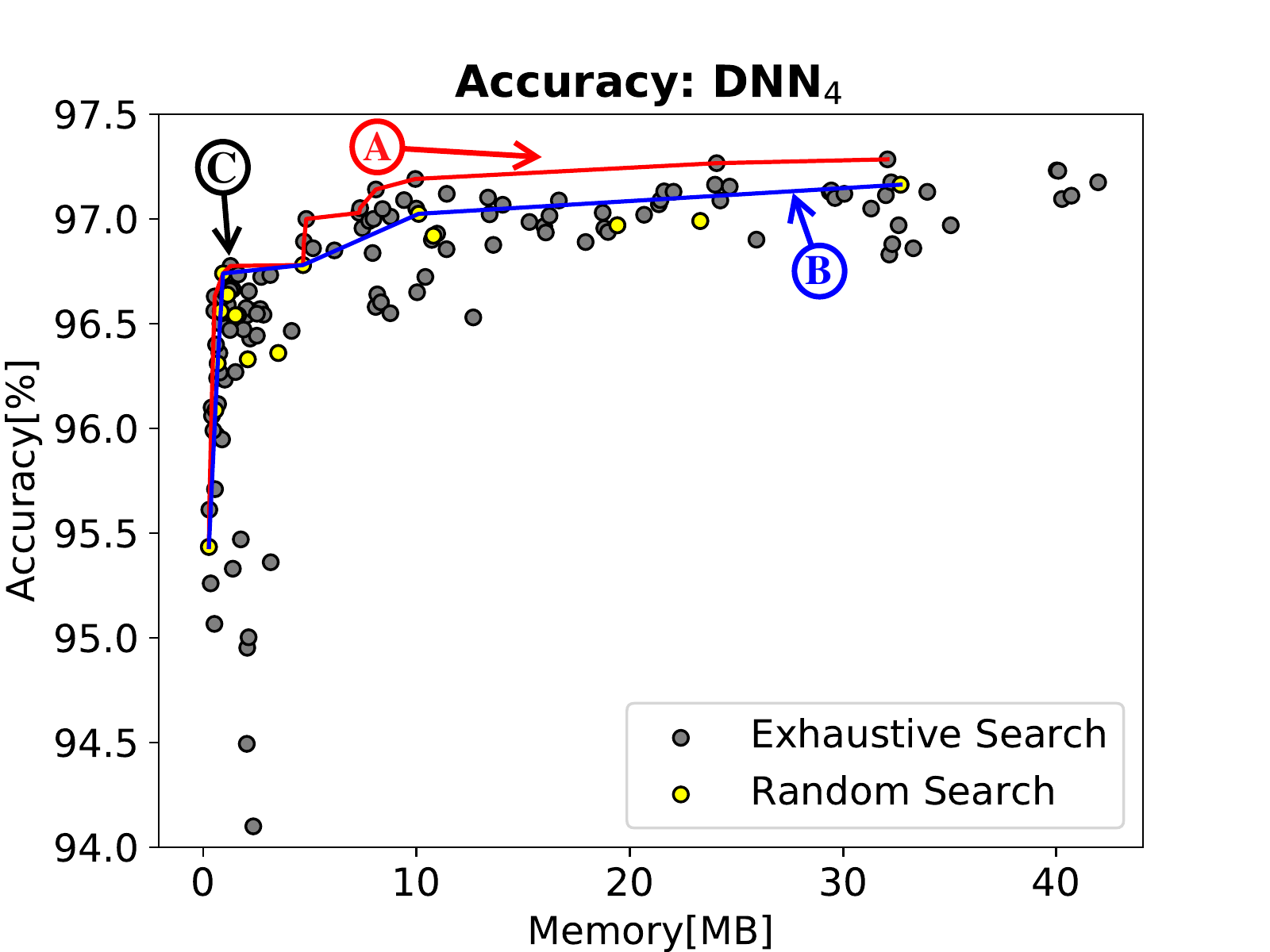}
\includegraphics[scale=0.36]{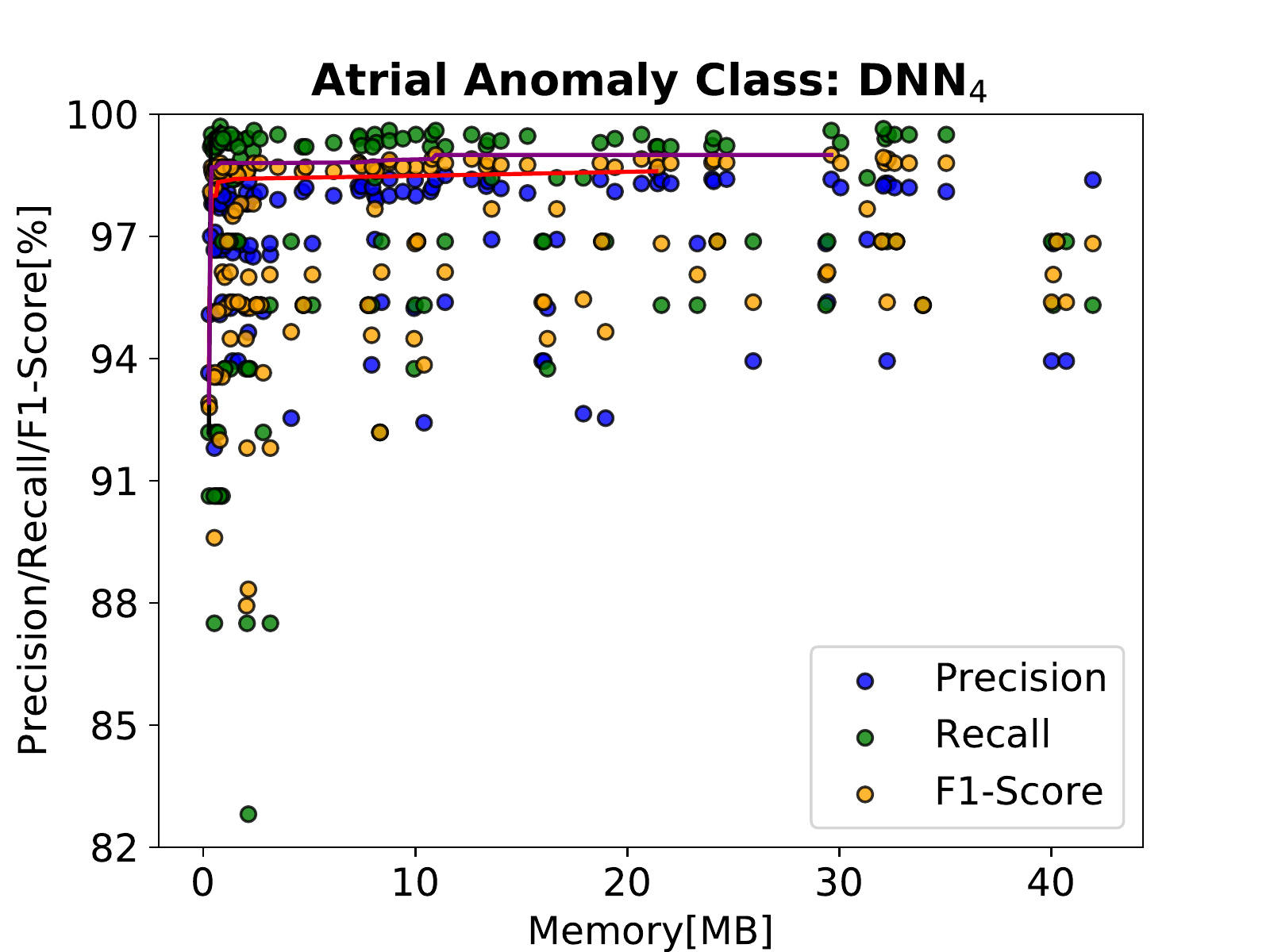}
\includegraphics[scale=0.36]{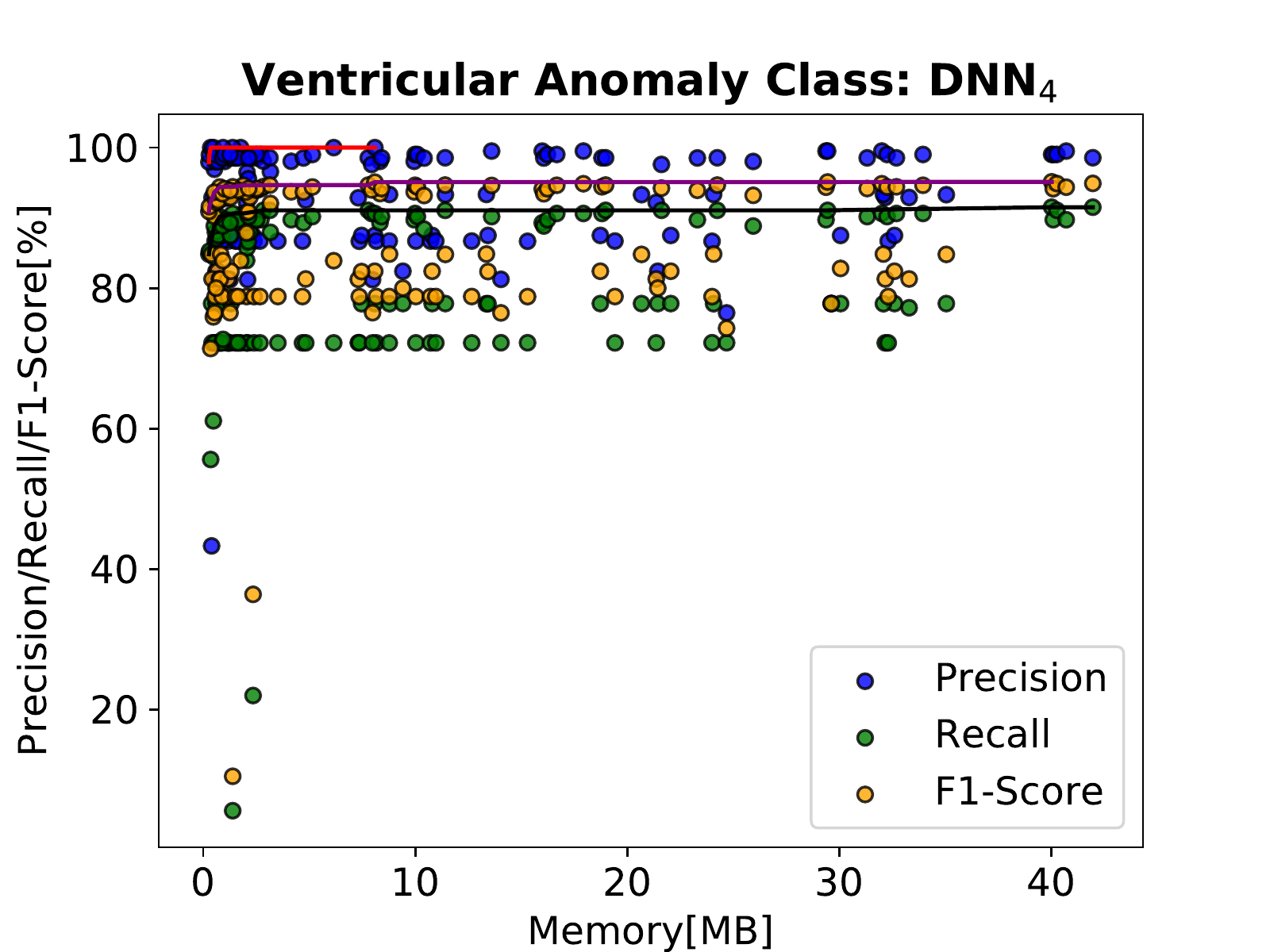}
\end{minipage}
\caption{\textbf{Exhaustive Exploration and Analysis of DNN$_4$ Architectures.}}
\label{fig:DNN4Ex}
\end{figure*}

\setcounter{figure}{10}
\begin{figure}[t]
    \centering
    \captionsetup{singlelinecheck=false}
    \includegraphics[width = \linewidth]{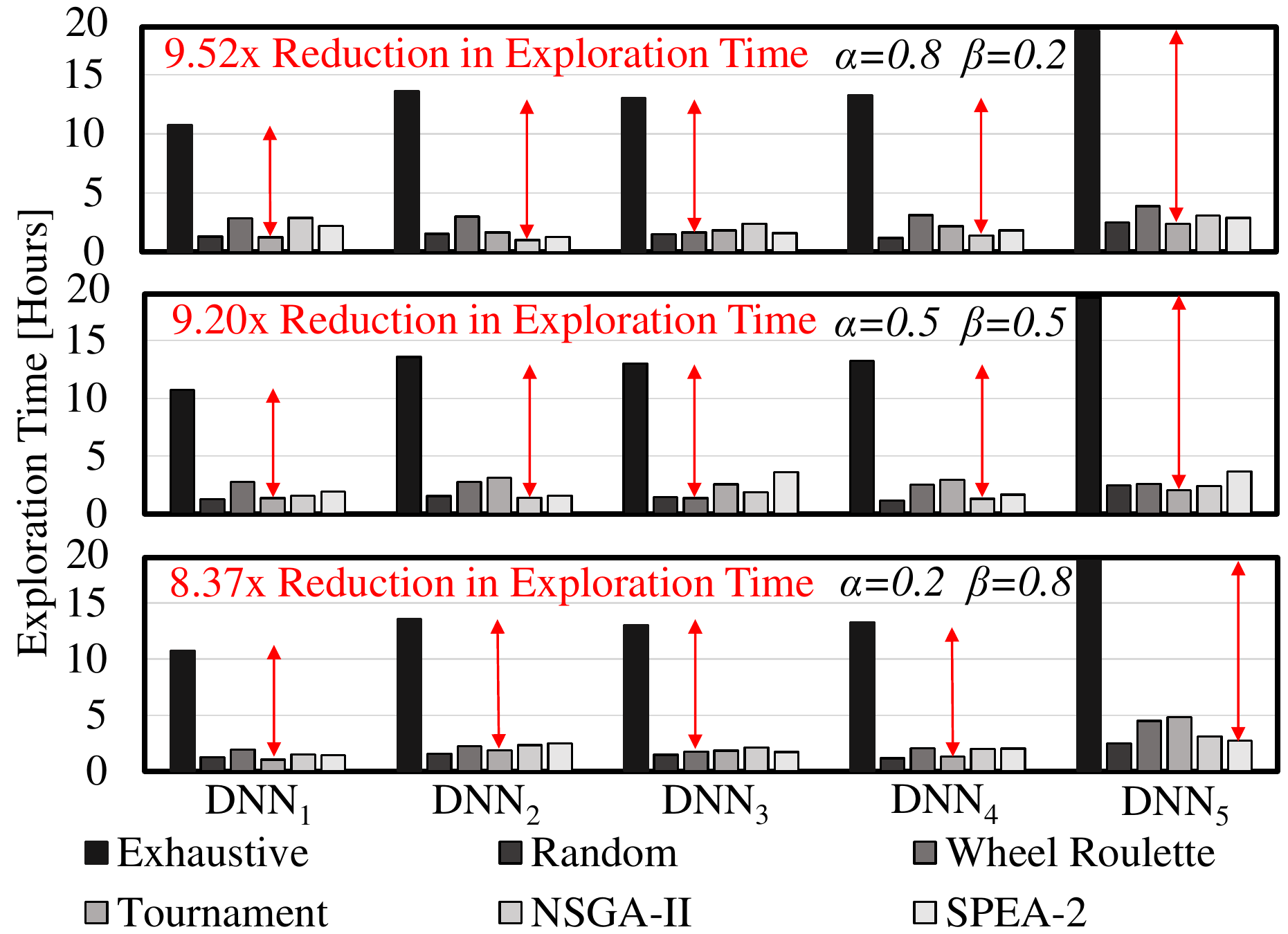}
    \caption{\textbf{Training Time Analysis of Exhaustive, Random, and Genetic Algorithm-Based Search.} The time required for training the random networks stay the same for all optimization goals. Genetic algorithms reduce the exploration time by \midtilde$9\times$ because they select and train a small-subset of the networks.}
    \label{fig:TimeGraph}
\end{figure}

% \setcounter{figure}{13}
% \begin{figure*}[!t]
% \captionsetup{singlelinecheck=false}
% \begin{minipage}{\textwidth}
% \centering
% \includegraphics[scale=0.36]{figures/DNN2_Ex_1Accuracy.pdf}
% \includegraphics[scale=0.36]{figures/DNN2_Ex_3Anomaly.pdf}
% \includegraphics[scale=0.36]{figures/DNN2_Ex_4PVC.pdf}
% \end{minipage}
% \caption{\textbf{\changed{Exhaustive Exploration and Analysis of DNN$_2$ Architectures.}}}
% \label{fig:DNN2Ex}
% \end{figure*}

Second, we illustrate the benefits obtained in training time of the DNN architecture-space exploration when a genetic algorithm is used instead of the exhaustive exploration for all five DNNs.
Fig.~\ref{fig:TimeGraph} presents the time required for selecting and training the Pareto-optimal (Exhaustive Search) or near-optimal (Genetic Algorithm-Based Search) DNNs for various values of $\alpha$ and $\beta$.
Similar to recent NAS studies~\cite{li2019random}, we also implement a random search as a potential search strategy. 
It randomly selects and trains $10\%$ of the architecture-space; see results in Fig.~\ref{fig:TimeGraph}.
Since random search has practically very low overhead and a fixed number of architectures to train for all optimization goals, the time required for training the DNN stays the same for all scenarios.
On average, the use of genetic algorithms reduces the exploration time by $9.03\times$.
Note, exhaustive exploration might not be a viable option in all scenarios. 
For example, in applications that require the use of deeper and complex neural networks, the training time for each network would lead to an exponential increase in the exploration time while requiring hundreds of GPU-hours.
Since genetic algorithms do not train all the networks in the architecture-space, the exploration time is limited to tens of GPU hours. Furthermore, as a back-up strategy, our system supports the use of random search that has a fixed-time complexity to provide a reasonably good solution.

% \begin{figure}[hbtp]
%     \centering
%     \includegraphics[width = 0.85\linewidth]{figures/DNN2_1GA_WR.pdf}
%     \includegraphics[width = 0.85\linewidth]{figures/DNN2_2GA_TS.pdf}
%     \includegraphics[width = 0.85\linewidth]{figures/DNN2_3GA_NS.pdf}
%     \includegraphics[width = 0.85\linewidth]{figures/DNN2_4GA_SP.pdf}
%     \caption{\textbf{Exploration of the \textbf{DNN$_2$} Architecture-Space Using Genetic Algorithms (Explored Designs are Highlighted).}}
%     \label{fig:DNN2GA}
% \end{figure}

% \begin{figure}[hbtp]
%     \centering
%     \includegraphics[width = 0.85\linewidth]{figures/DNN3_1GA_WR.pdf}
%     \includegraphics[width = 0.85\linewidth]{figures/DNN3_2GA_TS.pdf}
%     \includegraphics[width = 0.85\linewidth]{figures/DNN3_3GA_NS.pdf}
%     \includegraphics[width = 0.85\linewidth]{figures/DNN3_4GA_SP.pdf}
%     \caption{\textbf{Exploration of the \textbf{DNN$_3$} Architecture-Space Using Genetic Algorithms (Explored Designs are Highlighted).}}
%     \label{fig:DNN3GA}
% \end{figure}

\subsection{Exhaustive Exploration and Analysis of Deep Neural Networks from the \textbf{DNN$_1$} and \textbf{DNN$_4$} Architecture-Spaces}

Next, we exhaustively explore all the architectures of \textbf{DNN}$_1$ and \textbf{DNN}$_4$, \textit{i.e.}, train all the architectures using the constructed datasets.
We illustrate the quality-of-service and memory storage overhead trade-off of the \textbf{DNN}$_1$ and \textbf{DNN}$_4$ architecture-spaces in Figs.~\ref{fig:DNN1Ex}~and~\ref{fig:DNN4Ex}, respectively.
The Pareto-front obtained by random search (labeled \circled{B} in both figures) is similar to the Pareto-front obtained using exhaustive exploration (labeled \circled{A} in both figures) of the architecture-space, primarily due to \changed{a} large number of inter-dependent DNN parameters and their effects on the final accuracy.
The exhaustive exploration approach is able to successfully identify networks that trade-off between output quality and hardware overhead, as illustrated by the figures.
For instance, we have successfully identified a \textbf{DNN}$_4$ network that reduces the hardware overhead by \midtilde$30$MB for a quality loss of $0.5\%$ (labeled \circled{C} in Fig.~\ref{fig:DNN4Ex}).
The \textit{Precision}, \textit{Recall}, and \textit{F1-score} for the \textit{Anomaly} class is lower than that of the \textit{Normal} class. 
Therefore, a quality constraint can be imposed on any of the four quality metrics based on the application requirement.
The main drawback of this approach is the time required for training and evaluating all networks in the architecture-space of the DNN.

\setcounter{figure}{13}
\begin{figure*}[!t]
\captionsetup{singlelinecheck=false}
\begin{minipage}{\textwidth}
\centering
\includegraphics[scale=0.27]{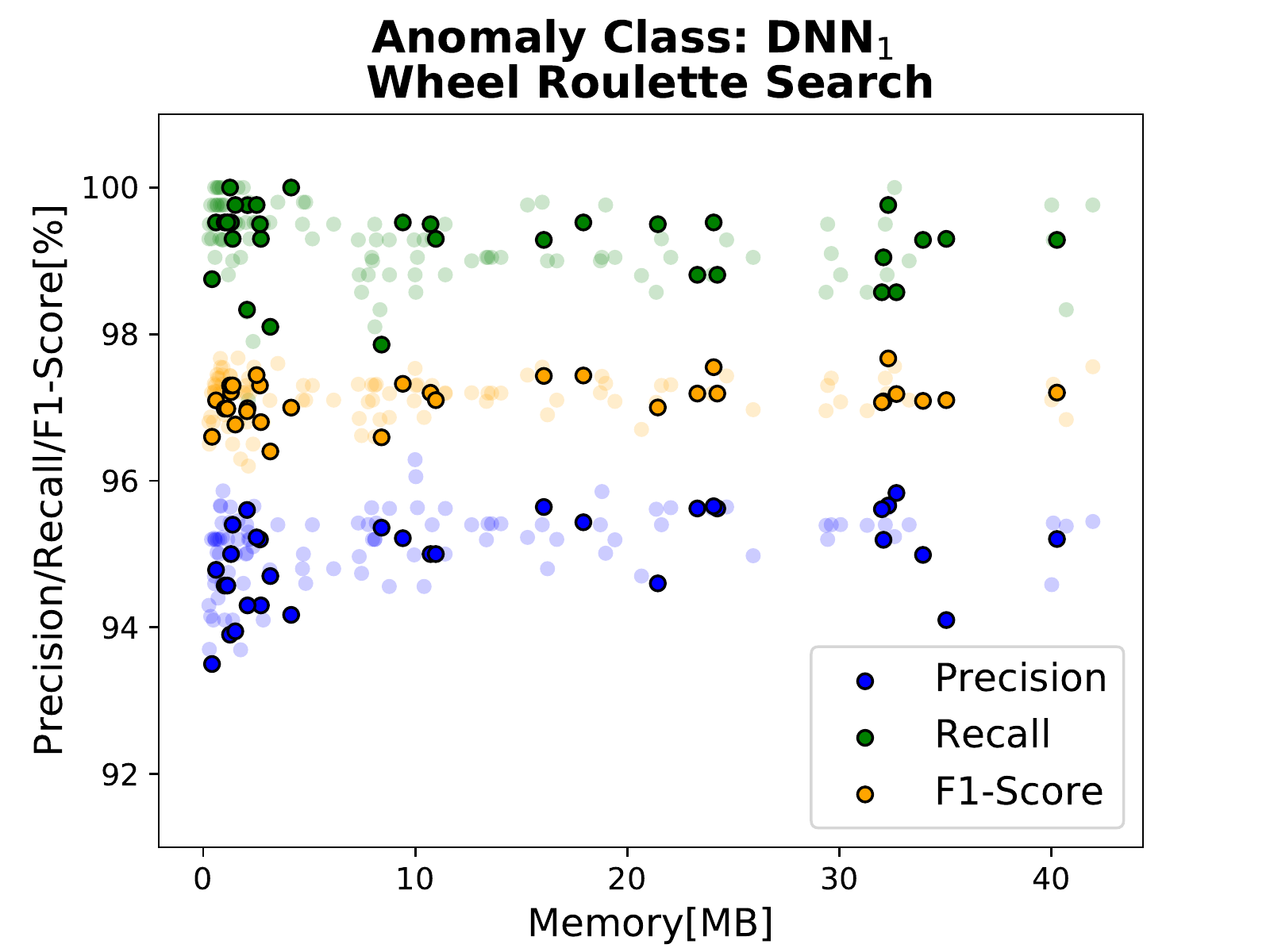}
\includegraphics[scale=0.27]{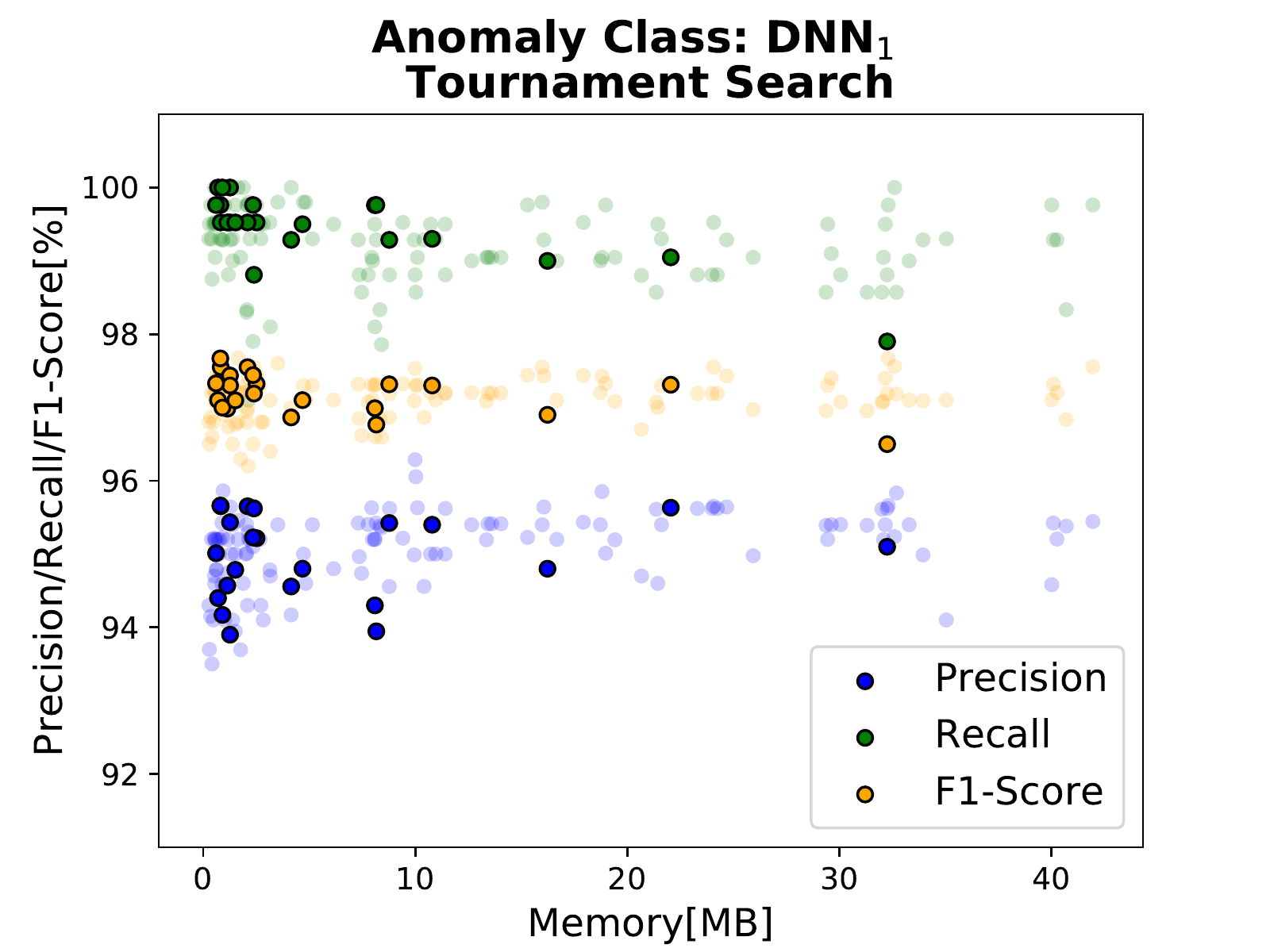}
\includegraphics[scale=0.27]{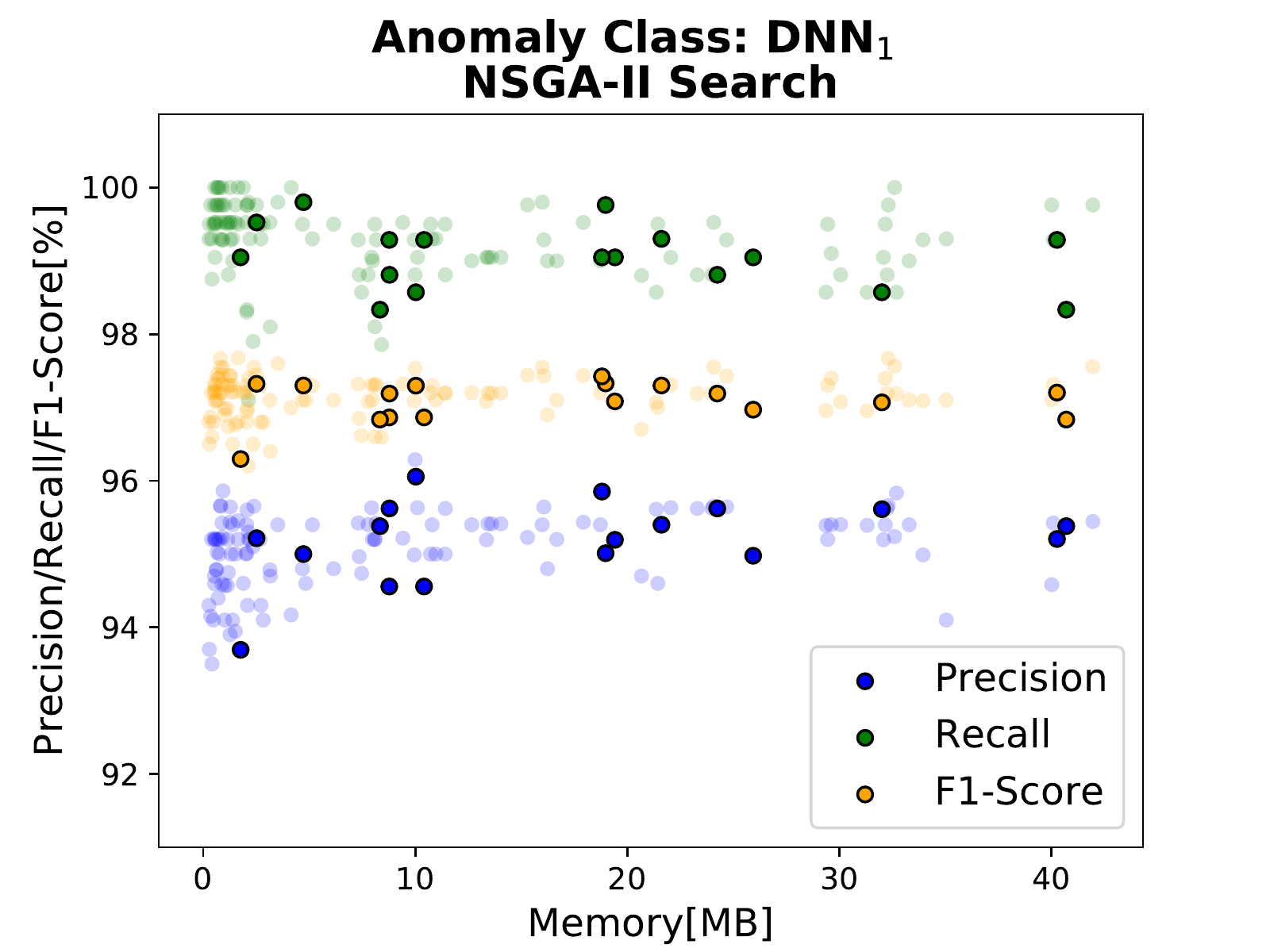}
\includegraphics[scale=0.27]{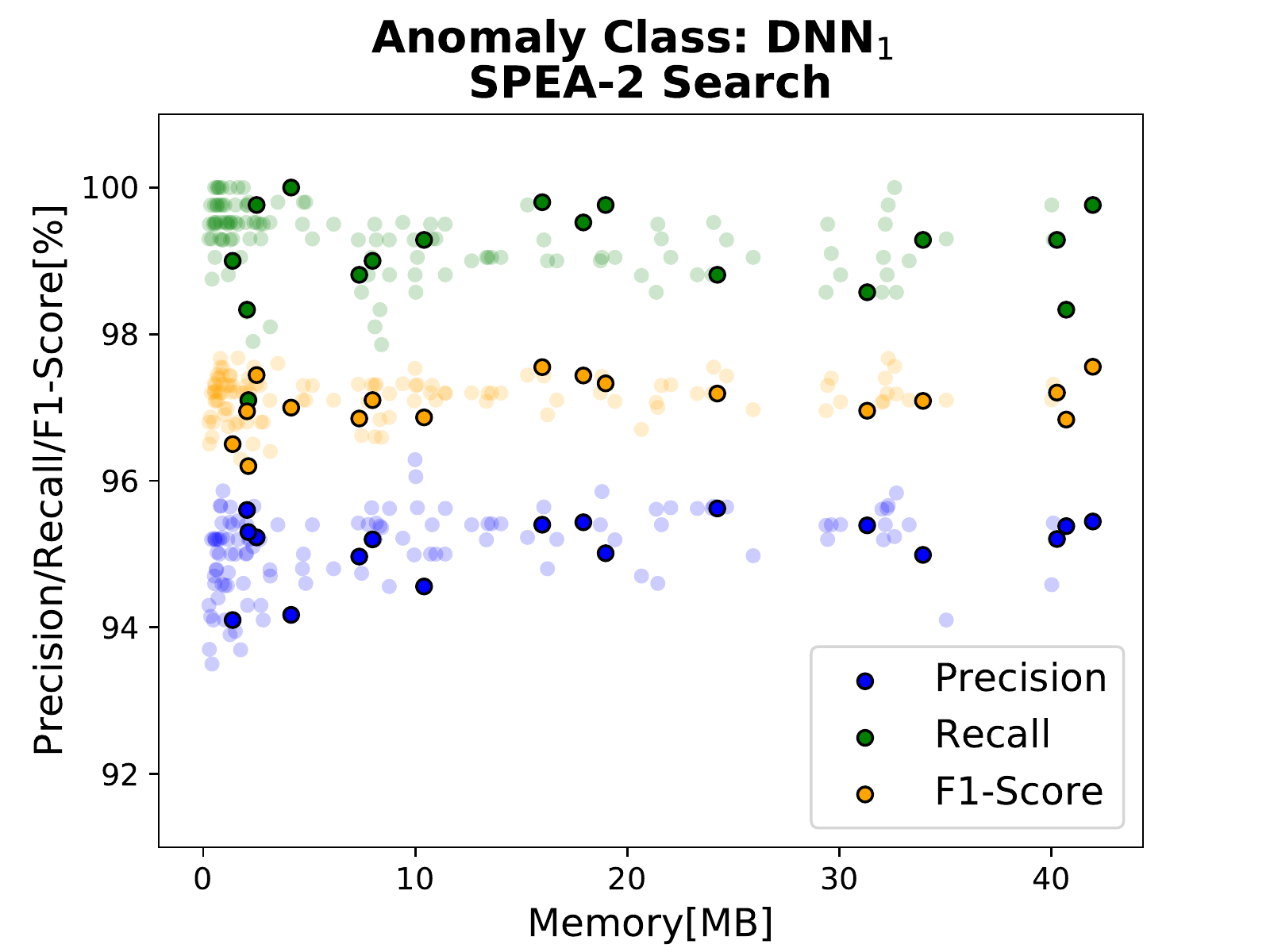}
\end{minipage}
\caption{\textbf{Exploration of the \textbf{DNN$_1$} Architecture-Space Using Genetic Algorithms (Explored Designs are Highlighted).}}
\label{fig:DNN1GA}
\end{figure*}

\begin{figure*}[!t]
\captionsetup{singlelinecheck=false}
\begin{minipage}{\textwidth}
\centering
\includegraphics[scale=0.27]{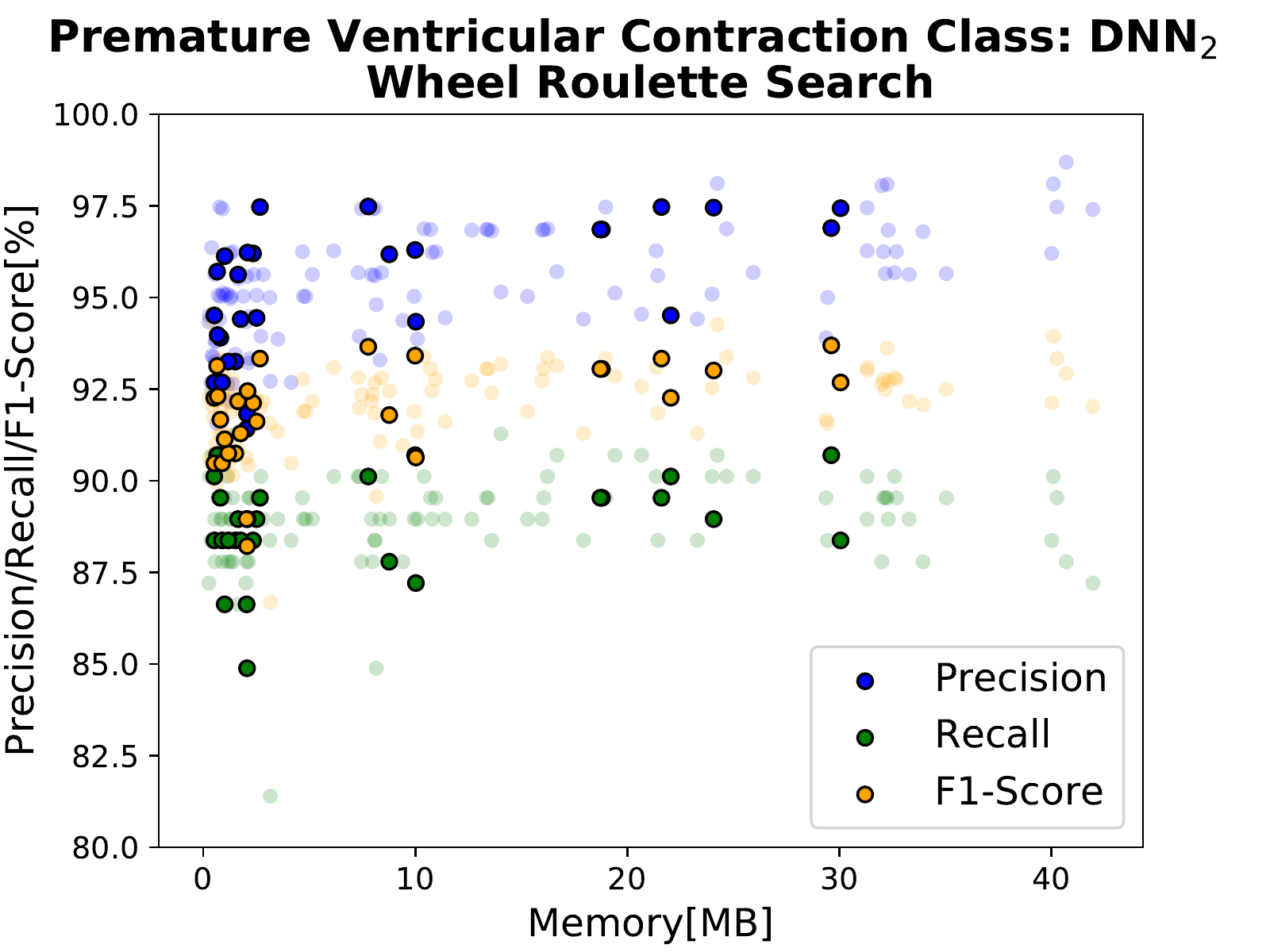}
\includegraphics[scale=0.27]{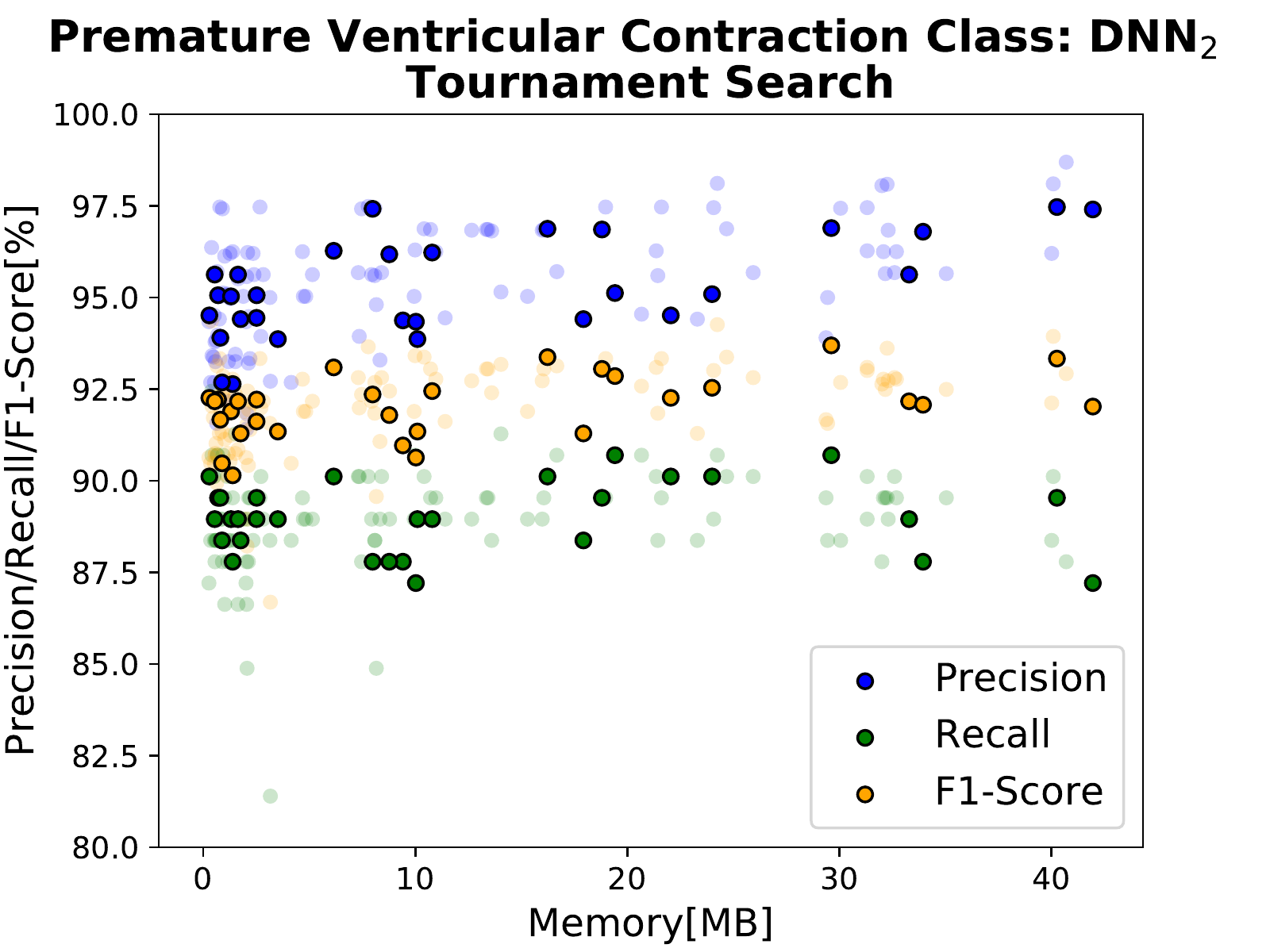}
\includegraphics[scale=0.27]{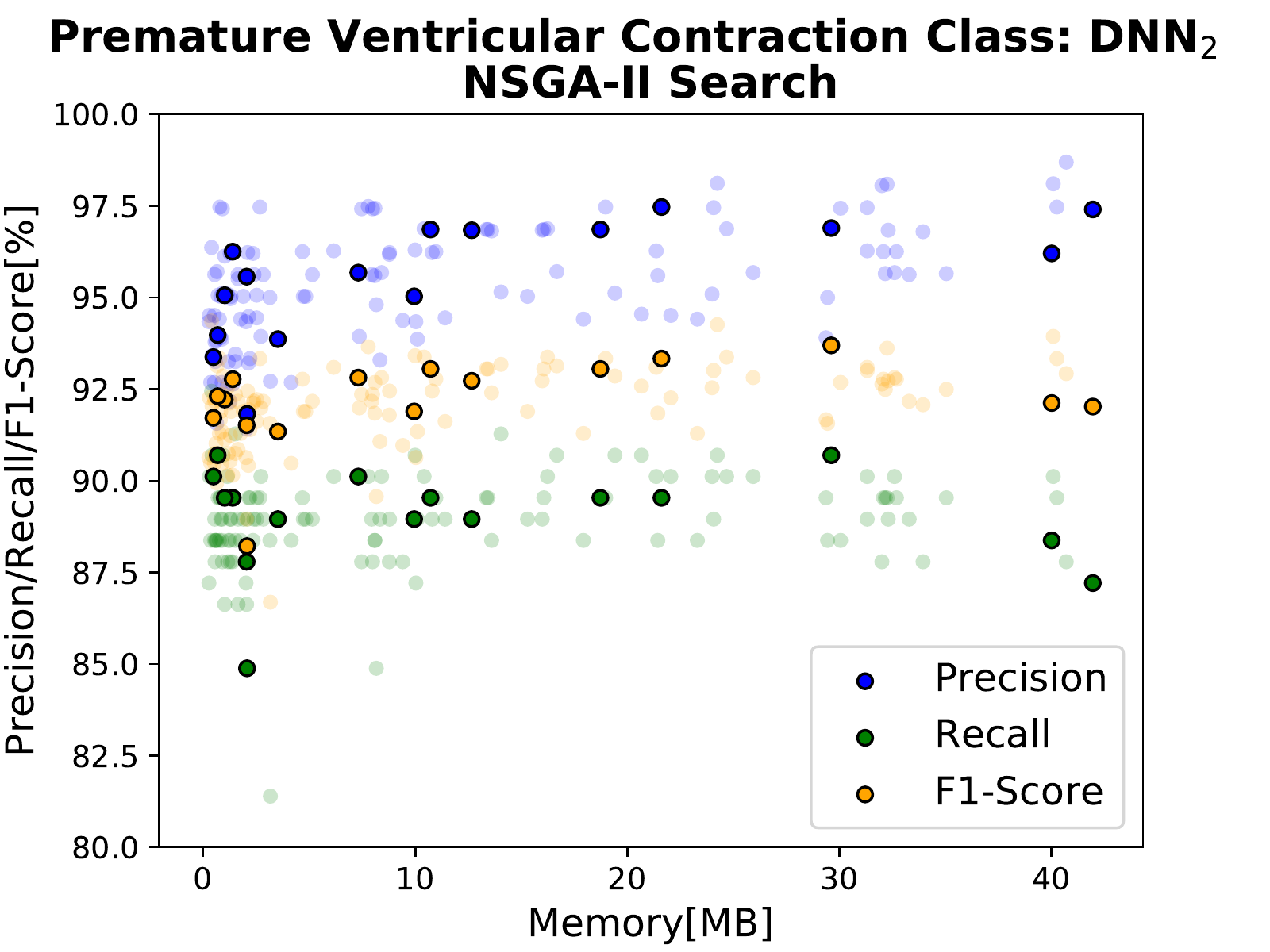}
\includegraphics[scale=0.27]{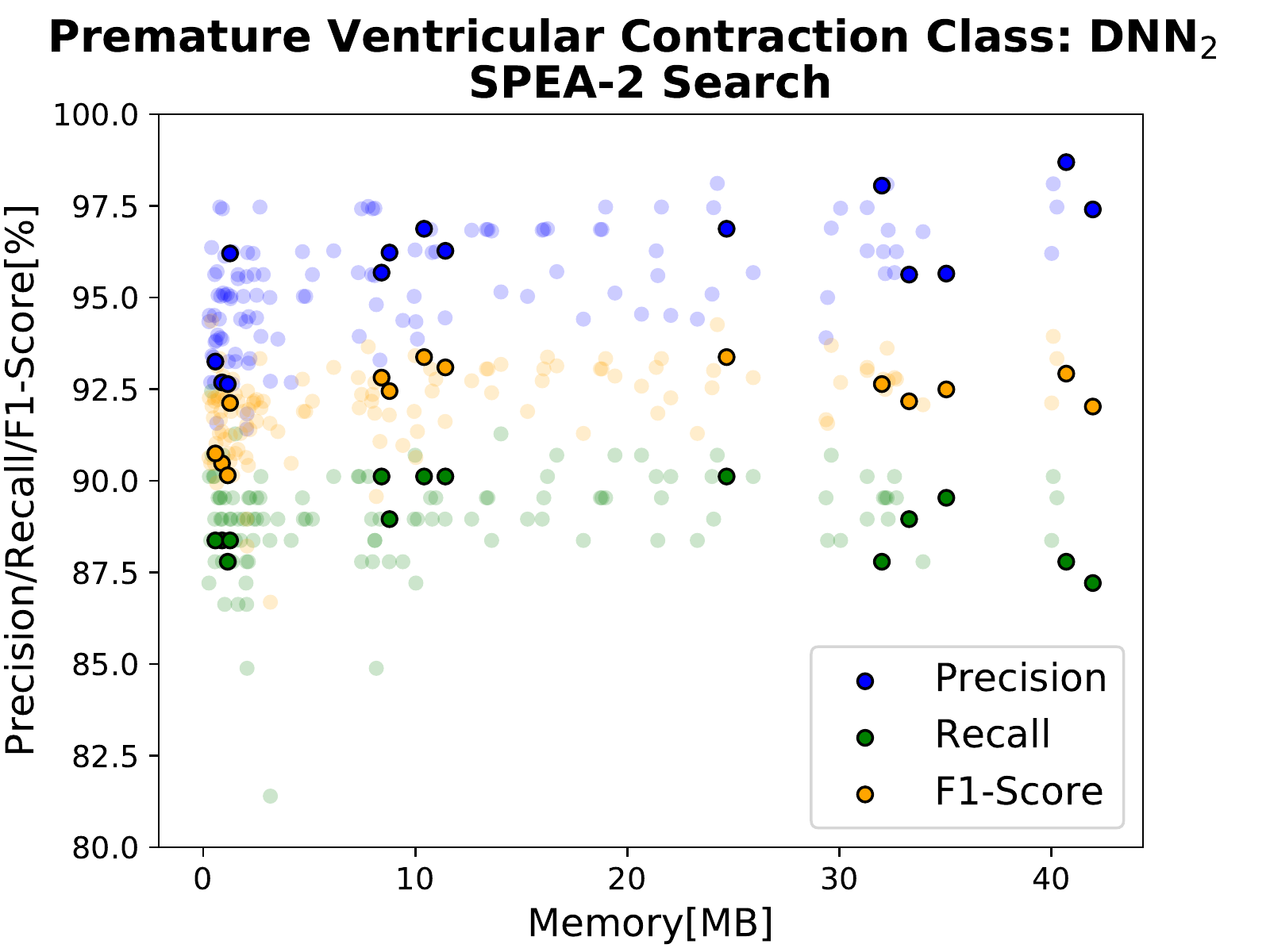}
\end{minipage}
\caption{\textbf{Exploration of the \textbf{DNN$_2$} Architecture-Space Using Genetic Algorithms (Explored Designs are Highlighted).}}
\label{fig:DNN2GA}
\end{figure*}

\begin{figure*}[!t]
\captionsetup{singlelinecheck=false}
\begin{minipage}{\textwidth}
\centering
\includegraphics[scale=0.27]{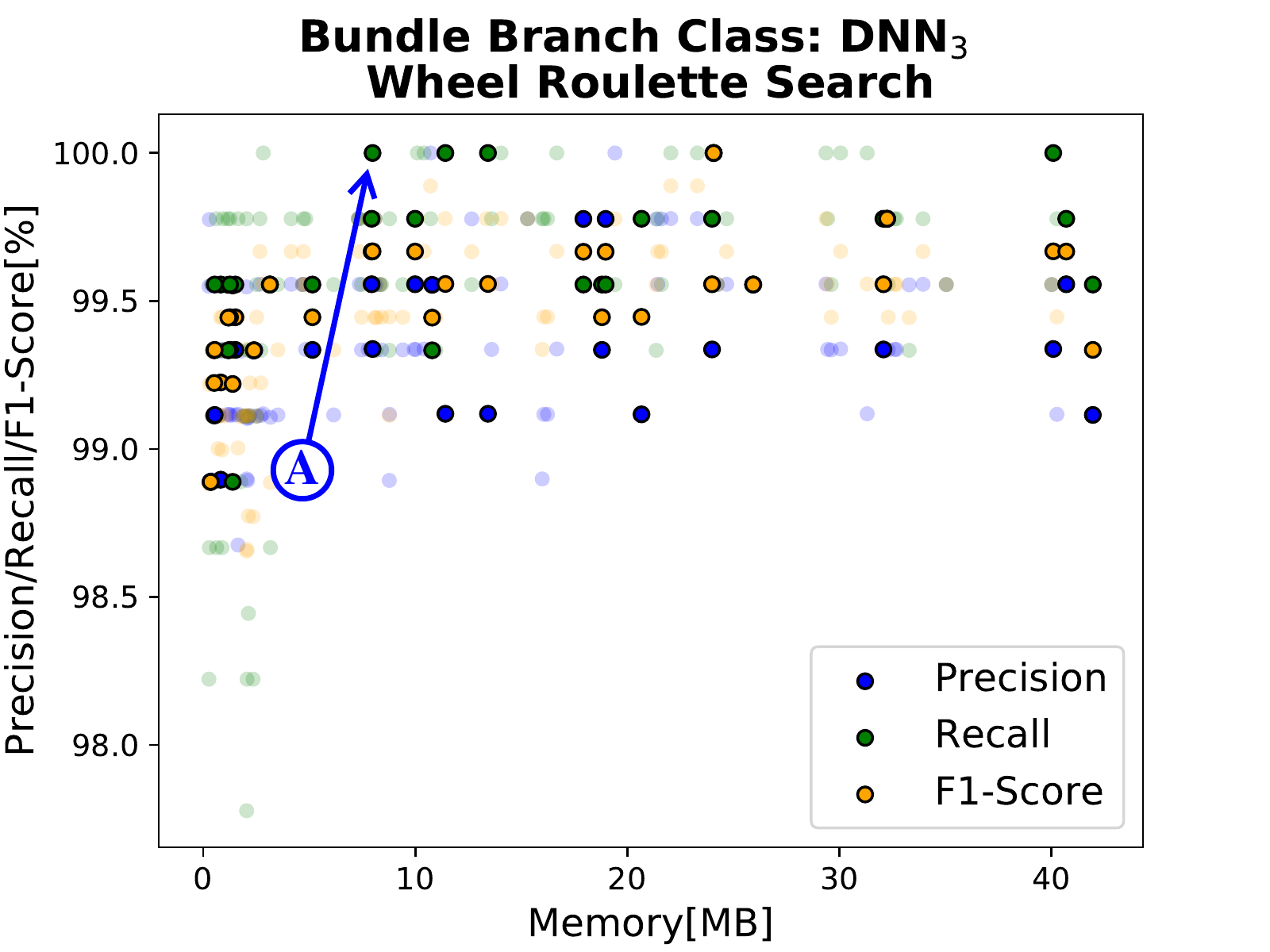}
\includegraphics[scale=0.27]{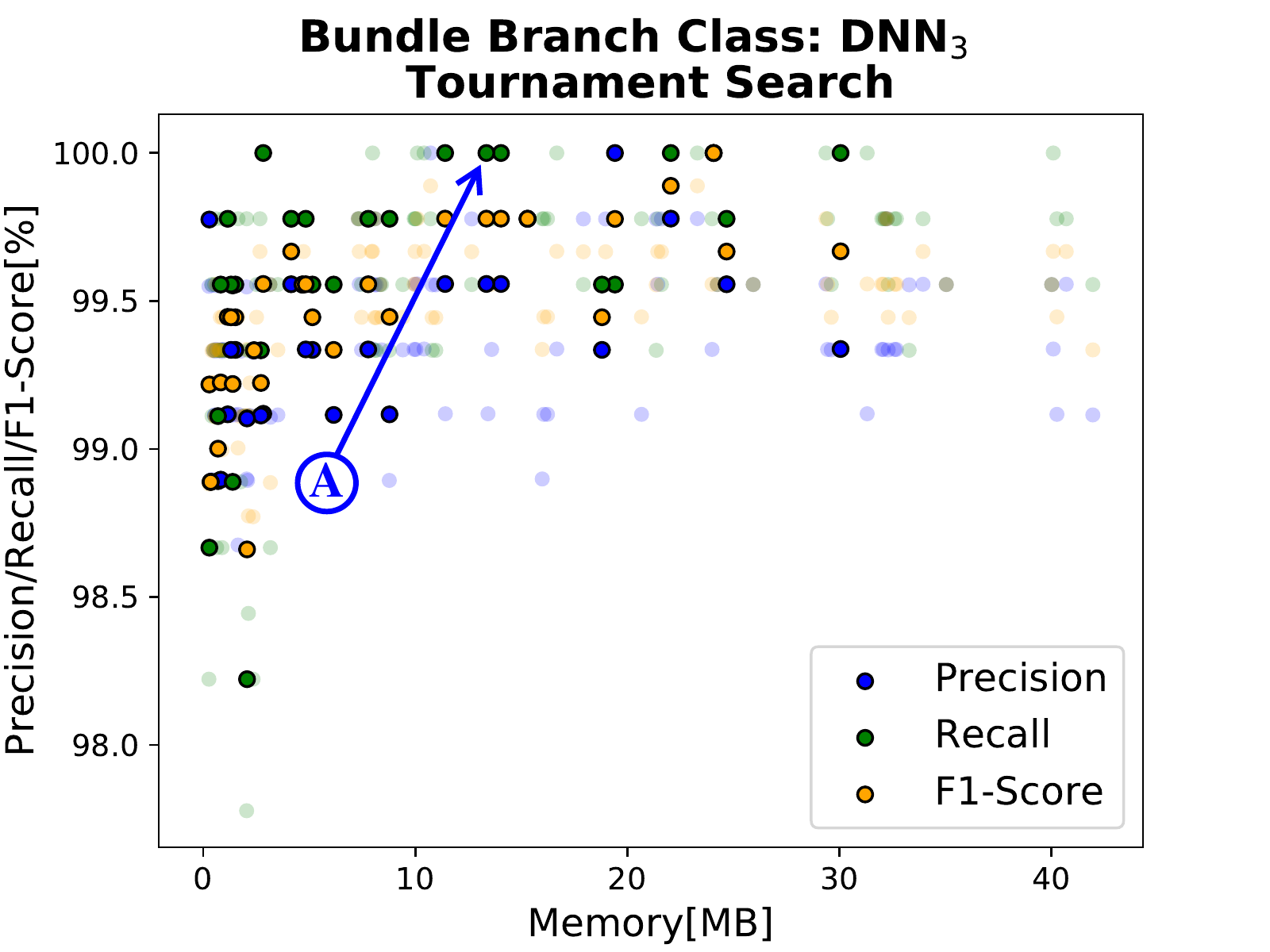}
\includegraphics[scale=0.27]{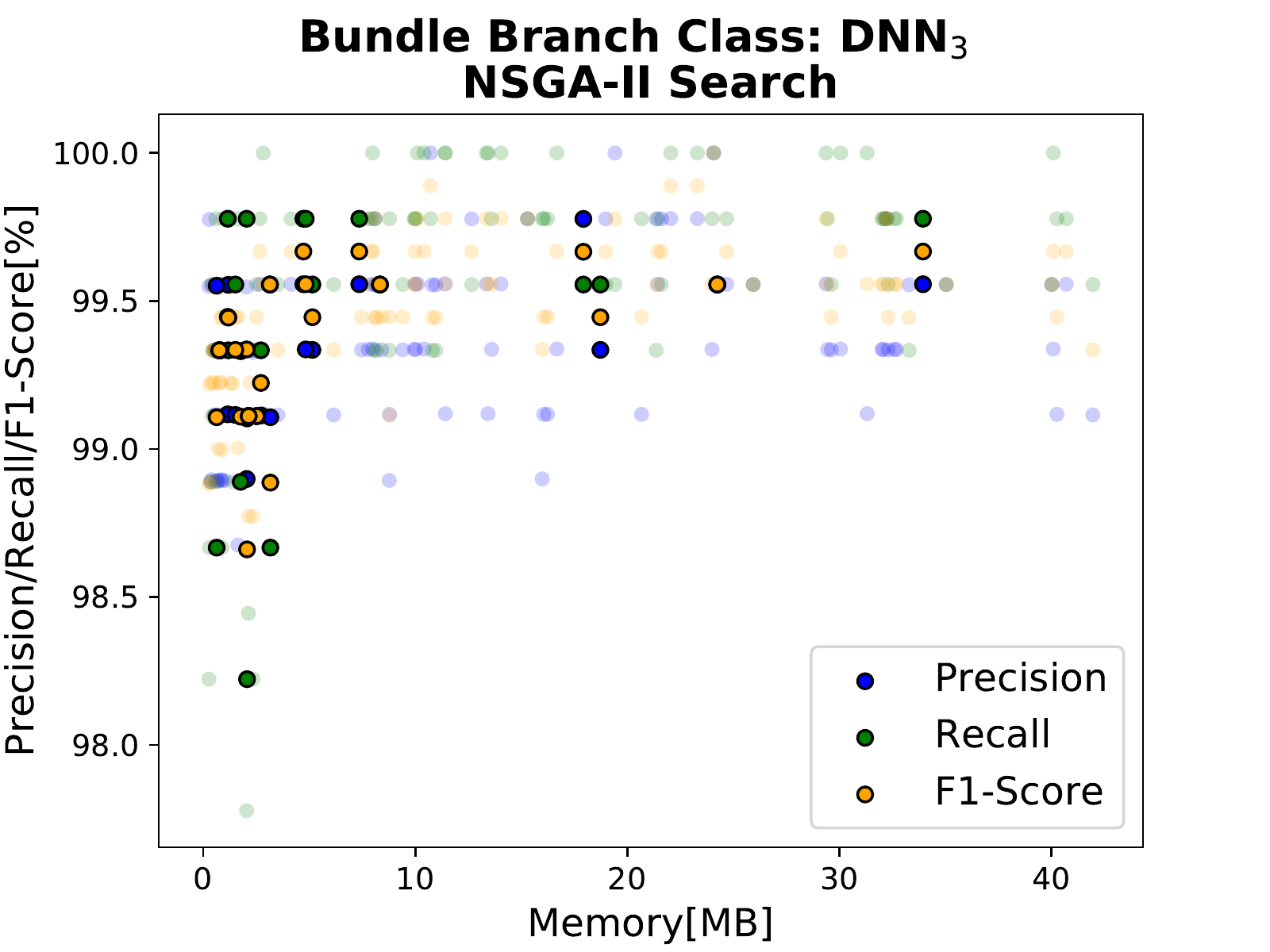}
\includegraphics[scale=0.27]{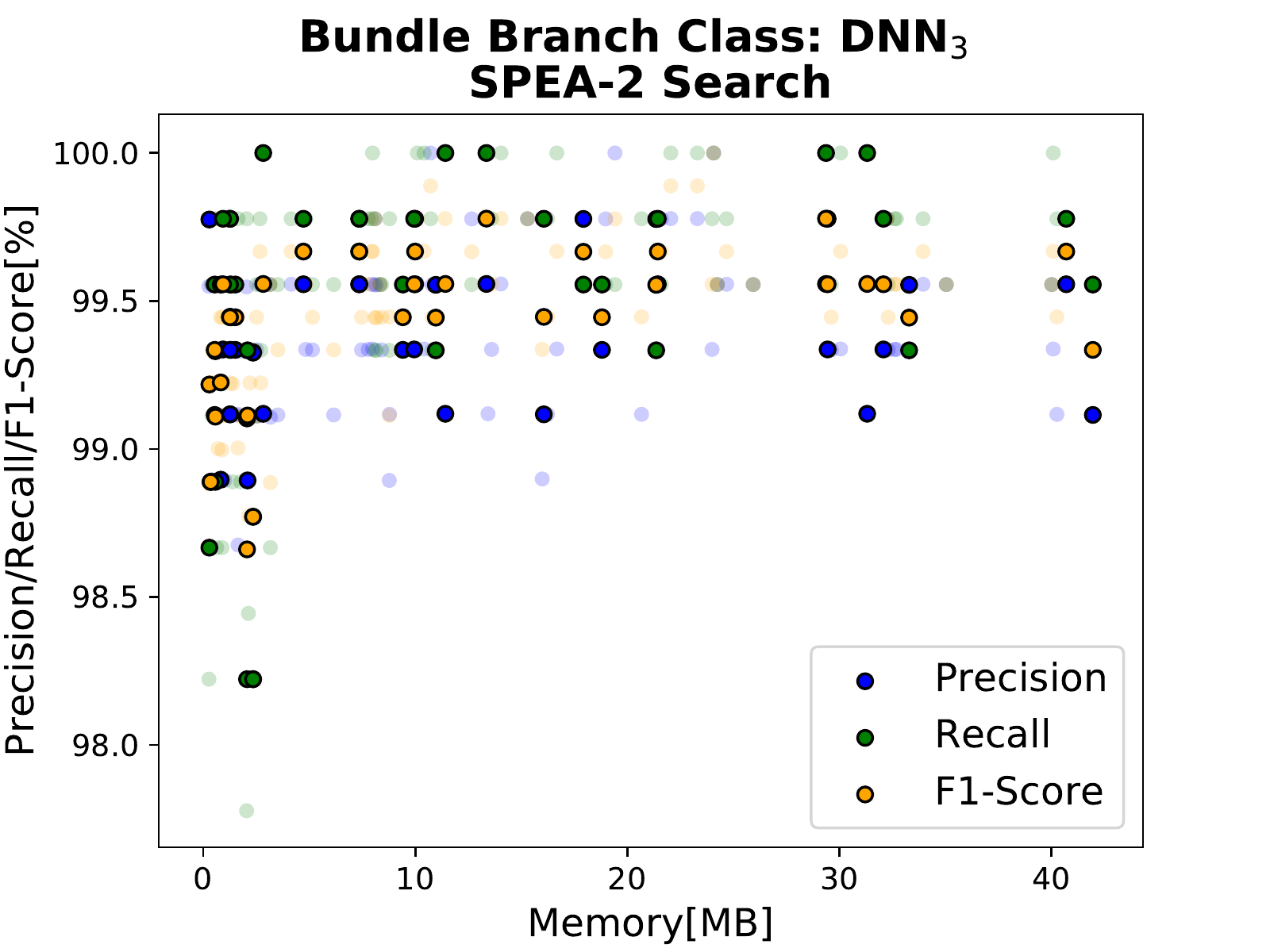}
\end{minipage}
\caption{\textbf{Exploration of the \textbf{DNN$_3$} Architecture-Space Using Genetic Algorithms (Explored Designs are Highlighted).}}
\label{fig:DNN3GA}
\end{figure*}

\begin{figure*}[!t]
\captionsetup{singlelinecheck=false}
\begin{minipage}{\textwidth}
\centering
\includegraphics[scale=0.27]{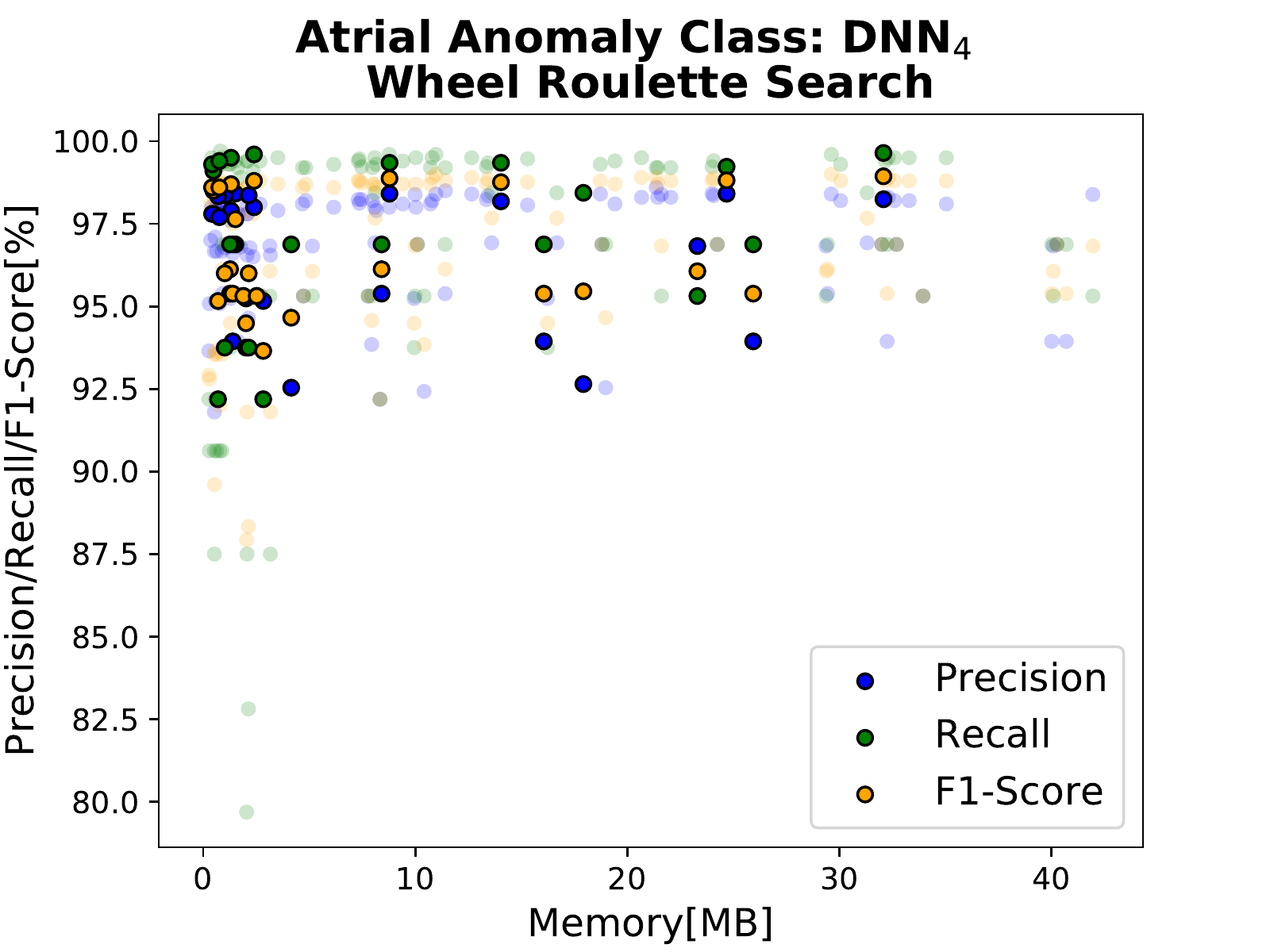}
\includegraphics[scale=0.27]{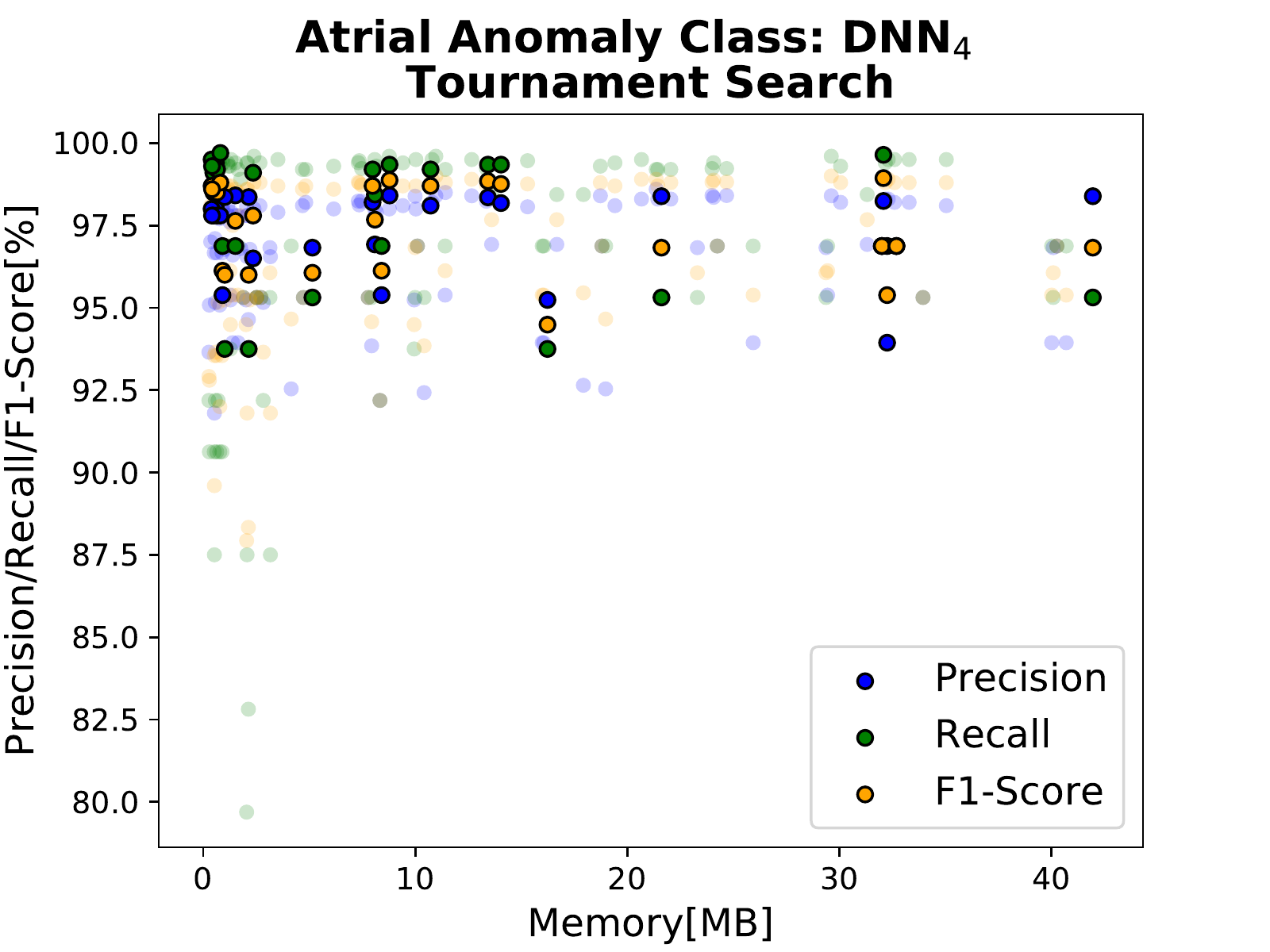}
\includegraphics[scale=0.27]{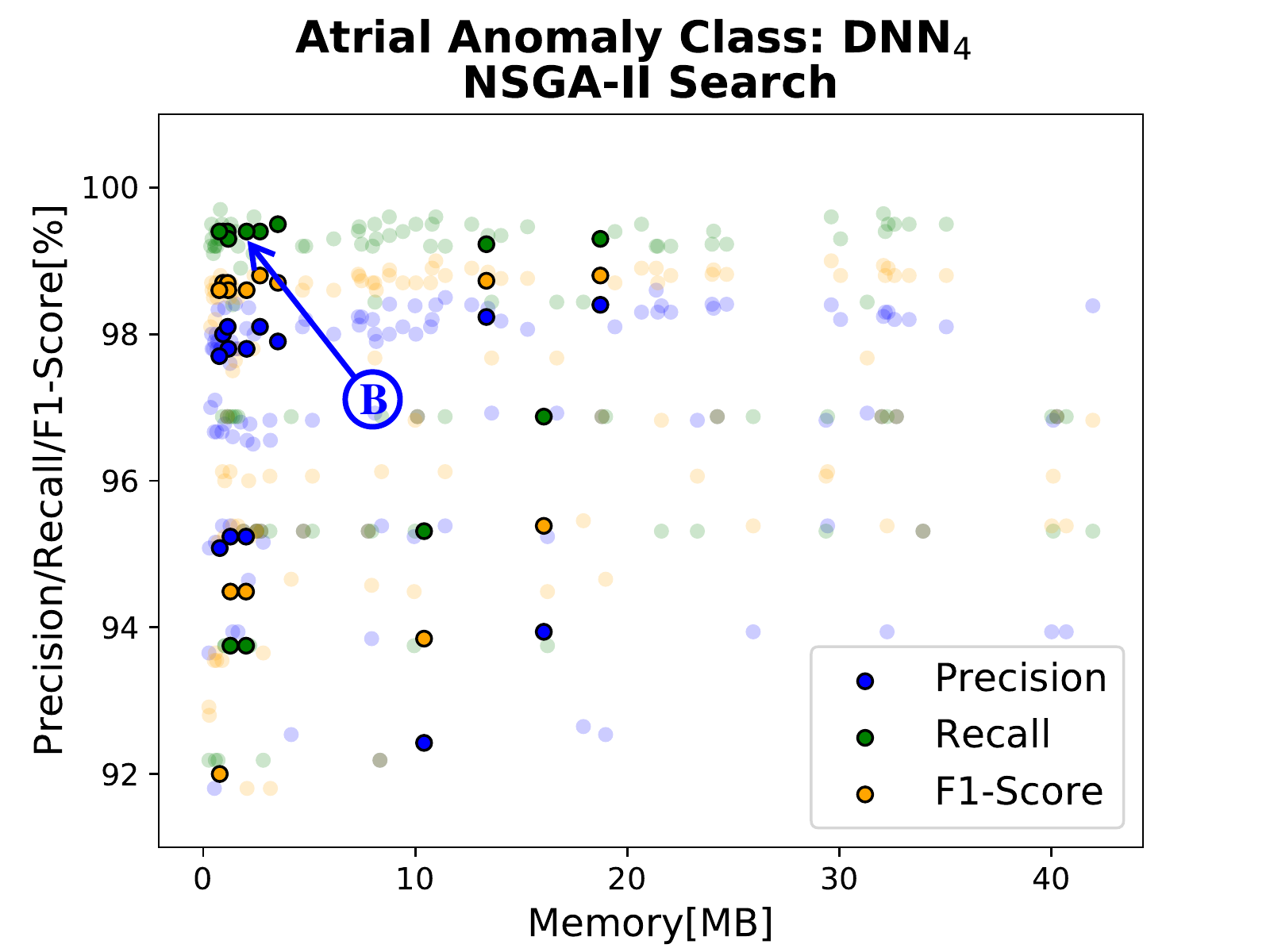}
\includegraphics[scale=0.27]{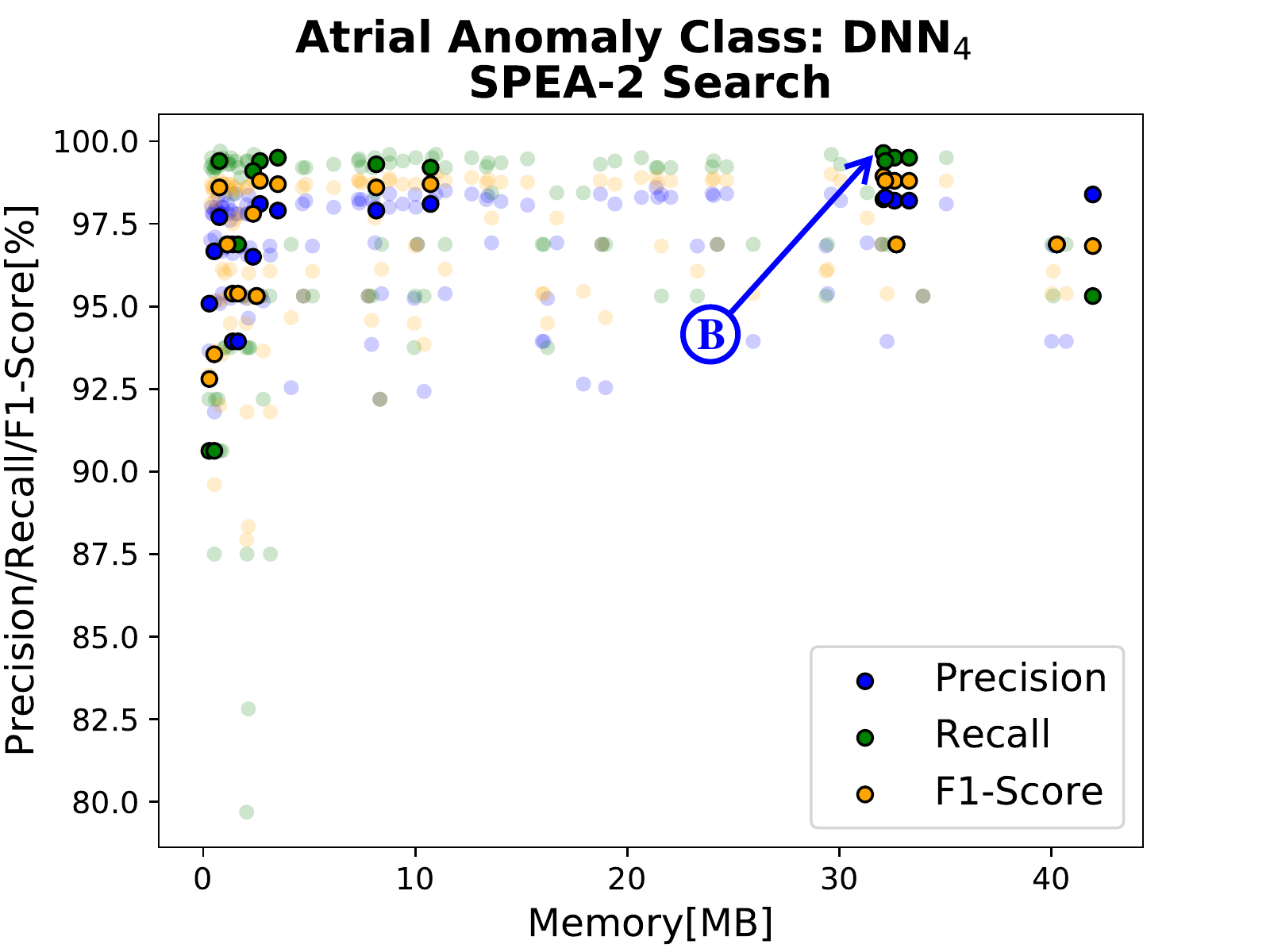}
\end{minipage}
\caption{\textbf{Exploration of the \textbf{DNN$_4$} Architecture-Space Using Genetic Algorithms (Explored Designs are Highlighted).}}
\label{fig:DNN4GA}
\end{figure*}

\subsection{Genetic Algorithm-Based Exploration and Analysis of Deep Neural Networks from the \textbf{DNN$_1$}, \textbf{DNN$_2$}, \textbf{DNN$_3$}, and \textbf{DNN$_4$} Architecture-Spaces}

As illustrated in Fig.~\ref{fig:TimeGraph}, the training time benefits obtained by using genetic algorithms to search and select points from the architecture-space is quite significant.
% However, the use of genetic algorithms is not \textit{ideal}, i.e., they do not identify all the Pareto-optimal designs like the exhaustive exploration technique, but instead identifies designs that are near-optimal. 
We use the four genetic algorithms (see Section~\ref{sec:Meth}) instead of exhaustive exploration to search the architecture-space for all five DNNs described at the start of Section~\ref{sec:ECG}, using equal weights ($\alpha=0.5,~\beta=0.5$) to construct the cost function. 

We illustrate a subset of the results obtained from these experiments in Figs.~\ref{fig:DNN1GA},~\ref{fig:DNN2GA},~\ref{fig:DNN3GA},~and~\ref{fig:DNN4GA}.
Genetic Algorithms identify near-optimal designs, without traversing the entire architecture-space of DNNs, thereby reducing the exploration time. 
It is important to note that a design selected by one algorithm might not be selected by another algorithm (for example, labels \circled{A} and \circled{B} in Figs.~\ref{fig:DNN3GA}~and~\ref{fig:DNN4GA}, respectively). 
Therefore, a genetic algorithm should be selected by the requirements of the application and the system designer.
This requires a preliminary analysis and evaluation of all genetic algorithms under consideration.
It is also interesting to note that NSGA-II and SPEA-2 select and train a small subset of the networks from the architecture-space as compared to the Roulette Wheel and Tournament Search techniques.

\begin{figure*}[t]
\captionsetup{singlelinecheck=false}
\begin{minipage}{\textwidth}
\centering
\includegraphics[scale=0.36]{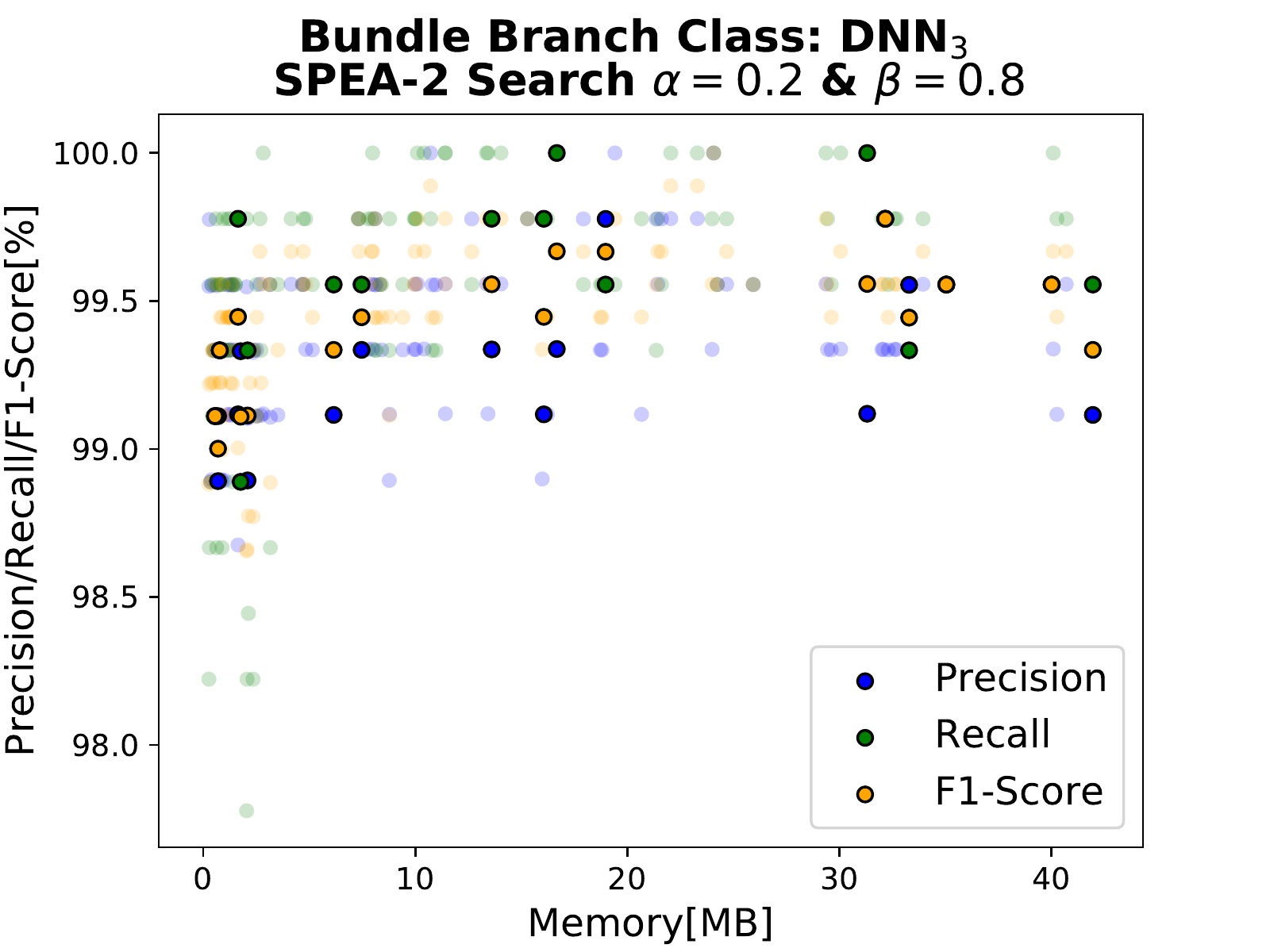}
\includegraphics[scale=0.36]{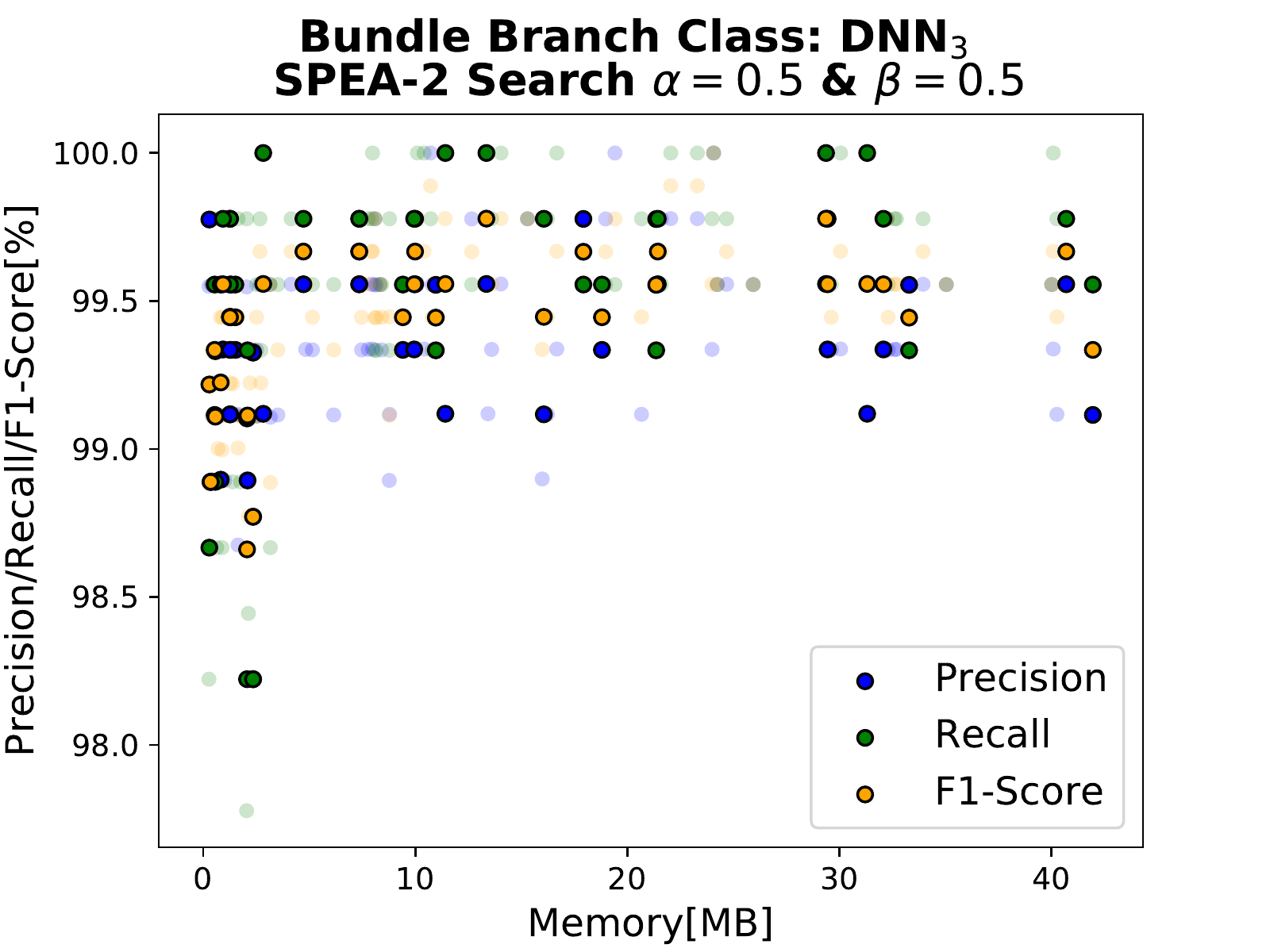}
\includegraphics[scale=0.36]{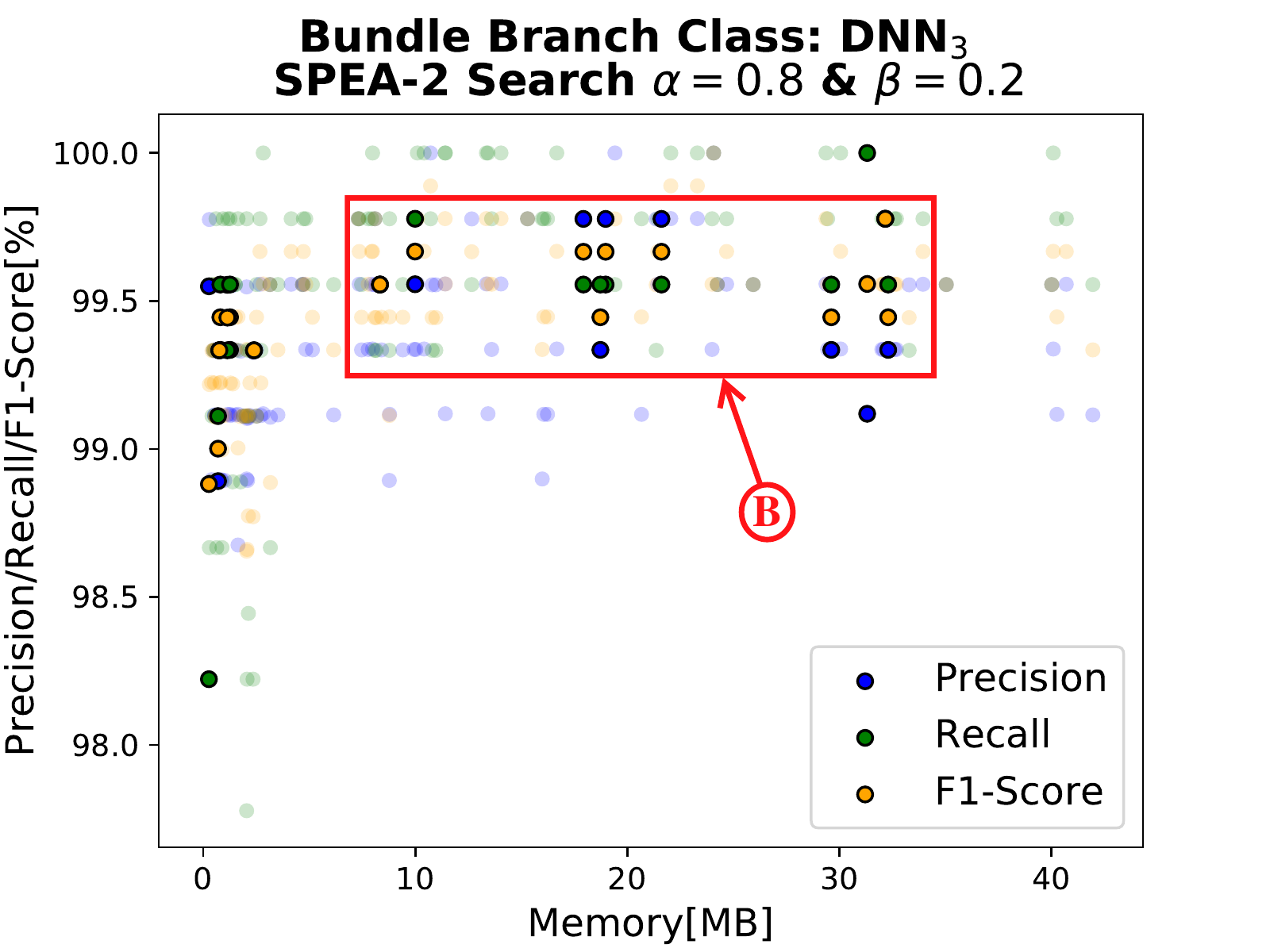}
\end{minipage}
\caption{\textbf{Weighted Architecture Search of \textbf{DNN$_3$} for the \textit{Bundle Branch} Class for Various Values of $\alpha$ and $\beta$.}}
\label{fig:DNN3WO}
\end{figure*}

\begin{figure*}[t]
\captionsetup{singlelinecheck=false}
\begin{minipage}{\textwidth}
\centering
\includegraphics[scale=0.36]{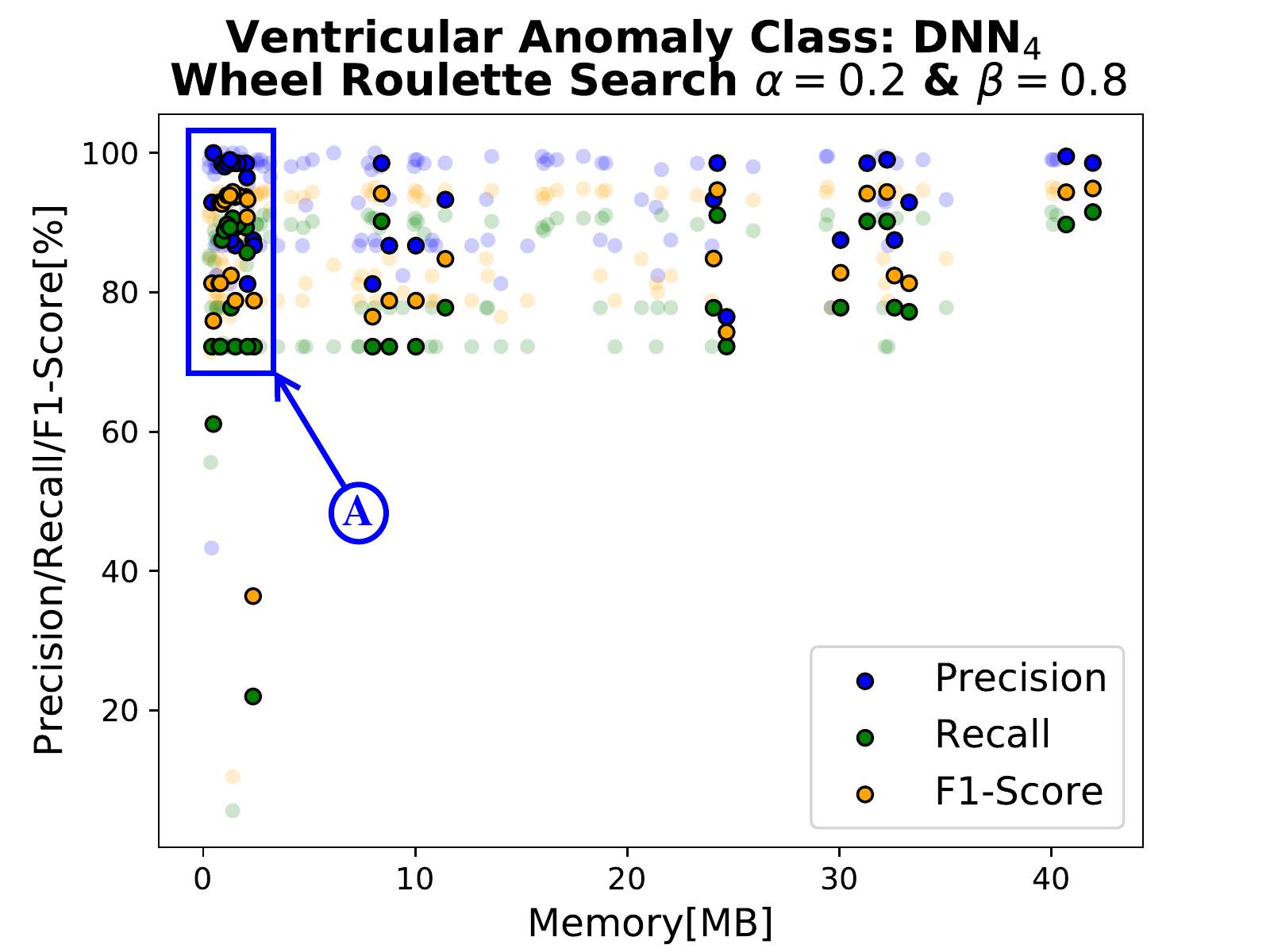}
\includegraphics[scale=0.36]{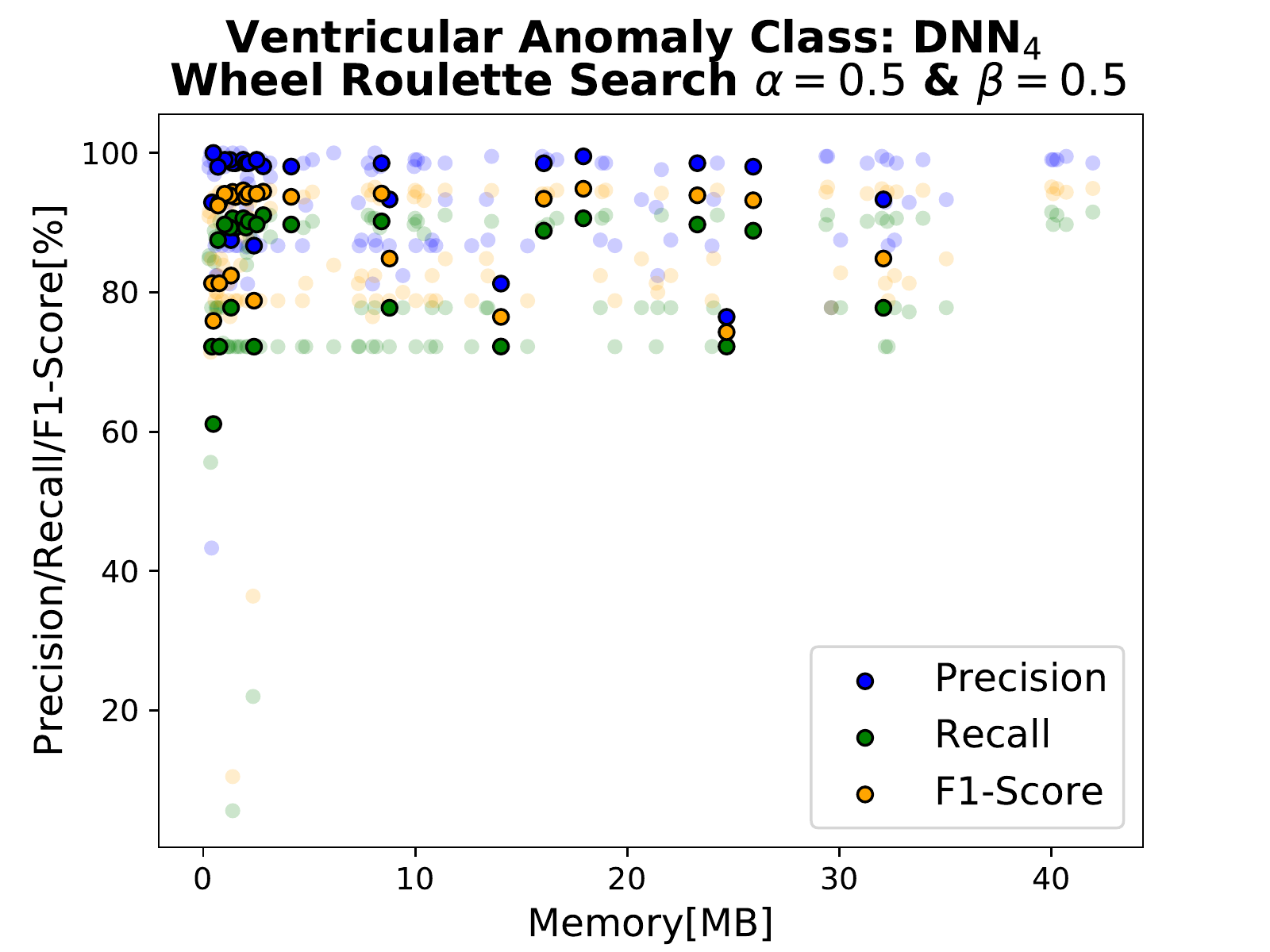}
\includegraphics[scale=0.36]{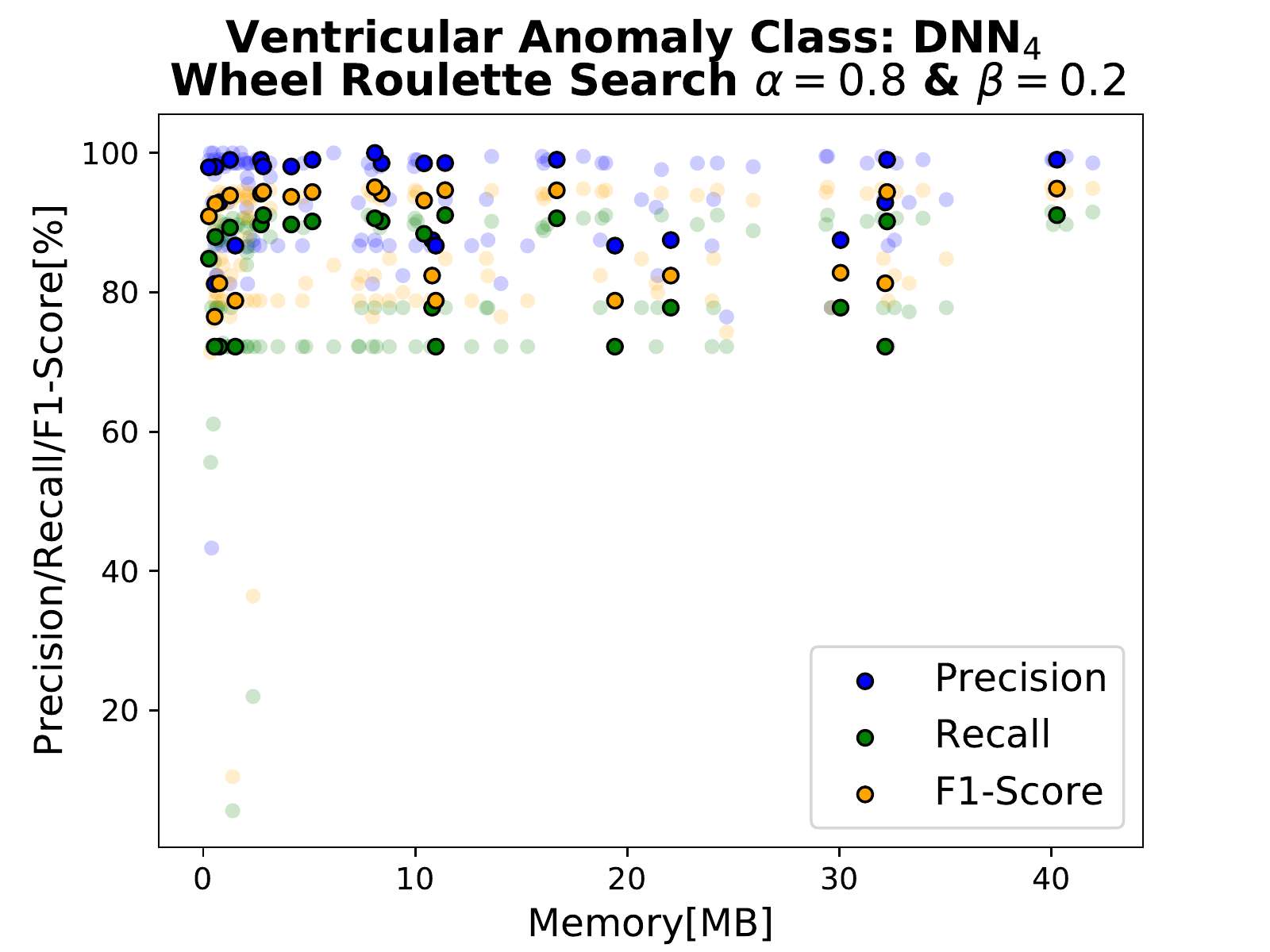}
\end{minipage}
\caption{\textbf{Weighted Architecture Search of \textbf{DNN$_4$} for the \textit{Ventricular Anomaly} Class for Various Values of $\alpha$ and $\beta$.}}
\label{fig:DNN4WO}
\end{figure*}

\begin{figure*}[t]
\captionsetup{singlelinecheck=false}
\begin{minipage}{\textwidth}
\centering
\includegraphics[scale=0.36]{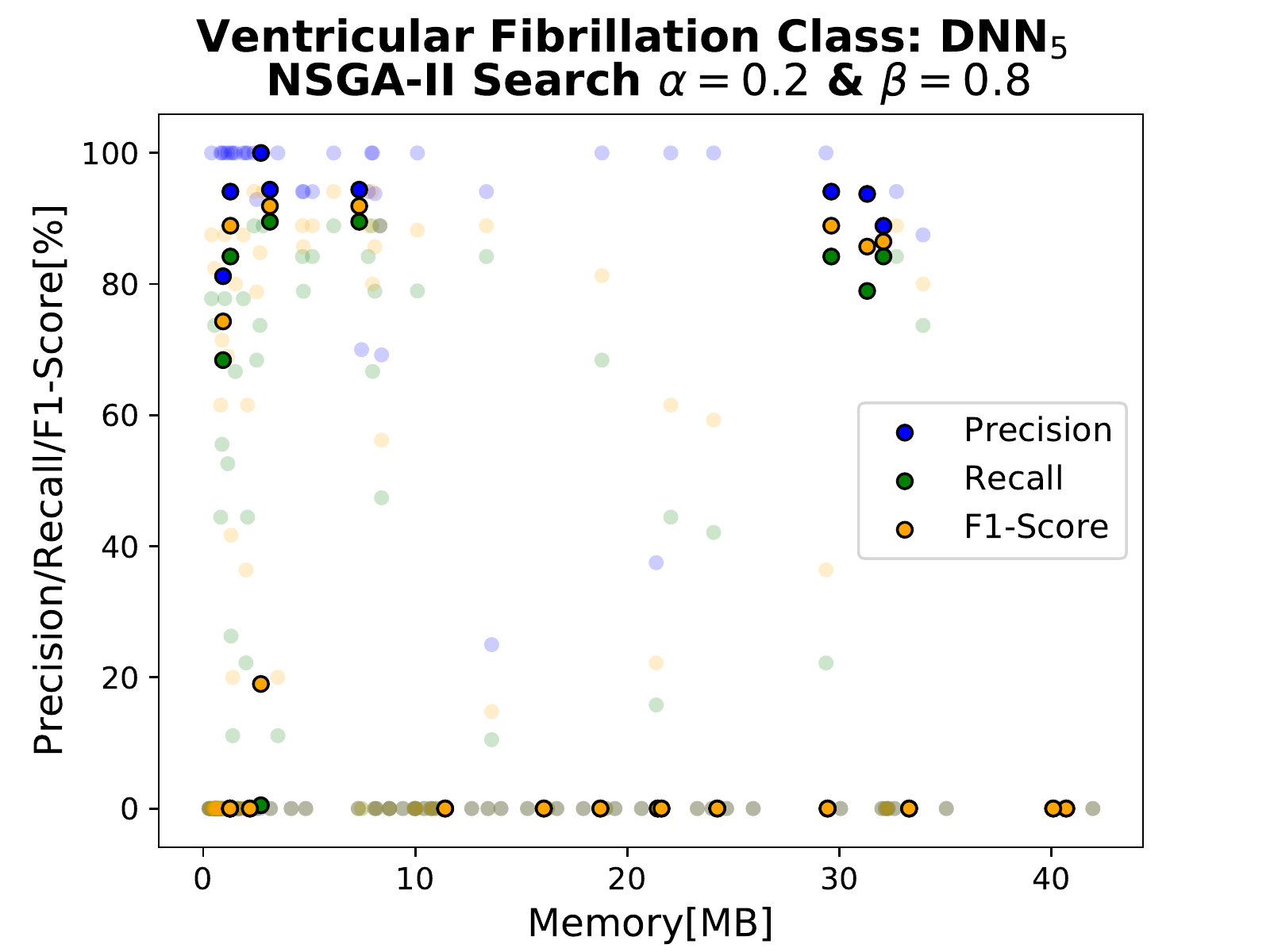}
\includegraphics[scale=0.36]{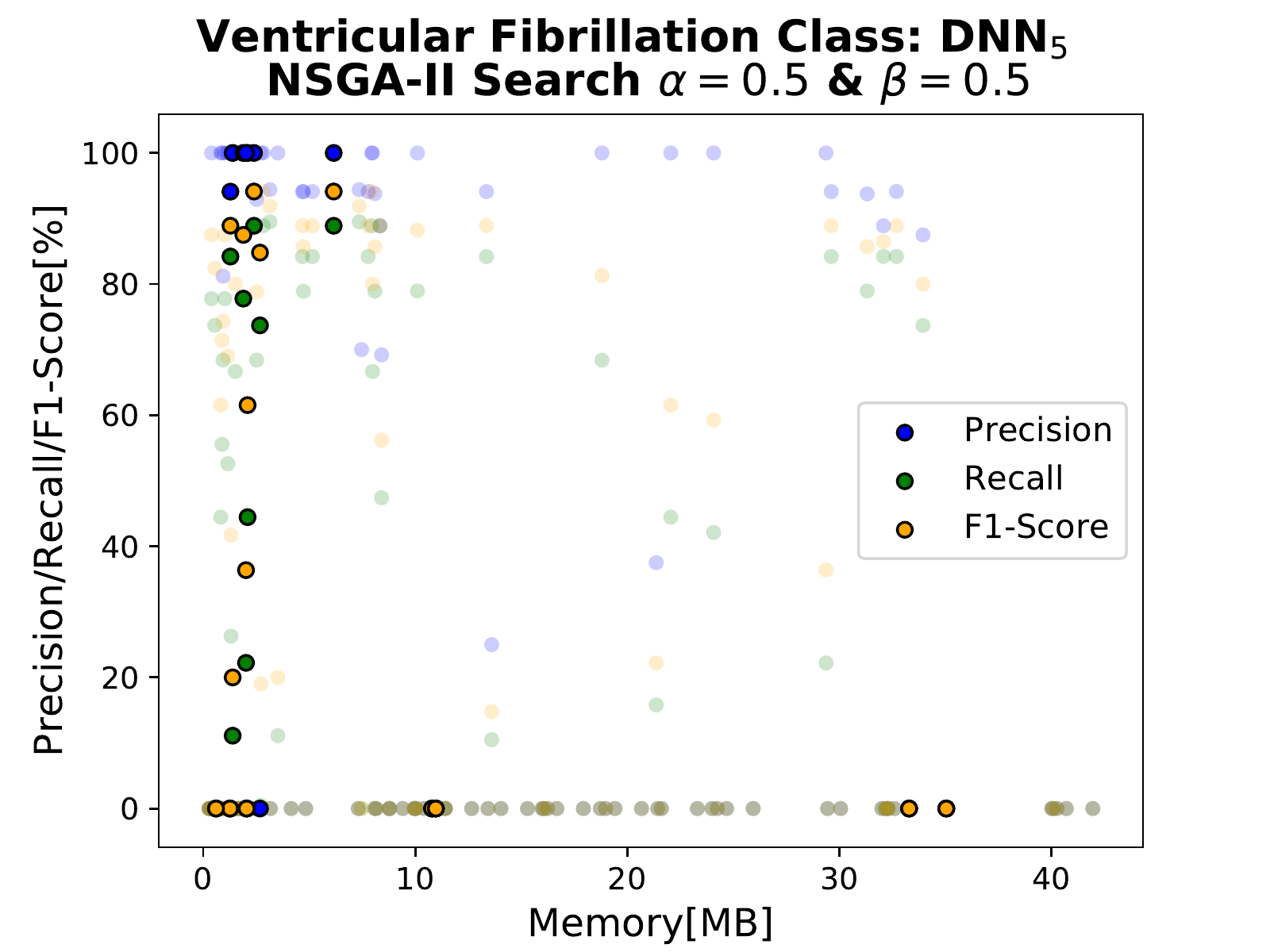}
\includegraphics[scale=0.36]{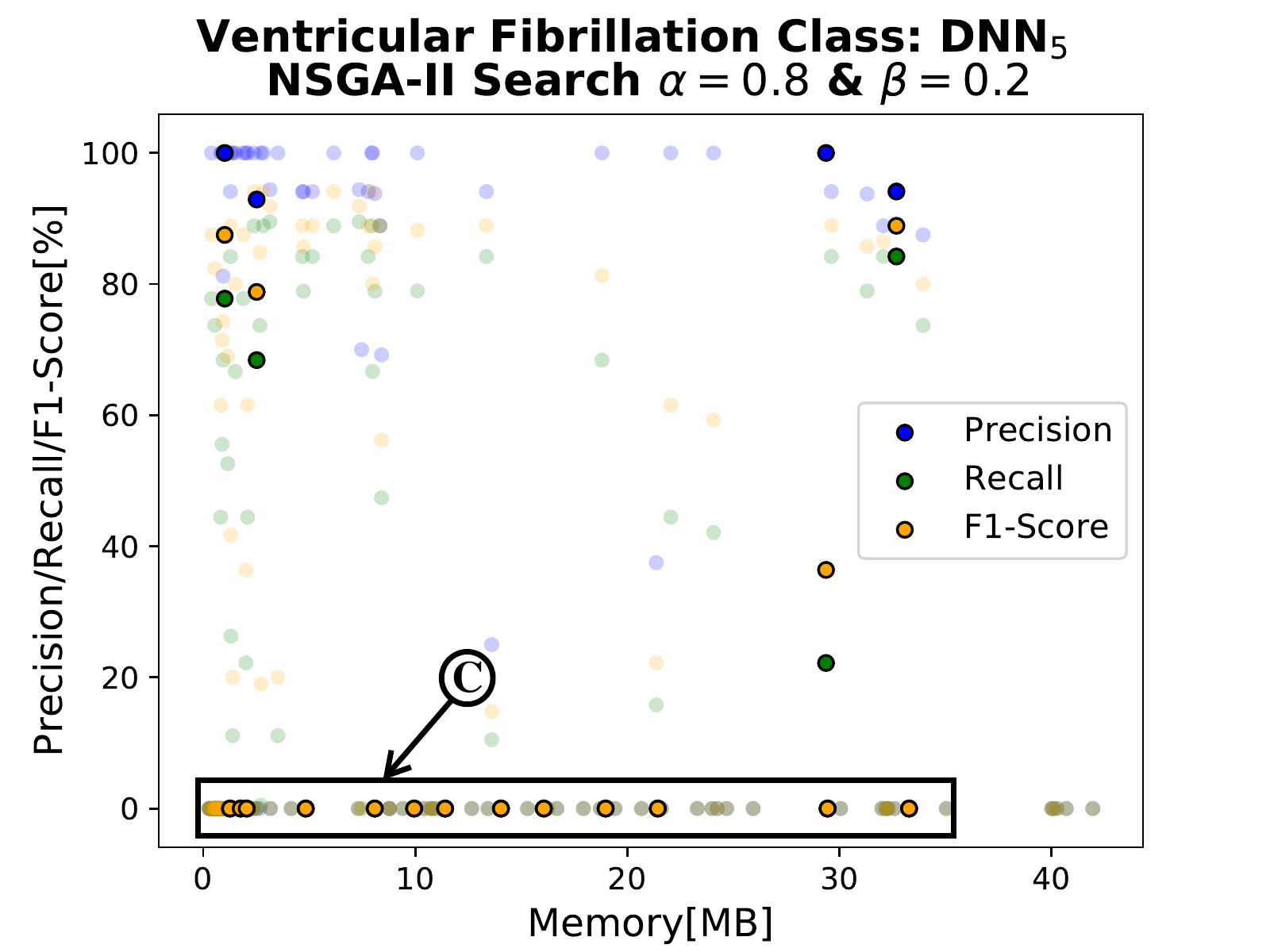}
\end{minipage}
\caption{\textbf{Weighted Architecture Search of \textbf{DNN$_5$} for the \textit{Ventricular Fibrillation} Class for Various Values of $\alpha$ and $\beta$.}}
\label{fig:DNN5WO}
\end{figure*}

\subsection{Proposed Weighted Architecture Search and Analysis of Networks from \textbf{DNN$_3$}, \textbf{DNN$_4$}, and \textbf{DNN$_5$} Architecture-Spaces}

Next, we explore the architecture-space of all DNNs using the proposed weighted DNN architecture search approach illustrated in Algorithm~\ref{Algo1}.
Although we have explored the architecture-space of all DNNs using all the genetic algorithms for various values of $\alpha$ and $\beta$, we illustrate a subset of these results for \textbf{DNN$_3$}, \textbf{DNN}$_4$, and \textbf{DNN}$_5$ architectures for three cases, namely,
\begin{inlinelist}
    \item focus on minimizing storage overhead ($\alpha=0.2$ and $\beta=0.8$),
    \item simultaneously optimize for quality and storage overhead ($\alpha=0.5$ and $\beta=0.5$), and
    \item focus on maximizing quality ($\alpha=0.8$ and $\beta=0.2$),
\end{inlinelist}
in Fig.~\ref{fig:DNN3WO},~\ref{fig:DNN4WO},~and~\ref{fig:DNN5WO}. %Figs.~\ref{fig:DNN4WO},~\ref{fig:DNN5WO},~and~\ref{fig:DNN2WO}.

The DNN architectures selected by the genetic algorithms clearly illustrate the effects of the cost function on the architecture-space exploration.
For example, in Case-(i), a large number of points are selected closer to the low-memory region, because that is the focus of exploration (for example, label \circled{A} in Fig.~\ref{fig:DNN4WO}). 
Similarly, in Case-(iii), since the search focuses more on \changed{the} accuracy, a large number of points with high accuracy are selected, as depicted by label \circled{B} in Fig.~\ref{fig:DNN3WO}.
A large percentage of the \textbf{DNN}$_5$ architectures have $0\%$ output quality (label \circled{C} in Fig.~\ref{fig:DNN5WO}) due to a bias against Ventricular Fibrillation in the number of samples present in the dataset.

\subsection{Results for Pruning and Quantization of Deep Neural Networks from DNN$_2$, DNN$_3$, and DNN$_4$ Architecture-Spaces}

Finally, we evaluate the benefits of applying pruning and quantization techniques on three different DNN architectures obtained from exploring the architecture-spaces of \textbf{DNN}$_2$, \textbf{DNN}$_3$, and \textbf{DNN}$_4$.
The first DNN architecture (\textbf{M1}), denoted by [$\alpha=1$], prioritizes the accuracy of the DNN with no consideration of the hardware overhead of the network.
The second network architecture (\textbf{M2}), depicted by [$\alpha=0.5,~\beta=0.5$], is the trade-off point obtained from the architecture-space of the DNN, which offers the best output quality for minimal hardware overhead.
The third DNN (\textbf{M3}), denoted by [$\beta=1$], prioritizes the hardware overhead of the network over its accuracy.

\begin{figure*}[!t]
    \centering
    \captionsetup{singlelinecheck=false}
    \includegraphics[width = 0.9\linewidth]{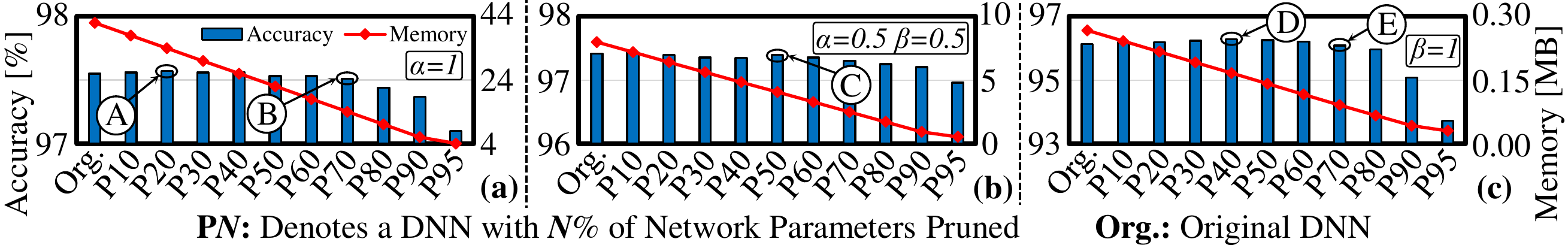}
    \caption{\textbf{Accuracy and Hardware Overhead Analysis of Pruning Pareto-Optimal \textbf{DNN$_3$}.} Fig.~(a) illustrates various levels of pruning in model \textbf{M1} ($\alpha=1$), which has the maximum accuracy from the architecture-space; Fig.~(b) analyzes various levels of pruning in model \textbf{M2} ($\alpha=0.5,~\beta=0.5$), which is the best trade-off between output quality and hardware overhead; Fig.~(c) illustrates various levels of pruning in model \textbf{M3} ($\beta=1$), which has the lowest hardware overhead in the architecture-space.}
    \label{fig:Prun}
\end{figure*}

\setcounter{figure}{22}
\begin{figure*}[!b]
    \centering
    \captionsetup{singlelinecheck=false}
    \includegraphics[width = \linewidth]{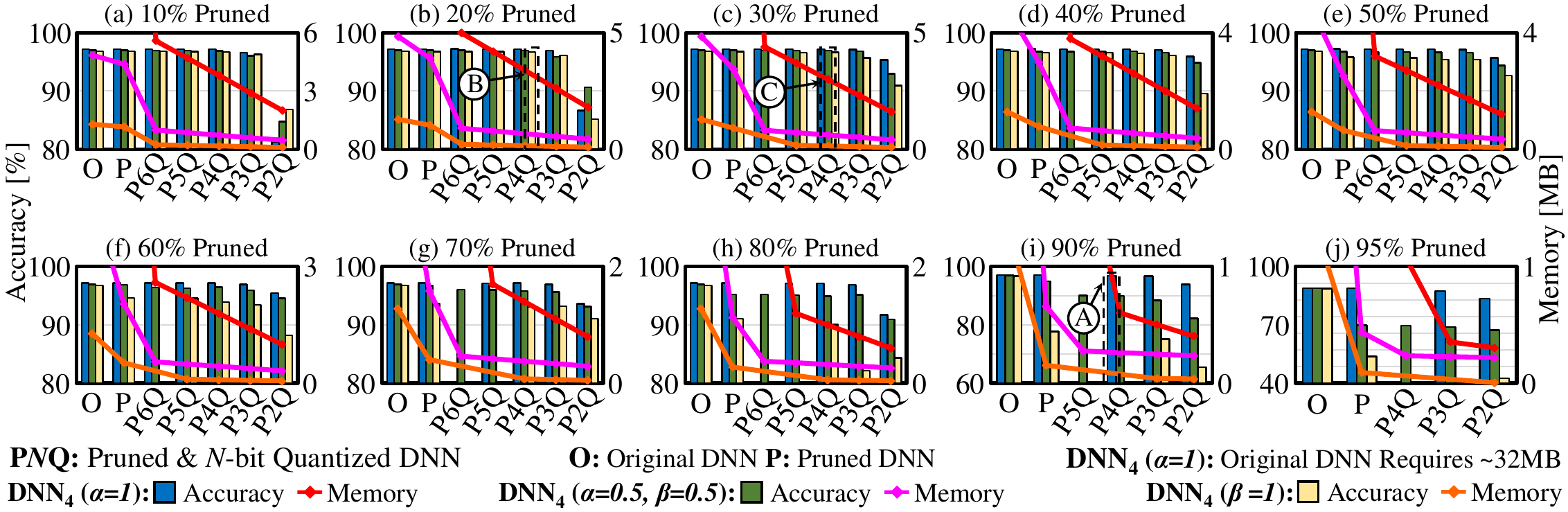}
    \caption{\textbf{Accuracy and Overhead Analysis of Pruning and Quantizing Pareto-Optimal DNN Architectures Obtained from the DNN$_4$ Architecture-Space.} Figs. (a) to (i) compare the original DNNs to networks that have been obtained after pruning X\% of network parameters and quantized, where X is increased in steps of $10$, starting from $10\%$ up to $90\%$. Fig. (j) illustrates the pruned and quantized network for $95\%$ parameters pruned. [\textbf{M1:} $\alpha=1$; \textbf{M2:} $\alpha=\beta=0.5$; \textbf{M3:} $\beta=1$].}
    \label{fig:P&Q}
\end{figure*}

\setcounter{figure}{21}
\begin{figure}[t]
    \centering
    \captionsetup{singlelinecheck=false}
    \includegraphics[width = \linewidth]{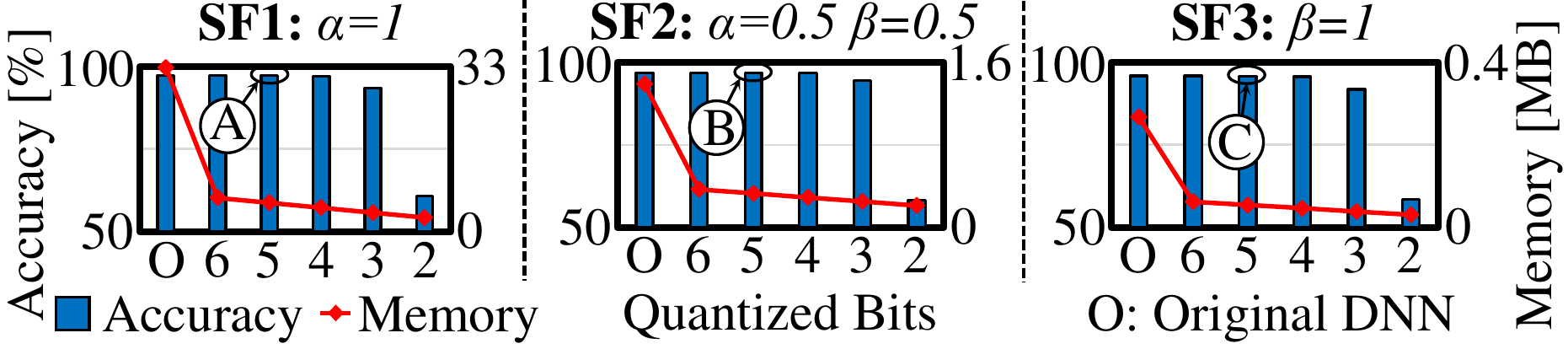}
    \caption{\textbf{Accuracy and Hardware Overhead Analysis of Quantizing Pareto-Optimal \textbf{DNN$_2$}.} The models \textbf{M1} in \textbf{SF1}, \textbf{M2} in \textbf{SF2}, and \textbf{M3} in \textbf{SF3} obtained from the \textbf{DNN}$_2$ architecture-space are quantized using $6,5,4,3,$ and $2$ bits. [\textbf{M1:} $\alpha=1$; \textbf{M2:} $\alpha=\beta=0.5$; \textbf{M3:} $\beta=1$]}
    \label{fig:Quan}
\end{figure}

We illustrate the benefits of pruning various percentages of DNN parameters on the above-discussed three different networks from the \textbf{DNN}$_3$ architecture-space using the technique presented in~\cite{marchisio2018prunet} (see Fig.~\ref{fig:Prun}).
Similarly, we illustrate the reduction in storage overhead for various levels of quantization for the networks obtained from the \textbf{DNN}$_2$ architecture-space (see Fig.~\ref{fig:Quan}).
Finally, we illustrate the benefits of combining pruning and quantization (discussed in~\cite{han2015deep} for image classification) on the networks obtained from the architecture-space of \textbf{DNN}$_4$ in Fig.~\ref{fig:P&Q}.

\setcounter{figure}{23}
\begin{figure}[t]
    \centering
    \captionsetup{singlelinecheck=false}
    \includegraphics[width = \linewidth]{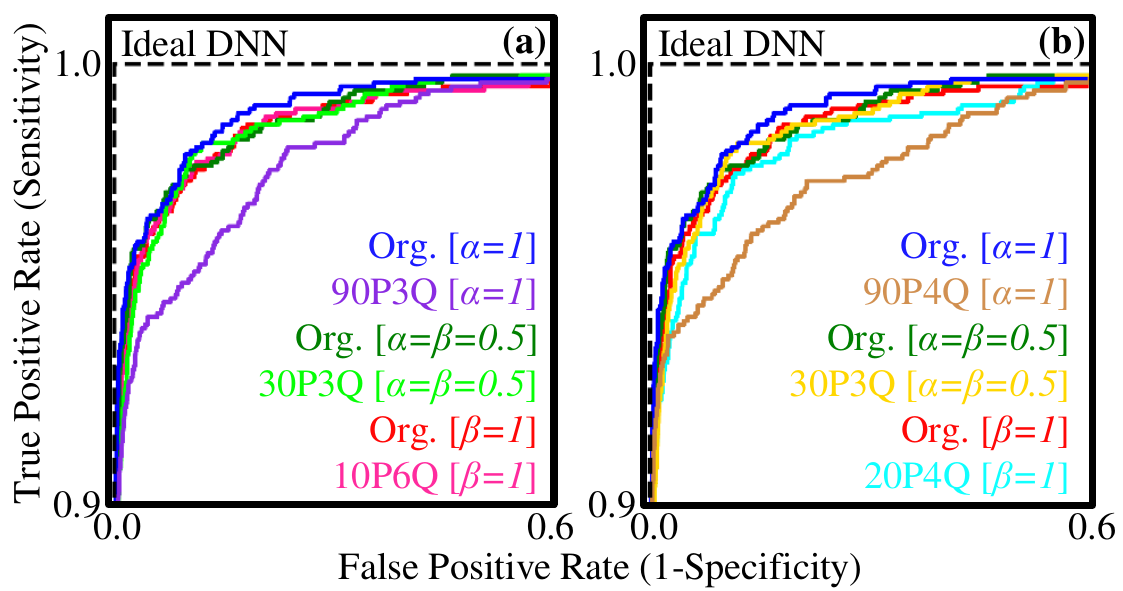}
    \caption{\textbf{ROC Graphs of Similar DNN$_4$ Architectures.} \textbf{M}P\textbf{N}Q depicts a network with \textbf{M\%} DNN parameters pruned and quantized using \textbf{N}-bits. [$\alpha=1$], [$\alpha=\beta=0.5$], and [$\beta=1$] depict models \textbf{M1}, \textbf{M2}, and \textbf{M3}, respectively, which have been discussed in the previous sub-section. The ideal curve illustrates the scenario where all classifications are correct with zero false-positives.}
    \label{fig:ROC}
\end{figure}

\textit{Pruning}, as a standalone technique, can be quite effective in reducing the hardware overhead by \midtilde$40\%$ while improving the accuracy by up to $0.15\%$ as illustrated by labels \circled{A} and \circled{D} in Fig.~\ref{fig:Prun}.
The labels \circled{B}, \circled{C}, and \circled{E} denote DNN architectures that have an output quality similar to the original network while minimizing the overhead by more than $50\%$.
Networks of model \textbf{M1} can endure the process of pruning to a significantly larger extent, without loss of accuracy, as compared to models \textbf{M2} and \textbf{M3}, primarily due to the higher number of non-essential parameters in the network. 
\textit{Quantization}, is similarly useful in reducing the hardware overhead of the network by \midtilde$5\times$ compared to the original DNNs for an output quality loss of $< 0.1\%$, as illustrated by labels \circled{A}, \circled{B}, and \circled{C} in Fig.~\ref{fig:Quan}.
A combination of pruning and quantization can be used to achieve a $53\times$ reduction in hardware overhead of the DNN for a quality loss of \midtilde$0.2\%$, as illustrated by \circled{A} in Fig.~\ref{fig:P&Q}.
Similarly, the best networks obtained from models \textbf{M2} and \textbf{M3} are depicted by labels \circled{B} and \circled{C}, similar to the output quality of the network \circled{A}.

\subsection{Receiver Operating Characteristics Analysis of the Pruned and Quantized DNNs}

Receiver operating characteristics (ROC) graphs are useful in visualizing and evaluating various types of classifiers.
They have been widely used in evaluating the quality of DNNs and ML-based classifiers since the $1990$s~\cite{spackman1989signal}.
In this work, we use ROC graphs to analyze the effectiveness of the DNNs, obtained after pruning and quantization, that have similar output quality and hardware overhead.
For this purpose, we explore the following two scenarios in each model of the DNN discussed in the previous section:
\begin{enumerate}[leftmargin=*, label=(\arabic*)]
    \item Maximize the DNN's accuracy with a constraint of $0.5$MB maximum hardware overhead (Fig.~\ref{fig:ROC}(a));
    \item Minimize the hardware overhead of the DNN with a minimum output quality of $96.7\%$ (Fig.~\ref{fig:ROC}(b)).
\end{enumerate}

The results of this analysis are presented in Fig.~\ref{fig:ROC}.
While model \textbf{M1} has the best ROC, \textbf{M2} and \textbf{M3} have similar, albeit slightly attenuated, characteristics, when compared to \textbf{M1}.
We can clearly observe that the operating characteristics of the DNN obtained after pruning $90\%$ of model \textbf{M1}'s parameters and quantizing them with $3/4$-bits \changed{have} severely deteriorated.
\textit{The pruned and quantized networks obtained from \textbf{M2} and \textbf{M3}, which have similar output quality to the pruned and quantized network obtained from \textbf{M1}, exhibit better operating characteristics, similar to the original uncompressed networks.} 
Therefore, these compressed DNNs are better suited for deployment in wearables.

To further re-emphasize the efficacy of our framework, we have included additional results in the supplementary material.

% \noindent The results of this analysis are presented in Fig.~\ref{fig:ROC}, for which the \textbf{\textit{key observations}} are:
% \begin{itemize}[leftmargin=*]
%     \item The original model \textbf{M1}, without pruning or quantization, illustrates the best characteristics and offers the best true-positive to false-positive ratio.
%     \item The original models \textbf{M2} and \textbf{M3} have similar characteristics. However, the false-positive rate of \textbf{M2} is a little less than \textbf{M3}.
%     \item The operating characteristics of the DNN obtained after pruning $90\%$ of model \textbf{M1}'s parameters and quantizing them with $3/4$-bits are severely deteriorated, as can be observed in both cases.
%     \item The pruned and quantized networks obtained from \textbf{M2} and \textbf{M3}, which have similar output quality to the pruned and quantized network obtained from \textbf{M1}, illustrate better operating characteristics, similar to the original uncompressed networks. Therefore, these compressed DNNs are better suited for deployment in wearables.
% \end{itemize}

% \lipsum[1-8]

% \begin{itemize}
%     \item Fig1: Illustrates the time benefits obtained with genetic algorithm search compared to traditional exhaustive search.
%     \item Fig2: Illustrates the exhaustive outputs for each of the different network types discusses
%     \item Fig3: Illustrates the GA search for the one of the output cases
%     \item Fig4: Different outputs for the different weights in optimization
%     \item Fig5: Drawback associated results
%     \item Fig6: Next Misc., etc.
% \end{itemize}
\section{Conclusion}
\label{sec:Conclusion}

To summarize and answer the questions raised in Section~\ref{sec:Intro}:

\noindent\textbf{[A1]} We presented the \textit{BioNetExplorer} framework, which can be used to systematically explore the architecture-space of Deep Neural Networks for bio-signal processing applications.
Based on the required output classes, user's quality requirements\changed{,} and hardware constraints of the target wearable device, we perform a genetic algorithm-based exploration of the architecture-space to identify (near-) optimal DNN architectures that can be deployed on the wearable and benchmark it against an exhaustive search, while reducing exploration time by $9\times$.

\noindent\textbf{[A2]} We proposed a weighted DNN architecture search algorithm with a modified cost function to simultaneously search for networks that offer high accuracy and minimum hardware overhead.
Based on these explorations, we have obtained a wide-range of (near-) optimal DNN architectures for all five cases that have been explored in Section~\ref{sec:ECG}.

\noindent\textbf{[A3]} We have investigated the applicability of model compression techniques such as pruning and quantization in our framework to further reduce the hardware overhead of the DNN with minimal loss in output quality.
The framework is successful in reducing the hardware overhead of the network by $53\times$ for a quality loss of less than $0.2\%$.
Furthermore, we have also illustrated that DNNs obtained through simultaneous quality and hardware overhead optimization \changed{that have} been subsequently pruned and quantized exhibit better operating characteristics, when compared to DNNs that have been optimized for quality, making them more suitable for deployment in wearables.

Our framework can also be used to evaluate DNNs for various bio-signal processing use-cases, besides anomaly detection in ECG signals, such as seizure detection and prediction using EEG signals, detecting anxiety attacks in users using a combination of ECG, blood pressure, and heartrate, hypoxia using SpO$_2$ measurements, etc.
The \textit{BioNetExplorer} framework and the explored DNN architectures will be open-sourced and available online at \textcolor{blue}{\url{https://bionetexplorer.sourceforge.io}} to ensure reproducibility.

\section*{Acknowledgement}
This work was partially supported by Doctoral College Resilient Embedded Systems which is run jointly by TU Wien's Faculty of Informatics and FH-Technikum Wien.

\bibliographystyle{IEEEtran}
\bibliography{IEEEfull,IEEEexample}

% Generated by IEEEtran.bst, version: 1.12 (2007/01/11)
\begin{thebibliography}{10}
\providecommand{\url}[1]{#1}
\csname url@samestyle\endcsname
\providecommand{\newblock}{\relax}
\providecommand{\bibinfo}[2]{#2}
\providecommand{\BIBentrySTDinterwordspacing}{\spaceskip=0pt\relax}
\providecommand{\BIBentryALTinterwordstretchfactor}{4}
\providecommand{\BIBentryALTinterwordspacing}{\spaceskip=\fontdimen2\font plus
\BIBentryALTinterwordstretchfactor\fontdimen3\font minus
  \fontdimen4\font\relax}
\providecommand{\BIBforeignlanguage}[2]{{%
\expandafter\ifx\csname l@#1\endcsname\relax
\typeout{** WARNING: IEEEtran.bst: No hyphenation pattern has been}%
\typeout{** loaded for the language `#1'. Using the pattern for}%
\typeout{** the default language instead.}%
\else
\language=\csname l@#1\endcsname
\fi
#2}}
\providecommand{\BIBdecl}{\relax}
\BIBdecl
\renewcommand{\BIBentryALTinterwordstretchfactor}{4}

\bibitem{IDCGrowth}
\BIBentryALTinterwordspacing
{International Data Corporation}. (2019) {The Growth in Connected IoT Devices}.
  [Online]. Available:
  \url{https://www.idc.com/getdoc.jsp?containerId=prUS45213219}
\BIBentrySTDinterwordspacing

\bibitem{Cisco}
\BIBentryALTinterwordspacing
{Cisco}. (2016) {Cisco Visual Networking Index: Global Mobile Data Traffic
  Forecast Update, 2015–2020}. [Online]. Available:
  \url{https://www.cisco.com/c/dam/m/en_in/innovation/enterprise/assets/mobile-white-paper-c11-520862.pdf}
\BIBentrySTDinterwordspacing

\bibitem{world2020coronavirus}
W.~H. Organization \emph{et~al.}, ``Coronavirus disease 2019 (covid-19):
  situation report, 72,'' 2020.

\bibitem{reutersGermany}
\BIBentryALTinterwordspacing
{Reuters}. (2020) {Germany Launches Smartwatch App to Monitor Coronavirus
  Spread}. [Online]. Available: \url{https://reut.rs/3b92buG}
\BIBentrySTDinterwordspacing

\bibitem{nia2015energy}
A.~M. Nia, M.~Mozaffari-Kermani, S.~Sur-Kolay, A.~Raghunathan, and N.~K. Jha,
  ``Energy-efficient long-term continuous personal health monitoring,''
  \emph{IEEE Transactions on Multi-Scale Computing Systems}, vol.~1, no.~2, pp.
  85--98, 2015.

\bibitem{stergiou2018secure}
C.~Stergiou, K.~E. Psannis, B.-G. Kim, and B.~Gupta, ``Secure integration of
  iot and cloud computing,'' \emph{Future Generation Computer Systems},
  vol.~78, pp. 964--975, 2018.

\bibitem{amodei2016deep}
D.~Amodei, S.~Ananthanarayanan, R.~Anubhai, J.~Bai, E.~Battenberg, C.~Case,
  J.~Casper, B.~Catanzaro, Q.~Cheng, G.~Chen \emph{et~al.}, ``Deep speech 2:
  End-to-end speech recognition in english and mandarin,'' in
  \emph{International conference on machine learning}, 2016, pp. 173--182.

\bibitem{wu2016google}
Y.~Wu, M.~Schuster, Z.~Chen, Q.~V. Le, M.~Norouzi, W.~Macherey, M.~Krikun,
  Y.~Cao, Q.~Gao, K.~Macherey \emph{et~al.}, ``Google's neural machine
  translation system: Bridging the gap between human and machine translation,''
  \emph{arXiv preprint arXiv:1609.08144}, 2016.

\bibitem{li2018harmonious}
W.~Li, X.~Zhu, and S.~Gong, ``Harmonious attention network for person
  re-identification,'' in \emph{Proceedings of the IEEE conference on computer
  vision and pattern recognition}, 2018, pp. 2285--2294.

\bibitem{huang2018condensenet}
G.~Huang, S.~Liu, L.~Van~der Maaten, and K.~Q. Weinberger, ``Condensenet: An
  efficient densenet using learned group convolutions,'' in \emph{Proceedings
  of the IEEE conference on computer vision and pattern recognition}, 2018, pp.
  2752--2761.

\bibitem{gehring2017convolutional}
J.~Gehring, M.~Auli, D.~Grangier, D.~Yarats, and Y.~N. Dauphin, ``Convolutional
  sequence to sequence learning,'' \emph{arXiv preprint arXiv:1705.03122},
  2017.

\bibitem{mao2017survey}
Y.~Mao, C.~You, J.~Zhang, K.~Huang, and K.~B. Letaief, ``A survey on mobile
  edge computing: The communication perspective,'' \emph{IEEE Communications
  Surveys \& Tutorials}, vol.~19, no.~4, pp. 2322--2358, 2017.

\bibitem{zhou2020near}
F.~Zhou and Y.~Chai, ``Near-sensor and in-sensor computing,'' \emph{Nature
  Electronics}, vol.~3, no.~11, pp. 664--671, 2020.

\bibitem{AppleWatch}
\BIBentryALTinterwordspacing
{Apple}. (2019) {Apple Watch Series 5}. [Online]. Available:
  \url{https://www.apple.com/apple-watch-series-5/}
\BIBentrySTDinterwordspacing

\bibitem{SamsungWatch}
\BIBentryALTinterwordspacing
{Samsung}. (2019) {Galaxy Watch}. [Online]. Available:
  \url{https://www.samsung.com/global/galaxy/galaxy-watch/}
\BIBentrySTDinterwordspacing

\bibitem{Fitbit}
\BIBentryALTinterwordspacing
{Fitbit}. (2019) {Fitness Tracker}. [Online]. Available:
  \url{https://www.fitbit.com/}
\BIBentrySTDinterwordspacing

\bibitem{seneviratne2017survey}
S.~Seneviratne, Y.~Hu, T.~Nguyen, G.~Lan, S.~Khalifa, K.~Thilakarathna,
  M.~Hassan, and A.~Seneviratne, ``A survey of wearable devices and
  challenges,'' \emph{IEEE Communications Surveys \& Tutorials}, vol.~19,
  no.~4, pp. 2573--2620, 2017.

\bibitem{turakhia2019rationale}
M.~P. Turakhia, M.~Desai, H.~Hedlin, A.~Rajmane, N.~Talati, T.~Ferris,
  S.~Desai, D.~Nag, M.~Patel, P.~Kowey \emph{et~al.}, ``Rationale and design of
  a large-scale, app-based study to identify cardiac arrhythmias using a
  smartwatch: The apple heart study,'' \emph{American heart journal}, vol. 207,
  pp. 66--75, 2019.

\bibitem{fu2018blood}
Y.~Fu and J.~Guo, ``Blood cholesterol monitoring with smartphone as
  miniaturized electrochemical analyzer for cardiovascular disease
  prevention,'' \emph{IEEE transactions on biomedical circuits and systems},
  vol.~12, no.~4, pp. 784--790, 2018.

\bibitem{sopic2018real}
D.~Sopic, A.~Aminifar, A.~Aminifar, and D.~Atienza, ``Real-time event-driven
  classification technique for early detection and prevention of myocardial
  infarction on wearable systems,'' \emph{IEEE transactions on biomedical
  circuits and systems}, vol.~12, no.~5, pp. 982--992, 2018.

\bibitem{amirshahi2019ecg}
A.~Amirshahi and M.~Hashemi, ``Ecg classification algorithm based on stdp and
  r-stdp neural networks for real-time monitoring on ultra low-power personal
  wearable devices,'' \emph{IEEE transactions on biomedical circuits and
  systems}, vol.~13, no.~6, pp. 1483--1493, 2019.

\bibitem{sun2020lightweight}
J.~Sun, H.~Xiong, X.~Liu, Y.~Zhang, X.~Nie, and R.~H. Deng, ``Lightweight and
  privacy-aware fine-grained access control for iot-oriented smart health,''
  \emph{IEEE Internet of Things Journal}, 2020.

\bibitem{liu2019wireless}
W.~Liu, P.~Popovski, Y.~Li, and B.~Vucetic, ``Wireless networked control
  systems with coding-free data transmission for industrial iot,'' \emph{IEEE
  Internet of Things Journal}, vol.~7, no.~3, pp. 1788--1801, 2019.

\bibitem{elsken2018neural}
T.~Elsken, J.~H. Metzen, F.~Hutter \emph{et~al.}, ``Neural architecture search:
  A survey.'' \emph{J. Mach. Learn. Res.}, vol.~20, no.~55, pp. 1--21, 2019.

\bibitem{snoek2012practical}
J.~Snoek, H.~Larochelle, and R.~P. Adams, ``Practical bayesian optimization of
  machine learning algorithms,'' \emph{Advances in neural information
  processing systems}, vol.~25, pp. 2951--2959, 2012.

\bibitem{zoph2016neural}
B.~Zoph and Q.~V. Le, ``Neural architecture search with reinforcement
  learning,'' \emph{arXiv preprint arXiv:1611.01578}, 2016.

\bibitem{liu2018progressive}
C.~Liu, B.~Zoph, M.~Neumann, J.~Shlens, W.~Hua, L.-J. Li, L.~Fei-Fei,
  A.~Yuille, J.~Huang, and K.~Murphy, ``Progressive neural architecture
  search,'' in \emph{Proceedings of the European Conference on Computer Vision
  (ECCV)}, 2018, pp. 19--34.

\bibitem{liu2019towards}
Y.~Liu \emph{et~al.}, ``Towards flops-constrained face recognition,'' in
  \emph{Proceedings of the IEEE International Conference on Computer Vision
  Workshops}, 2019, pp. 0--0.

\bibitem{deng2009imagenet}
J.~Deng, W.~Dong, R.~Socher, L.-J. Li, K.~Li, and L.~Fei-Fei, ``Imagenet: A
  large-scale hierarchical image database,'' in \emph{2009 IEEE conference on
  computer vision and pattern recognition}.\hskip 1em plus 0.5em minus
  0.4em\relax IEEE, 2009, pp. 248--255.

\bibitem{tan2019mnasnet}
M.~Tan, B.~Chen, R.~Pang, V.~Vasudevan, M.~Sandler, A.~Howard, and Q.~V. Le,
  ``Mnasnet: Platform-aware neural architecture search for mobile,'' in
  \emph{Proceedings of the IEEE Conference on Computer Vision and Pattern
  Recognition}, 2019, pp. 2820--2828.

\bibitem{wu2019fbnet}
B.~Wu, X.~Dai, P.~Zhang, Y.~Wang, F.~Sun, Y.~Wu, Y.~Tian, P.~Vajda, Y.~Jia, and
  K.~Keutzer, ``Fbnet: Hardware-aware efficient convnet design via
  differentiable neural architecture search,'' in \emph{Proceedings of the IEEE
  Conference on Computer Vision and Pattern Recognition}, 2019, pp.
  10\,734--10\,742.

\bibitem{dai2019chamnet}
X.~Dai, P.~Zhang, B.~Wu, H.~Yin, F.~Sun, Y.~Wang, M.~Dukhan, Y.~Hu, Y.~Wu,
  Y.~Jia \emph{et~al.}, ``Chamnet: Towards efficient network design through
  platform-aware model adaptation,'' in \emph{Proceedings of the IEEE
  Conference on computer vision and pattern recognition}, 2019, pp.
  11\,398--11\,407.

\bibitem{howard2019searching}
A.~Howard, M.~Sandler, G.~Chu, L.-C. Chen, B.~Chen, M.~Tan, W.~Wang, Y.~Zhu,
  R.~Pang, V.~Vasudevan \emph{et~al.}, ``Searching for mobilenetv3,'' in
  \emph{Proceedings of the IEEE International Conference on Computer Vision},
  2019, pp. 1314--1324.

\bibitem{han2015deep}
S.~Han, H.~Mao, and W.~J. Dally, ``Deep compression: Compressing deep neural
  networks with pruning, trained quantization and huffman coding,'' \emph{arXiv
  preprint arXiv:1510.00149}, 2015.

\bibitem{anwar2017structured}
S.~Anwar, K.~Hwang, and W.~Sung, ``Structured pruning of deep convolutional
  neural networks,'' \emph{ACM Journal on Emerging Technologies in Computing
  Systems (JETC)}, vol.~13, no.~3, pp. 1--18, 2017.

\bibitem{luo2017thinet}
J.-H. Luo, J.~Wu, and W.~Lin, ``Thinet: A filter level pruning method for deep
  neural network compression,'' in \emph{Proceedings of the IEEE international
  conference on computer vision}, 2017, pp. 5058--5066.

\bibitem{marchisio2018prunet}
A.~Marchisio, M.~A. Hanif, M.~Martina, and M.~Shafique, ``Prunet: Class-blind
  pruning method for deep neural networks,'' in \emph{2018 International Joint
  Conference on Neural Networks (IJCNN)}.\hskip 1em plus 0.5em minus
  0.4em\relax IEEE, 2018, pp. 1--8.

\bibitem{lin2017runtime}
J.~Lin, Y.~Rao, J.~Lu, and J.~Zhou, ``Runtime neural pruning,'' in
  \emph{Advances in neural information processing systems}, 2017, pp.
  2181--2191.

\bibitem{wang2019haq}
K.~Wang, Z.~Liu, Y.~Lin, J.~Lin, and S.~Han, ``Haq: Hardware-aware automated
  quantization with mixed precision,'' in \emph{Proceedings of the IEEE
  conference on computer vision and pattern recognition}, 2019, pp. 8612--8620.

\bibitem{jacob2018quantization}
B.~Jacob, S.~Kligys, B.~Chen, M.~Zhu, M.~Tang, A.~Howard, H.~Adam, and
  D.~Kalenichenko, ``Quantization and training of neural networks for efficient
  integer-arithmetic-only inference,'' in \emph{Proceedings of the IEEE
  Conference on Computer Vision and Pattern Recognition}, 2018, pp. 2704--2713.

\bibitem{zhu2016trained}
C.~Zhu, S.~Han, H.~Mao, and W.~J. Dally, ``Trained ternary quantization,''
  \emph{arXiv preprint arXiv:1612.01064}, 2016.

\bibitem{zhang2018lq}
D.~Zhang, J.~Yang, D.~Ye, and G.~Hua, ``Lq-nets: Learned quantization for
  highly accurate and compact deep neural networks,'' in \emph{Proceedings of
  the European conference on computer vision (ECCV)}, 2018, pp. 365--382.

\bibitem{jain2018compensated}
S.~Jain, S.~Venkataramani, V.~Srinivasan, J.~Choi, P.~Chuang, and L.~Chang,
  ``Compensated-dnn: energy efficient low-precision deep neural networks by
  compensating quantization errors,'' in \emph{2018 55th ACM/ESDA/IEEE Design
  Automation Conference (DAC)}.\hskip 1em plus 0.5em minus 0.4em\relax IEEE,
  2018, pp. 1--6.

\bibitem{kumar2017resource}
A.~Kumar, S.~Goyal, and M.~Varma, ``Resource-efficient machine learning in 2 kb
  ram for the internet of things,'' in \emph{International Conference on
  Machine Learning}, 2017, pp. 1935--1944.

\bibitem{gural2019memory}
A.~Gural and B.~Murmann, ``Memory-optimal direct convolutions for maximizing
  classification accuracy in embedded applications.'' in \emph{ICML}, 2019, pp.
  2515--2524.

\bibitem{xia2018automatic}
Y.~Xia, H.~Zhang, L.~Xu, Z.~Gao, H.~Zhang, H.~Liu, and S.~Li, ``An automatic
  cardiac arrhythmia classification system with wearable electrocardiogram,''
  \emph{IEEE Access}, vol.~6, pp. 16\,529--16\,538, 2018.

\bibitem{yildirim2018arrhythmia}
{\"O}.~Y{\i}ld{\i}r{\i}m, P.~P{\l}awiak, R.-S. Tan, and U.~R. Acharya,
  ``Arrhythmia detection using deep convolutional neural network with long
  duration ecg signals,'' \emph{Computers in biology and medicine}, vol. 102,
  pp. 411--420, 2018.

\bibitem{hannun2019cardiologist}
A.~Y. Hannun, P.~Rajpurkar, M.~Haghpanahi, G.~H. Tison, C.~Bourn, M.~P.
  Turakhia, and A.~Y. Ng, ``Cardiologist-level arrhythmia detection and
  classification in ambulatory electrocardiograms using a deep neural
  network,'' \emph{Nature medicine}, vol.~25, no.~1, p.~65, 2019.

\bibitem{hochreiter1997long}
S.~Hochreiter and J.~Schmidhuber, ``Long short-term memory,'' \emph{Neural
  computation}, vol.~9, no.~8, pp. 1735--1780, 1997.

\bibitem{he2016identity}
K.~He, X.~Zhang, S.~Ren, and J.~Sun, ``Identity mappings in deep residual
  networks,'' in \emph{European conference on computer vision}.\hskip 1em plus
  0.5em minus 0.4em\relax Springer, 2016, pp. 630--645.

\bibitem{ioffe2015batch}
S.~Ioffe and C.~Szegedy, ``Batch normalization: Accelerating deep network
  training by reducing internal covariate shift,'' \emph{arXiv preprint
  arXiv:1502.03167}, 2015.

\bibitem{srivastava2014dropout}
N.~Srivastava, G.~Hinton, A.~Krizhevsky, I.~Sutskever, and R.~Salakhutdinov,
  ``Dropout: a simple way to prevent neural networks from overfitting,''
  \emph{The journal of machine learning research}, vol.~15, no.~1, pp.
  1929--1958, 2014.

\bibitem{sastry2005genetic}
K.~Sastry, D.~Goldberg, and G.~Kendall, ``Genetic algorithms,'' in \emph{Search
  methodologies}.\hskip 1em plus 0.5em minus 0.4em\relax Springer, 2005, pp.
  97--125.

\bibitem{glover1998tabu}
F.~Glover and M.~Laguna, ``Tabu search,'' in \emph{Handbook of combinatorial
  optimization}.\hskip 1em plus 0.5em minus 0.4em\relax Springer, 1998, pp.
  2093--2229.

\bibitem{van1987simulated}
P.~J. Van~Laarhoven and E.~H. Aarts, ``Simulated annealing,'' in
  \emph{Simulated annealing: Theory and applications}.\hskip 1em plus 0.5em
  minus 0.4em\relax Springer, 1987, pp. 7--15.

\bibitem{dorigo2006ant}
M.~Dorigo, M.~Birattari, and T.~Stutzle, ``Ant colony optimization,''
  \emph{IEEE computational intelligence magazine}, vol.~1, no.~4, pp. 28--39,
  2006.

\bibitem{reeves2002genetic}
C.~Reeves and J.~E. Rowe, \emph{Genetic algorithms: principles and
  perspectives: a guide to GA theory}.\hskip 1em plus 0.5em minus 0.4em\relax
  Springer Science \& Business Media, 2002, vol.~20.

\bibitem{zames1981genetic}
G.~Zames, N.~Ajlouni, N.~Ajlouni, N.~Ajlouni, J.~Holland, W.~Hills, and
  D.~Goldberg, ``Genetic algorithms in search, optimization and machine
  learning,'' \emph{Information Technology Journal}, vol.~3, no.~1, pp.
  301--302, 1981.

\bibitem{miller1995genetic}
B.~L. Miller, D.~E. Goldberg \emph{et~al.}, ``Genetic algorithms, tournament
  selection, and the effects of noise,'' \emph{Complex systems}, vol.~9, no.~3,
  pp. 193--212, 1995.

\bibitem{deb2002fast}
K.~Deb, A.~Pratap, S.~Agarwal, and T.~Meyarivan, ``A fast and elitist
  multiobjective genetic algorithm: Nsga-ii,'' \emph{IEEE transactions on
  evolutionary computation}, vol.~6, no.~2, pp. 182--197, 2002.

\bibitem{zitzler2001spea}
E.~Zitzler, M.~Laumanns, and L.~Thiele, ``Spea2: Improving the strength pareto
  evolutionary algorithm,'' \emph{TIK-report}, vol. 103, 2001.

\bibitem{han2015learning}
S.~Han, J.~Pool, J.~Tran, and W.~Dally, ``Learning both weights and connections
  for efficient neural network,'' \emph{Advances in neural information
  processing systems}, vol.~28, pp. 1135--1143, 2015.

\bibitem{he2015delving}
K.~He, X.~Zhang, S.~Ren, and J.~Sun, ``Delving deep into rectifiers: Surpassing
  human-level performance on imagenet classification,'' in \emph{Proceedings of
  the IEEE international conference on computer vision}, 2015, pp. 1026--1034.

\bibitem{kingma2014adam}
D.~P. Kingma and J.~Ba, ``Adam: A method for stochastic optimization,''
  \emph{arXiv preprint arXiv:1412.6980}, 2014.

\bibitem{goldberger2000physiobank}
A.~L. Goldberger, L.~A. Amaral, L.~Glass, J.~M. Hausdorff, P.~C. Ivanov, R.~G.
  Mark, J.~E. Mietus, G.~B. Moody, C.-K. Peng, and H.~E. Stanley, ``Physiobank,
  physiotoolkit, and physionet: components of a new research resource for
  complex physiologic signals,'' \emph{circulation}, vol. 101, no.~23, pp.
  e215--e220, 2000.

\bibitem{moody2001impact}
G.~B. Moody and R.~G. Mark, ``The impact of the mit-bih arrhythmia database,''
  \emph{IEEE Engineering in Medicine and Biology Magazine}, vol.~20, no.~3, pp.
  45--50, 2001.

\bibitem{nolle1986crei}
F.~Nolle, F.~Badura, J.~Catlett, R.~Bowser, and M.~Sketch, ``Crei-gard, a new
  concept in computerized arrhythmia monitoring systems,'' \emph{Computers in
  Cardiology}, vol.~13, pp. 515--518, 1986.

\bibitem{rainville2012deap}
F.-M. De~Rainville, F.-A. Fortin, M.-A. Gardner, M.~Parizeau, and C.~Gagn{\'e},
  ``Deap: A python framework for evolutionary algorithms,'' in
  \emph{Proceedings of the 14th annual conference companion on Genetic and
  evolutionary computation}, 2012, pp. 85--92.

\bibitem{li2019random}
L.~Li and A.~Talwalkar, ``Random search and reproducibility for neural
  architecture search,'' in \emph{Uncertainty in Artificial
  Intelligence}.\hskip 1em plus 0.5em minus 0.4em\relax PMLR, 2020, pp.
  367--377.

\bibitem{spackman1989signal}
K.~A. Spackman, ``Signal detection theory: Valuable tools for evaluating
  inductive learning,'' in \emph{Proceedings of the sixth international
  workshop on Machine learning}.\hskip 1em plus 0.5em minus 0.4em\relax
  Elsevier, 1989, pp. 160--163.

\end{thebibliography}

\newpage
\begin{IEEEbiography}[{\includegraphics[width=1in,height=1.25in,clip,keepaspectratio]{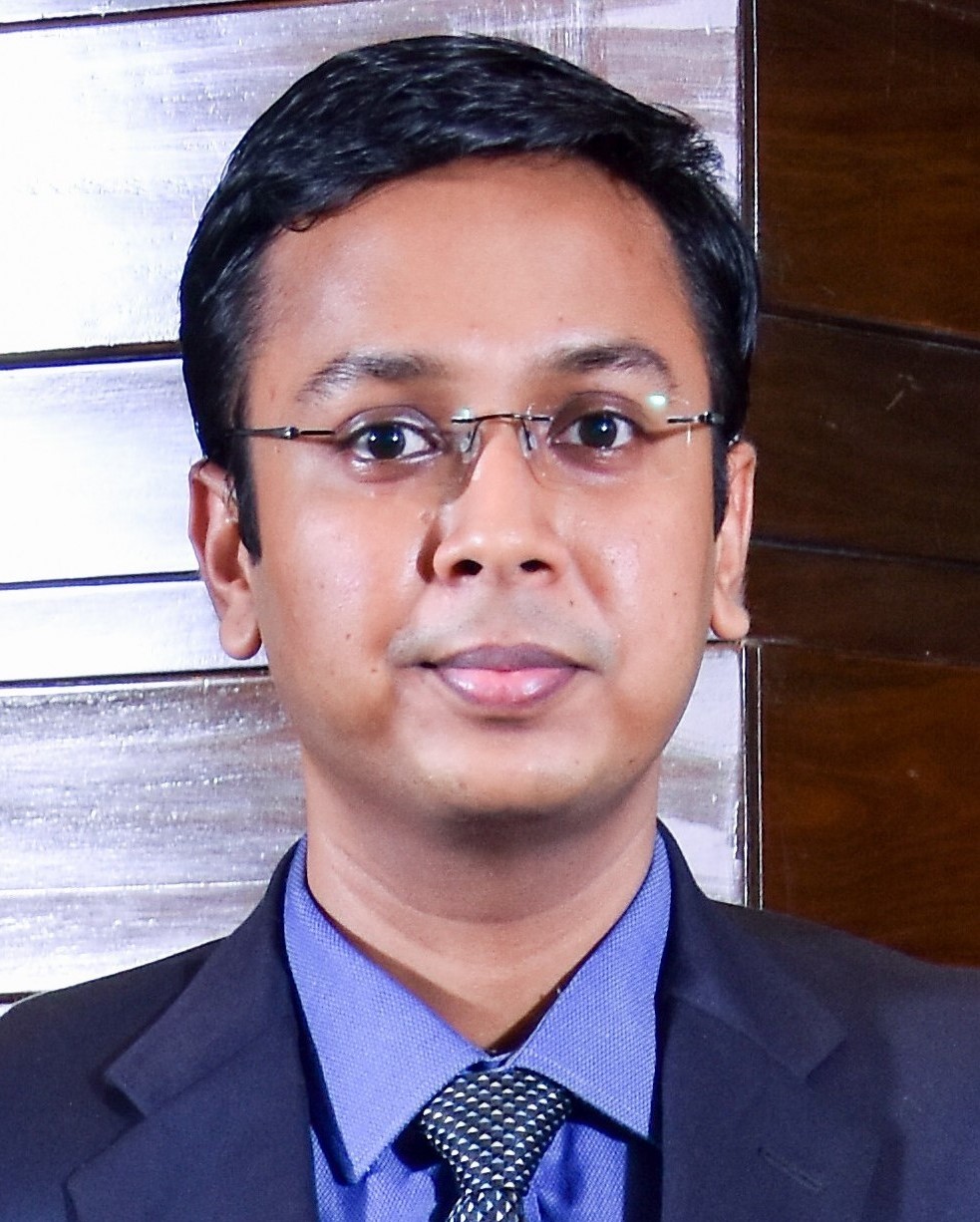}}]{Bharath Srinivas Prabakaran} (S'19) is a PhD Student at Doctoral College - Resilient Embedded Systems, Faculty of Informatics, TU Wien, Austria. 
He is working in the Computer Architecture and Robust Energy-Efficient Technologies (CARE-Tech.) research group, which is led by Prof. Muhammad Shafique.
He graduated with a Bachelor of Engineering in Electrical and Electronics and a Master of Science in Biological Sciences from the Birla Institute of Technology and Science (BITS), Pilani, India in 2017. 
He was as a visiting researcher at TU Dresden, Germany for a span of 1 year from 2016, where he completed his master thesis.
His research interests include fault-tolerant computing, wearable architectures, healthcare systems, energy-efficient technologies, and embedded machine learning.
\end{IEEEbiography}

\begin{IEEEbiography}[{\includegraphics[width=1in,height=1.25in,clip,keepaspectratio]{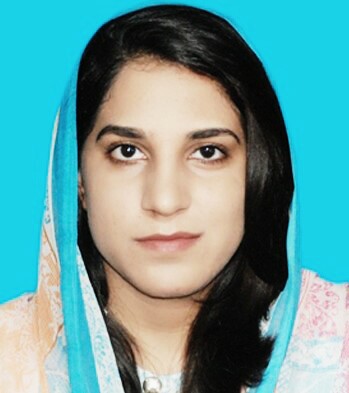}}]{Asima Akhtar} received her Bachelor of Science degree in Electronic Engineering from University of Engineering and Technology (UET)-Taxila, Pakistan, in 2016. 
She is currently completing her Master’s in Electrical Engineering from School of Electrical Engineering and Computer Science, National University of Science and Technology (NUST), Pakistan. 
Her main research interests are Embedded Machine Learning, Neural Architecture Search Applications, and Smart Healthcare Systems. 
She was also a recipient of the prestigious DAC 2020 Young Student Fellow Program.
\end{IEEEbiography}

\begin{IEEEbiography}[{\includegraphics[width=1in,height=1.25in,clip,keepaspectratio]{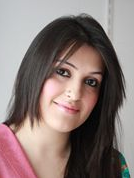}}]{Semeen Rehman} is currently with the Faculty of Electrical Engineering, Technische Universität Wien (TU Wien) as a tenure-track Assistant Professor. 
In October 2020, she received her habilitation in the area of Embedded Systems from TU Wien. 
Before that, she was a Postdoctoral Researcher with Technische Universität Dresden (TU Dresden) and Karlsruhe Institute of Technology (KIT), Germany, since 2015. 
In July 2015, she received her Ph.D. from Karlsruhe Institute of Technology (KIT), Germany. 
She has co-authored one book, multiple book chapters, and more than 50 publications in premier journals and conferences. 
Her main research interests include dependable systems, cross-layer design for error resiliency with a focus on run-time adaptations, emerging computing paradigms, such as approximate computing, hardware security, energy-efficient computing, embedded systems, MPSoCs, Internet of Things, and Cyber Physical Systems. 
She has received the CODES+ISSS 2011 and 2015 Best Paper Awards, DATE 2017 Best Paper Award Nomination, several HiPEAC Paper Awards, Richard Newton Young Student Fellow Award at DAC 2015, and Research Student Award at KIT, in 2012. 
She has served as the TPC track chair for the ISVLSI 2020 conference and on the TPC of multiple premier conferences on design automation and embedded systems (such as DAC, DATE, CASES, ASP-DAC). 
She has (co-)chaired sessions at the DATE 2019, 2018, and 2017 conferences.
\end{IEEEbiography}

\begin{IEEEbiography}[{\includegraphics[width=1in,height=1.25in,clip,keepaspectratio]{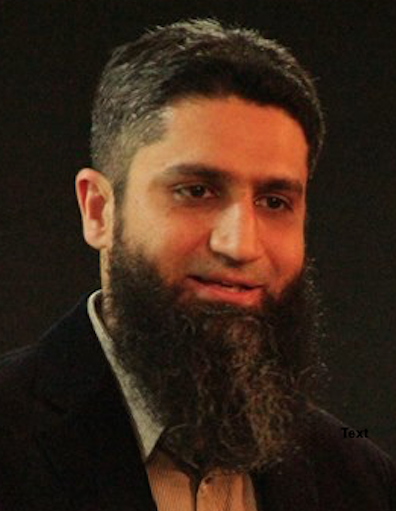}}]{Osman Hasan} received his BEng (Hons) degree from the University of Engineering and Technology, Peshawar Pakistan in 1997, and the MEng and PhD degrees from Concordia University, Montreal, Quebec, Canada in 2001 and 2008, respectively. 
Before his PhD, he worked as an ASIC Design Engineer from 2001 to 2004 at LSI Logic. He worked as a postdoctoral fellow at the Hardware Verification Group (HVG) of Concordia University for one year until August 2009. 
Currently, he is an Associate Professor and the Dean of the School of Electrical Engineering and Computer Science of National University of Science and Technology (NUST), Islamabad, Pakistan. 
He is the founder and director of System Analysis and Verification (SAVe) Lab at NUST, which mainly focuses on the design and formal verification of energy, embedded and e-health related systems. 
He has received several awards and distinctions, including the Pakistan’s Higher Education Commission’s Best University Teacher (2010) and Best Young Researcher Award (2011) and the President’s gold medal for the best teacher of the University from NUST in 2015. 
Dr. Hasan is a senior member of IEEE, member of the ACM, Association for Automated Reasoning (AAR) and the Pakistan Engineering Council. 
\end{IEEEbiography}

\begin{IEEEbiography}[{\includegraphics[width=1in,height=1.25in,clip,keepaspectratio]{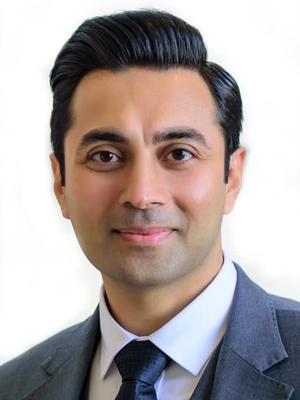}}]{Muhammad Shafique} (M'11 - SM'16) received the Ph.D. degree in computer science from the Karlsruhe Institute of Technology (KIT), Germany, in 2011. 
Afterwards, he established and led a highly recognized research group at KIT for several years as well as conducted impactful collaborative R\&D activities across the globe. 
In Oct.2016, he joined the Institute of Computer Engineering at the Faculty of Informatics, Technische Universität Wien (TU Wien), Vienna, Austria as a Full Professor of Computer Architecture and Robust, Energy-Efficient Technologies. 
Since Sep.2020, he is with the Division of Engineering, New York University Abu Dhabi (NYU-AD), United Arab Emirates, and is a Global Network faculty at the NYU Tandon School of Engineering, USA. 
His research interests are in design automation and system level design for brain-inspired computing, AI \& machine learning hardware, wearable healthcare devices and systems, autonomous systems, energy-efficient systems, robust computing, hardware security, emerging technologies, FPGAs, MPSoCs, and embedded systems. 
His research has a special focus on cross-layer analysis, modeling, design, and optimization of computing and memory systems. 
The researched technologies and tools are deployed in application use cases from Internet-of-Things (IoT), smart Cyber-Physical Systems (CPS), and ICT for Development (ICT4D) domains. 
Dr. Shafique has given several Keynotes, Invited Talks, and Tutorials, as well as organized many special sessions at premier venues. 
He has served as the PC Chair, General Chair, Track Chair, and PC member for several prestigious IEEE/ACM conferences. 
Dr. Shafique holds one U.S. patent has (co-)authored 6 Books, 10+ Book Chapters, and over 300 papers in premier journals and conferences. 
He received the 2015 ACM/SIGDA Outstanding New Faculty Award, AI 2000 Chip Technology Most Influential Scholar Award in 2020, six gold medals, and several best paper awards and nominations at prestigious conferences.
\end{IEEEbiography}

\end{document}